\documentclass[useAMS,usenatbib]{mnras}

\usepackage{natbib}
\usepackage{tabu,booktabs}
\usepackage{amsmath}
\usepackage{xcolor}
\usepackage[english]{babel}
\usepackage{graphicx}
\usepackage{xspace}

\usepackage{longtable}
\usepackage{threeparttablex}
\usepackage{array}
\usepackage{enumitem}

\def \etal {et~al.~}

\newcommand{\hMpc}{{\ifmmode{h^{-1}{\rm Mpc}}\else{$h^{-1}$Mpc}\fi}}
\newcommand{\Mpc}{{\ifmmode{{\rm Mpc}}\else{Mpc}\fi}}
\newcommand{\hkpc}{{\ifmmode{h^{-1}{\rm kpc}}\else{$h^{-1}$kpc}\fi}}
\newcommand{\kpc}{{\ifmmode{ {\rm kpc} }\else{{\rm kpc}}\fi}}
\newcommand{\kms}{{\ifmmode{ {\rm km\,s^{-1}} }\else{ ${\rm km\,s^{-1}}$ }\fi}}
\newcommand{\hMsun}{{\ifmmode{h^{-1}{\rm {M_{\odot}}}}\else{$h^{-1}{\rm{M_{\odot}}}$}\fi}}
\newcommand{\Msun}{{\ifmmode{{\rm M}_{\odot}}\else{${\rm M}_{\odot}$}\fi}}
\newcommand{\Mhalo}{{\ifmmode{M_{\rm halo}}\else{$M_{\rm halo}$}\fi}}
\newcommand{\Rvir}{{\ifmmode{R_{\rm vir}}\else{$R_{\rm vir}$}\fi}}
\newcommand{\Mvir}{{\ifmmode{M_{\rm vir}}\else{$M_{\rm vir}$}\fi}}
\newcommand{\Mstar}{{\ifmmode{M_{\rm star}}\else{$M_{\rm star}$}\fi}}
\newcommand{\Vrot}{{\ifmmode{V_{\rm rot}}\else{$V_{\rm rot}$}\fi}}
\newcommand{\ltsima}{$\; \buildrel < \over \sim \;$}
\newcommand{\gtsima}{$\; \buildrel > \over \sim \;$}
\newcommand{\lsim}{\lower.5ex\hbox{\ltsima}}
\newcommand{\gsim}{\lower.5ex\hbox{\gtsima}}

\def\lesssim{\mathrel{\hbox{\rlap{\hbox{\lower4pt\hbox{$\sim$}}}\hbox{$<$}}}}
\def\gtrsim{\mathrel{\hbox{\rlap{\hbox{\lower4pt\hbox{$\sim$}}}\hbox{$>$}}}}

\newcommand{\beq}{\begin{equation}}
\newcommand{\eeq}{\end{equation}}
\def\beqa{\begin{eqnarray}}
\def\eeqa{\end{eqnarray}}
\def\LCDM{\ensuremath{\Lambda}CDM}

\def\head{ \vbox to 0pt{\vss \hbox to 0pt{\hskip 440pt\rm
      LA-UR-10-07069\hss} \vskip 25pt}}

\def \kms {\ifmmode  \,\rm km\,s^{-1} \else $\,\rm km\,s^{-1}  $ \fi }
\def \kpc {\ifmmode  {\rm kpc}  \else ${\rm  kpc}$ \fi  }  
\def \hkpc {\ifmmode  {h^{-1}\rm kpc}  \else ${h^{-1}\rm kpc}$ \fi  }  
\def \hMpc {\ifmmode  {h^{-1}\rm Mpc}  \else ${h^{-1}\rm Mpc}$ \fi  }  
\def \Mpch {\ifmmode  {h^{-1}\rm Mpc}  \else ${h^{-1}\rm Mpc}$ \fi  }  
\def \Msun {\ifmmode {\rm M}_{\odot} \else ${\rm M}_{\odot}$ \fi} 
\def \hMsun {\ifmmode h^{-1}\,\rm M_{\odot} \else $h^{-1}\,\rm M_{\odot}$ \fi}

\def \LCDM {\ifmmode \Lambda{\rm CDM} \else $\Lambda{\rm CDM}$ \fi}
\def \sig8 {\ifmmode \sigma_8 \else $\sigma_8$ \fi} 
\def \OmegaM {\ifmmode \Omega_{\rm m} \else $\Omega_{\rm m}$ \fi} 
\def \Omegab {\ifmmode \Omega_{\rm b} \else $\Omega_{\rm b}$ \fi} 
\def \OmegaL {\ifmmode \Omega_{\rm \Lambda} \else $\Omega_{\rm \Lambda}$\fi} 
\def \Deltavir {\ifmmode \Delta_{\rm vir} \else $\Delta_{\rm vir}$ \fi}
\def \rhocrit {\ifmmode \rho_{\rm crit} \else $\rho_{\rm crit}$ \fi}
\def \rhou {\ifmmode \rho_{\rm u} \else $\rho_{\rm u}$ \fi}
\def \zc {\ifmmode z_{\rm c} \else $z_{\rm c}$ \fi}

\def\lcdm{\ensuremath{\Lambda\textrm{CDM}}\xspace}

\def\aj{AJ}%
\def\apj{ApJ}%
\def\apjl{ApJ}%
\def\apjs{ApJS}%
\def\ao{Appl.~Opt.}%
\def\aap{A\&A}%
\def\mnras{MNRAS}%
\def\na{New A}%
\def\pasa{PASA}%
\def\pasp{PASP}%
\def\nat{Nature}%
\def\head{ .ps \vbox to 0pt{\vss \hbox to 0pt{\hskip 440pt\rm
      LA-UR-10-07069\hss} \vskip 25pt}} 

\def \spose#1{\hbox  to 0pt{#1\hss}}  
\def \lta{\mathrel{\spose{\lower 3pt\hbox{$\sim$}}\raise 2.0pt\hbox{$<$}}}
\def \gta{\mathrel{\spose{\lower 3pt\hbox{$\sim$}}\raise 2.0pt\hbox{$>$}}}

\def\lcdm{\ensuremath{\Lambda\textrm{CDM}}\xspace}

\title[Simulated Milky Way Satellites]{NIHAO XV: The environmental impact of the host galaxy on galactic satellite and field dwarf galaxies}

\author[T. Buck \etal] {Tobias Buck$^{1}$\thanks{E-mail:
    buck@mpia.de}\thanks{Member of the International Max Planck Research School for Astronomy and Cosmic Physics at the University of Heidelberg, IMPRS-HD, Germany.},
    Andrea V. Macci\`o$^{2,1}$, Aaron A. Dutton$^{2}$, Aura Obreja$^{3}$, \newauthor{Jonas Frings$^{1,4}$}\\
$^1$Max-Planck-Institut f\"ur Astronomie, K\"onigstuhl 17, 69117 Heidelberg, Germany\\
$^2$New York University Abu Dhabi, PO Box 129188, Saadiyat Island, Abu Dhabi, United Arab Emirates\\
$^3$Universit\"ats-Sternwarte M\"unchen, Scheinerstraße 1, D-81679 M\"unchen, Germany\\
$^4$Astronomisches Recheninstitut, Zentrum f\"ur Astronomie der Universit\"at Heidelberg, Philosophenweg 12, D-69120 Heidelberg, Germany
}

\begin{document}

\date{Accepted 2018 October 25. Received 2018 August 31; in original form 2018 March 23}

\pagerange{\pageref{firstpage}--\pageref{lastpage}} \pubyear{2018}

\maketitle

\label{firstpage}

\begin{abstract}
We study the impact of the host on dwarf galaxy properties using four new Milky Way-like, ultra high-resolution simulations, ($N_{part} >10^7$) from the NIHAO project. 
We split our sample in satellite ($R<R_{200}$), nearby ($1<R/R_{200}<2.5$), and field ($R>2.5 R_{\rm 200}$) galaxies.
Simulated galaxies from all three groups are in excellent agreement with Local Group dwarf galaxies in terms of: stellar mass-velocity dispersion, stellar mass-metallicity relation, star formation histories, and stellar mass functions.
Satellites and nearby galaxies show lower velocity dispersions and gas fractions compared to field galaxies. While field galaxies follow global abundance matching relations, satellites and nearby galaxies deviate from them, showing lower dark matter masses for given stellar mass.
The reason for this deficit in dark matter mass is substantial mass loss experienced by satellites and $\sim80$\% of the nearby galaxies, while orbiting inside $R_{200}$ at earlier times. 
However, both satellites and nearby objects fall back onto the relation for field galaxies if we use the maximum of their virial mass instead of the present-day value. This allows us to provide estimates for the peak masses of observed Local Group galaxies.
Finally, using radial velocities, distances, and the velocity dispersion-stellar mass relation from our simulations, we derive a metric to distinguish between galaxies harassed by the central object and unaffected ones.
Applying this metric to observed objects we find that even far away dwarf galaxies like Eri II ($D\approx$ 370 kpc) have a strong probability ($\approx 83$\%) of having been affected by the Milky Way in the past. This naturally explains the lack of gas and recent star formation seen in Eri II.
\end{abstract}

\noindent
\begin{keywords}

Galaxy: formation --- galaxies: Local Group --- galaxies: dwarfs --- galaxies: kinematics and dynamics --- dark matter --- methods: numerical

 \end{keywords}


\section{Introduction} \label{sec:introduction}

\begin{table*}
\label{tab:res}
\begin{center}
\caption{Simulation parameters for the dark matter, gas and stellar particles in our simulation runs.}
\begin{tabular}{c c c c c}
		\hline\hline
		property & particle mass & Force soft. & Smoothing length & particles within $R_{200}$\\
		  & [$10^5 \Msun$] & [pc] & (median, min.) [pc] & \\
		\hline
        \multicolumn{5}{c}{g2.79e12}\\
        \hline
		DM & 5.141 & 620 & - & 5,413,017\\
		GAS & 0.938 & 265 & (155, 20) & 2,224,231\\
		STARS & 0.313 & 265 & - & 8,249,934\\
        \hline
        \multicolumn{5}{c}{g1.12e12}\\
        \hline
        DM & 1.523 & 414 & - & 7,454,484\\
		GAS & 0.278 & 177 & (200, 11) & 3,041,329\\
		STARS & 0.093 & 177 & - & 11,087,938\\
		\hline
        \multicolumn{5}{c}{g8.26e11}\\
        \hline
        DM & 2.168 & 466 & - & 3,742,713\\
		GAS & 0.396 & 199 & (194, 32) & 1,665,173\\
		STARS & 0.132 & 199 & - & 4,171,413\\
        \hline
        \multicolumn{5}{c}{g7.08e11}\\
        \hline
        DM & 1.11 & 373 & - & 4,439,242\\
		GAS & 0.203 & 159 & (117, 28) & 2,004,656\\
		STARS & 0.068 & 159 & - & 4,861,964\\
        \hline
\end{tabular}
\end{center}
\end{table*}

\begin{table*}
\label{tab:props}
\begin{center}
\caption{Simulation properties of the main galaxies: \normalfont{We state the total virial mass $M_{200}$, the virial radius $R_{200}$ and the total stellar mass, $M_{\rm star}$, as well as the total amount of gas, $M_{\rm gas}$, within the virial radius. The last column indicates the numbers of \textit{satellites} (luminous sub-halos within $R_{200}$), \textit{nearby dwarfs} (luminous halos within one to $2.5$ virial radii) and \textit{field dwarfs} which are defined as luminous halos outside $2.5$ virial radii.}}
\begin{tabular}{c c c c c c c c}
		\hline\hline
		simulation & $M_{200}$ & $R_{200}$ & $M_{\rm star}$ & $M_{\rm gas}$ & $N_{\rm sat}$ & $N_{\rm nearby}$ & $N_{\rm field}$ \\
		  & [$10^{12}\Msun$] & [kpc] & [$10^{10}\Msun$] & [$10^{10}\Msun$] & $r<R_{200}$ & $R_{200}<r<2.5R_{200}$ & $2.5R_{200}<r<3$ Mpc \\
		\hline
		g2.79e12 & 3.13 & 306 & 15.9 & 18.48 & 20&16&26\\
		g1.12e12 & 1.28 & 234 & 6.32 & 7.93 & 8&13&6\\
		g8.26e11 & 0.91 & 206 & 3.40 & 6.09 & 7&7&13\\
        g7.08e11 & 0.55 & 174 & 2.00 & 3.74 & 8&6&24\\
        \hline
\end{tabular}
\end{center}
\end{table*}

\begin{figure*}
\begin{center}
\includegraphics[width=\textwidth]{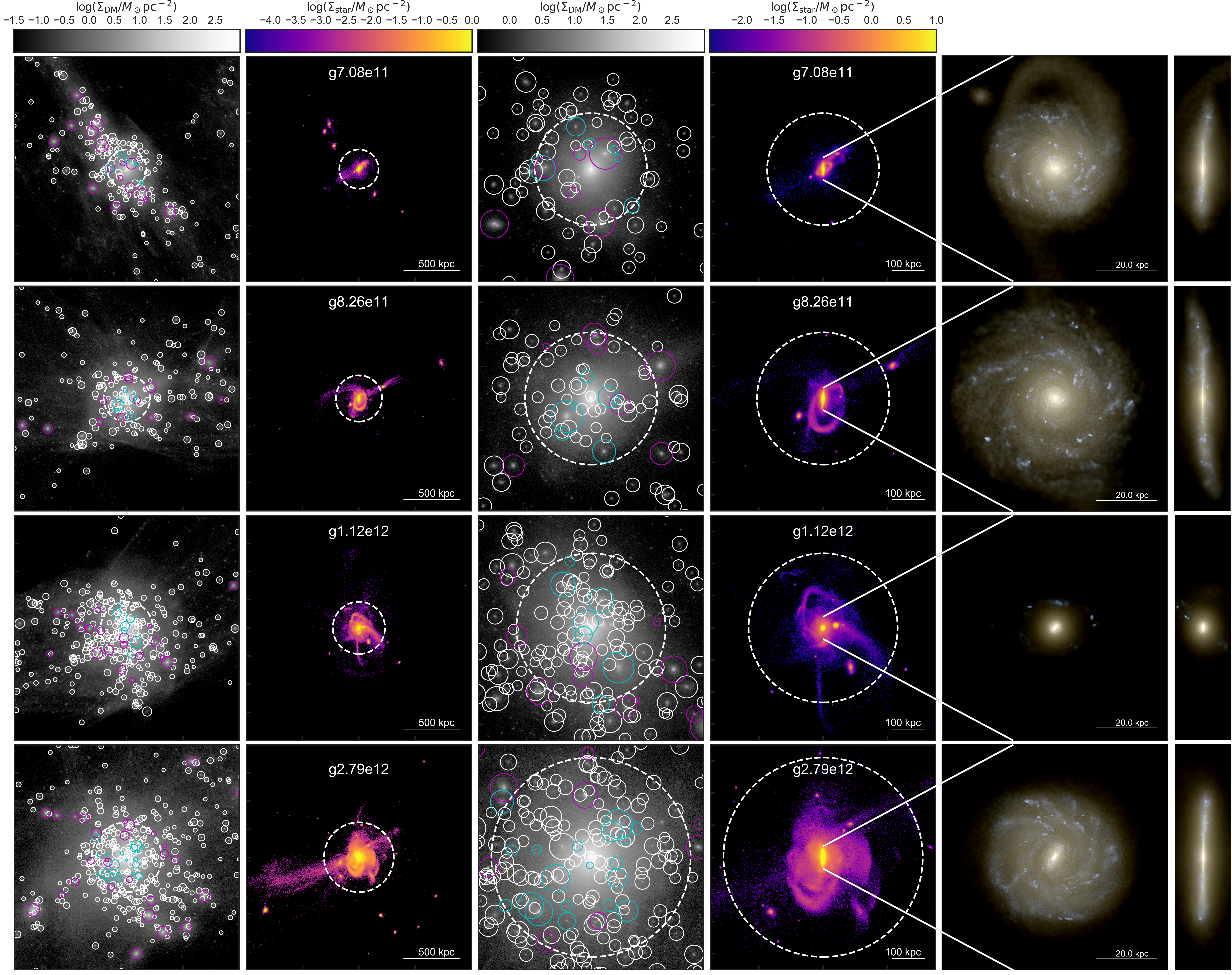}
\end{center}
\vspace{-.35cm}
\caption{Projections of all four host haloes and their surrounding low mass haloes at redshift zero. The two left most columns show a view of 2 Mpc side length; the first column shows the dark matter surface density, the second column shows the corresponding stellar surface density. The third and fourth column show the same but for a close-in view that is 700 kpc across. The depth of each projection is equal to the width. In each panel, a white dashed circle shows the virial radius $R_{200}$. The projections are such that the disk of the host halo is viewed edge-on. We circle every subhalo with a total mass greater than $10^8\Msun$. Luminous subhaloes are circled in color: cyan for subhaloes within the host's virial radius, and magenta for subhaloes beyond that\protect\footnotemark[1]. We chose the diameter of each subhalo circle to be proportional to the logarithm of the subhalo’s mass. The dark matter surface density projections reveal that there is a large number of low mass haloes out to large distances ($\gsim$1 Mpc) where many of these small haloes are completely devoid stars. The two right most panels show an RGB image of the main galaxy in face-on and edge-on views where we render the stellar particles using their luminosity in the wavelength bands $i,v,u$ to create the r, g and b channels.}
\label{fig:impression}
\end{figure*}

The standard cosmological model of  Dark Energy
\citep[$\Lambda$,][]{Riess1998,Perlmutter1999} and Cold Dark Matter \citep[CDM,][]{Peebles1984},
is in outstanding
accordance with observations on large ($>$Mpc) scales
\citep{Tegmark2004,Springel2005}. This model provides a successful
framework to describe the formation and evolution of dark matter
haloes, which are the hosts of galaxies.
Given a set of cosmological parameters \citep[e.g.][]{Planck}, it makes clear predictions for the
number densities of these haloes
\citep{Jenkins2001,Tinker2008,Angulo2012}.  For Milky Way mass haloes
and above, the predictions are in good agreement with observations,
while for dwarf galaxies the theory nominally predicts too many luminous
field and sub-haloes
\citep[e.g.][]{Klypin1999,Moore1999a,Trujillo2011,Papastergis2012}. Solutions
to these problems involve the suppression of gas cooling in low mass
haloes \citep[e.g.][]{Bullock2000}, and the disordered nature of dwarf
galaxy kinematics \citep{Maccio2016}.  The dynamics of dwarf galaxies
around the host, showing nearly planar motions, also presents an
apparent inconsistency with LCDM \citep{Pawlowski2012a,Ibata2013}, but
see \citep{Cautun2015,Buck2015,Buck2016} for possible resolutions to this
puzzle.

The theory of structure formation in the \lcdm cosmology predicts
hierarchical merging of small dark matter haloes and their
galaxies. Thus a galaxy like our Milky Way was formed by accreting a large
number of small galaxies. Depending on the masses of the accreted dark
matter haloes they either merge quickly with the main galaxy or they
spend a long time orbiting around it
\citep{Chandrasekhar1943,Binney2008,Boylan2008,Jiang2008}. While
orbiting the main galaxy these haloes and the galaxies they
contain experience strong tidal forces eventually leading to (dark
matter) mass loss
\citep{Mayer2001,Klimentowski2007,Penarrubia2008,Kazantzidis2011,Chang2013,Frings2017}. Additionally, due to the hot gas halo of the host ram pressure stripping is acting
on their gas content \citep{Einasto1974,Faber1983,Grebel2003} removing
additional gaseous mass. 

More and more observations have confirmed the scenario of tidal
stripping which in simulations results in streams and complex stellar
structures such as shells and cusps (see also Fig. 1 for some of these
features). Indeed, such structures are observed in our own Galaxy
\citep{Ibata1994,Yanny2000,
  Newberg2002,Majewski2003,Martin2004,Martinez2005,Belokurov2006}, in
the Andromeda galaxy \citep{Ferguson2005,Kalirai2006,Ibata2007} and
beyond the Local Group in more distant galaxies
\citep{Malin1997,Forbes2003,Pohlen2004,Martinez2012,Martinez2015}. Furthermore,
the origin of the stellar halo of galaxies such as our MW is thought
to be the unbound stars from disrupted satellite galaxies
\citep{Bullock2005,Bell2008,Carollo2010,Cooper2010}.

Several authors have studied the environmental effects on the
properties of the local dwarf spheroidal galaxies and found severe effects \citep[e.g.][]{Kravtsov2004, Mayer2006, Mayer2007,  Klimentowski2009,Klimentowski2009a,Brooks2014,Hirschmann2014,Wetzel2015,Kazantzidis2017,Simpson2017,Frings2017}. In the Local Group the dwarf galaxies within the virial radius of MW or M31 ($\sim300$ kpc)
show a sharp evolution in their morphology and properties compared to
dwarf galaxies outside that radius. Crossing the virial radius inwards, galaxies:
\begin{itemize}[leftmargin=*,labelindent=5pt]
\itemsep-.15em 
\item decrease their molecular gas fraction (sometimes to zero)
\item become more spheroidal (morphologically)
\item become quiescent as a consequence of the drop in the molecular gas fraction
\end{itemize}
\citep[e.g.][]{McConnachie2012}.
There are only a few exceptions to
this behavior. Outside the virial radii of MW and M31 there exist
only few quiescent and gas-poor spheroidal galaxies like Cetus,
Tucana, KKR 25, KKs 3 \citep{Karachentsev2015} and Andromeda XVIII.
\cite{Teyssier2012} argued via the distance and radial velocities
of Tucana and Cetus that they have been inside the virial radius of MW
at earlier times. However, KKR 25 and KKs 3 are located much further away at
$\sim2$ Mpc and thus are not likely to have orbited inside the virial
radius of any host. On the other hand, the only gas-rich, star forming
irregular galaxies within the virial radius of MW are LMC and SMC and
of M31 are LGS 3 and IC 10 which are all most probably on their first
infall \citep{Besla2007}.

\footnotetext[1]{The two luminous satellites of galaxy g7.08e11 and g1.12e12 not circled in cyan can not be identified by AHF in the redshift $z=0$ snapshot. We include data for these two satellites from a snapshot $70$ Myr earlier.}

Except for the above mentioned examples most local dwarf galaxies
outside the virial radius of the MW or M31 are assumed to be similar
to field objects, taken to be representative of unaffected
objects. However, many others before \citep{Moore2004,Gill2005,Warnick2008,Ludlow2009,Knebe2011,Wetzel2014,Behroozi2014,Barber2014,Diemer2017} have noted, that there exists a prominent population of dwarf galaxies, termed ``backsplash" galaxies, that are located outside the virial radius of their host at the present day, but their orbits took them inside it at earlier times. 
\citet[][]{More2015} showed that for
dark matter haloes the region in which these ``backsplash" galaxies
can be found is $\sim2.5R_{\rm 200}$. This finding renders the
traditional virial radius meaningless as a physical boundary between
satellite and field galaxies.

Since ``backsplash" galaxies had, at one point in their lives, a close encounter with their host galaxy, the tidal influence of the host must
have been stronger for them compared to field galaxies. What is the
effect of this difference in tidal forces on the properties of the
galaxies, and can we discriminate between the unaffected field objects
and ``backsplash" galaxies? \citet[e.g.][]{Gill2005} showed that
``backsplash" galaxies lose about 40 per cent of their initial mass when
brushing their host galaxy and \cite{Knebe2011} found that their
total mass-to-light ratio is lower than for field
galaxies, but higher than for bound satellites. This raises the
question whether or not the observed sample of field galaxies consist
(at least partly) of these ``backsplash" galaxies and how one would distinguish them from unaffected dwarf
galaxies.

In this work we elaborate on these previous efforts and
investigate the effects of the host on its satellites and nearby dwarf
galaxies. Quantifying the effects of tidal stripping of satellites and
backsplash galaxies is crucial to understand galaxy evolution in
general and observations of Local Group dwarf galaxies in
particular. Thus, we will study how the dark and luminous components
of the galaxies are effected, what influence can potentially be
measured. Finally, we try to separate observed Local Group galaxies
into unaffected galaxies and potentially disturbed ones.

This paper is organized as follows: We start by introducing the set of
simulations in \S2 and discuss our host and sub-halo selection in \S3
where we also define three spatially different populations of dwarf
galaxies. In \S4 we present the properties of our satellite and dwarf galaxies in comparison to observations. We focus on star formation histories, gas fractions, the mass-metallicity relation and the stellar mass-velocity dispersion relation. We then move on to assess the structural effects of the host on its satellites and dwarfs in \S5 where we put our simulated galaxies onto the abundance matching relation and quantify the amount of mass lost from their dark matter haloes. We use our findings from this section to make predictions for the peak masses of Local Group galaxies. Finally, in \S6 we discuss our results and show that there exists a tidally affected population of dwarf galaxies outside the virial radius of the Milky Way and make an attempt to separate tidally effected galaxies
from pure field objects. Further, we give probabilities for Local Group field galaxies to be affected by their host. We conclude the paper with a final summary in \S7.  

\section{Cosmological Zoom-in Simulations} \label{sec:simulation}

The four simulations analyzed in this work, g2.79e12 (M31 analogue), g8.26e11, g7.08e11 (MW analogues) and g1.12e12 are higher-resolution versions of MW mass galaxies taken from the Numerical Investigation of a Hundred Astronomical Objects (NIHAO) simulation suite \citep{Wang2015}. NIHAO galaxies have been proven to match remarkably well many of the properties of observed galaxies, as for example the local velocity function \citep{Maccio2016}, distribution of metals in the Circum
Galactic Medium \citep{Gutcke2016}, Tully-Fisher relations \citep{Dutton2017}, and the structure of stellar \citep{Obreja2016} and gaseous discs \citep{Buck2017}. The g2.79e12 simulation has recently been used to study the build-up of a boxy/peanut shaped bulge in a cosmological framework \citep[][Buck et al. 2018 in prep.]{Buck2018}. 
Most notably for this work, as a result of dark halo expansion, the dwarf galaxies in the NIHAO sample have been shown to resolve the so-called Too-big-to-fail problem for field dwarfs \citep{Dutton2016}, and successfully reproduce  the diversity in
rotation curves of observed dwarf galaxies \citep{Santos-Santos2017}. This means the employed feedback scheme of NIHAO has been proven to create realistic galaxies and high-resolution NIHAO galaxies are well suited to study satellite systems similar to the Milky Way or Andromeda. An impression of the galaxies and their surroundings used in this paper is given in figure \ref{fig:impression}.

In order to resolve the satellite system of these galaxies we
increased the mass resolution by a factor of 8 to 16 compared to the standard NIHAO resolution (see table \ref{tab:res}) such that it ranges between $m_{\rm dark}\sim1.1 - 5.1\times10^5 \Msun$ for dark matter particles and $m_{\rm gas}\sim2.0 - 9.4\times10^4 \Msun$ for the  gas particles. Stellar particles are born with an intial mass of
$1/3\times m_{\rm{gas}}$ and then lose mass according to the stellar
evolution models applied. The corresponding force softenings are
$\epsilon_{\rm dark}=373 - 620$ pc for the dark matter particles and
$\epsilon_{\rm gas}=160 - 265$ pc for the gas and star particles (see
also table 1). However, the smoothing length of the gas
particles can be much smaller, e.g.  as  low  as  $h_{\rm  smooth}\sim
20$ pc in the disc mid-plane.

The galaxies have been run using cosmological parameters from the
\cite{Planck}, namely: \OmegaM=0.3175,   \OmegaL=0.6825,
\Omegab=0.049,  H${_0}$   =   67.1\kms\Mpc$^{-1}$, \sig8  =
0.8344. Initial conditions are created the same way as for the
original NIHAO runs \citep[see][]{Wang2015} using a modified version
of the \texttt{GRAFIC2} package \citep{Bertschinger2001,
  Penzo2014}. The hydrodynamics, star formation recipe and feedback
schemes exploited are the same as for the original NIHAO runs and are
shortly summarized below.

\subsection{Hydrodynamics}

\begin{figure}
\includegraphics[width=\columnwidth]{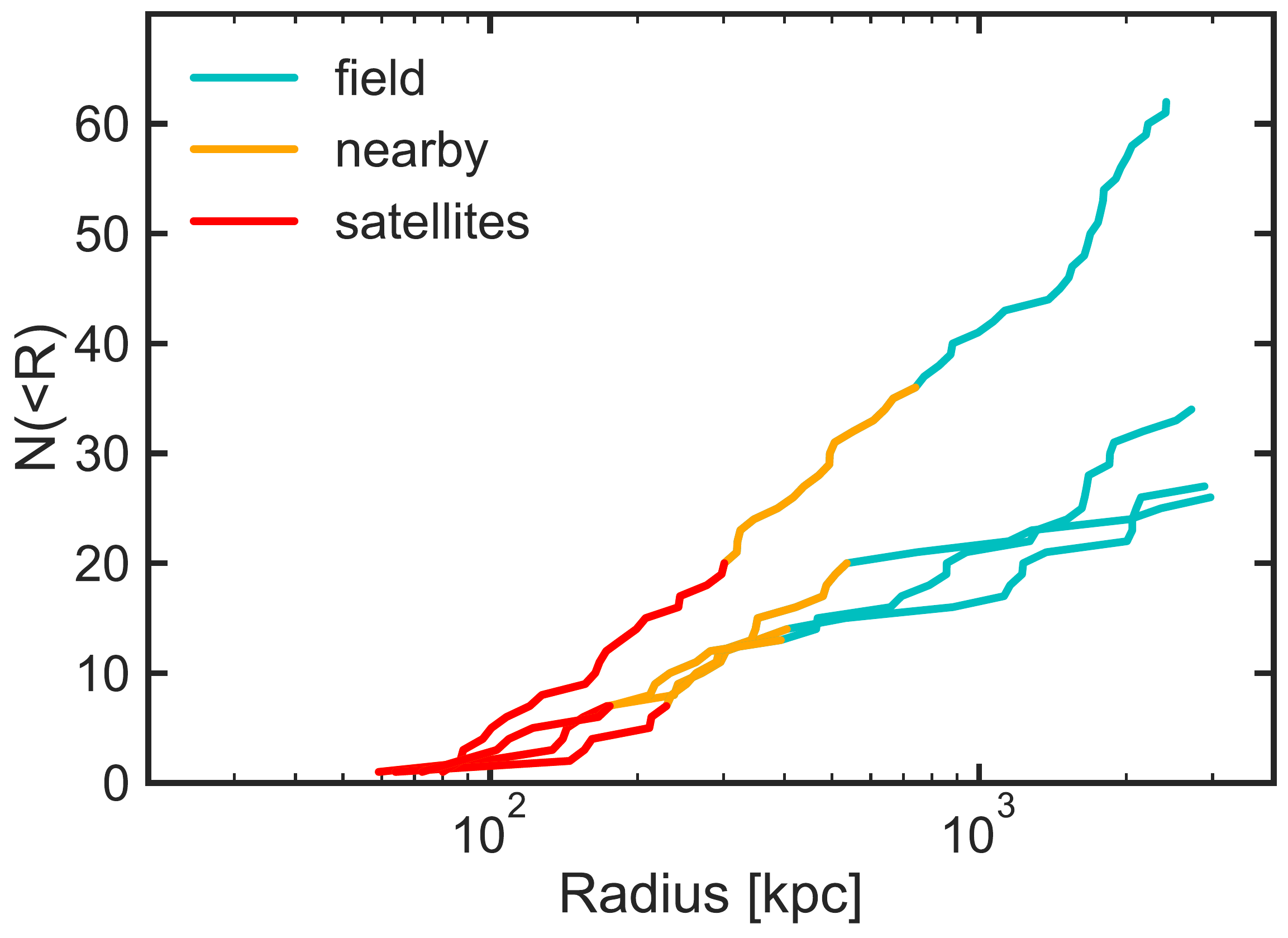}
\vspace{-.35cm}
\caption{Radial profiles of {\it{satellites}} and {\it{nearby}} and
  {\it{field}} galaxies in the simulations. The colored lines show the
  profiles for the three simulations where red color denotes the
  {\it{satellites}} within the virial radius\protect\footnotemark[2], orange color the
  {\it{nearby}} galaxies in the region $1\rm{R}_{\rm
    200}<R<2.5\rm{R}_{\rm 200}$ and cyan color marks the {\it{field}}
  galaxies with distances between $2.5\rm{R}_{\rm 200}<R<3$ Mpc from
  the host.}
\label{fig:rad_dist}
\end{figure}

The high-resolution simulations were run with a modified version of the smoothed particle hydrodynamics (SPH) solver {\texttt{GASOLINE2}} \citep{Wadsley2004,Wadsley2017} with substantial updates to the hydrodynamics as described in \citep{Keller2014}. We will refer to this version of {\texttt{GASOLINE2}} as \texttt{ESF-GASOLINE2}. The  modifications of the hydrodynamics improve multi-phase mixing and remove spurious numerical surface tension by calculating $P/\rho^2$ as a geometrical average over the particles in the smoothing  kernel as proposed  by \cite{Ritchie2001}. The treatment of artificial viscosity has  been modified to use the signal velocity as described  in \cite{Price2008} and the \cite{Saitoh2009} timestep limiter was implemented  so that cool particles behave correctly when a hot blastwave hits them. All runs of \texttt{ESF-GASOLINE2} use the Wendland C2 smoothing kernel \citep{Dehnen2012} to avoid pairing instabilities and a number of 50 neighbor particles  used  in  the  calculation  of  the  smoothed  hydrodynamic properties was employed. 

We adopted a metal diffusion algorithm between particles as described in \cite{Wadsley2008}. Cooling via hydrogen, helium, and various  metal-lines is included as described in \cite{Shen2010} and was calculated using \texttt{cloudy} \citep[version 07.02;][]{Ferland1998}. These  calculations include photo ionization and heating from  the \cite{Haardt2005} UV background  and Compton cooling in a temperature range from 10 to $10^9$ K.

\footnotetext[2]{This figure is mainly to demonstrate the separation into the three populations thus we do not include the two satellites not appearing in the redshift $z=0$ halo catalogue here.}

\subsection{Star Formation and Feedback}
The  simulation  employ the star formation recipe as described in \cite{Stinson2006} which is summarized below. Gas is eligible to form stars when it is dense ($n_{\rm  th}  >  10.3$cm$^{-3}$) and cold (T $< 15,000$K) such that the Kennicutt-Schmidt Law is reproduced.  The threshold number  density $n_{\rm th}$ of gas is set to the maximum density at which gravitational instabilities can be resolved in  the  simulation:  n$_{\rm th}$  $=$  50$m_{\rm  gas}/\epsilon_{\rm gas}^3 = 10.3$ cm$^{-3}$, where $m_{\rm gas}$ denotes the gas particle mass and $\epsilon_{\rm gas}$ the gravitational  softening of the gas and the value of 50 denotes the number of neighboring particles.

Two modes of stellar feedback are implemented as described in \cite{Stinson2013}. The  first  mode  models the energy  input from  stellar  winds and  photo ionization from  luminous young  stars. This mode happens  before any supernovae explode and consists of the total stellar flux,  $2 \times  10^{50}$  erg of  thermal energy  per $M_{\odot}$ of the entire stellar population. The efficiency parameter for coupling the energy input is set to $\epsilon_{\rm ESF}=13\%$ \citep{Wang2015}.

The second mode models the energy input via supernovae and starts 4 Myr after the formation of the star particle. It is implemented using the blastwave formalism as described in \cite{Stinson2006} and applies a delayed cooling formalism for particles inside  the blast region. The reason for this is that in the simulations the interstellar gas surrounding the region of the supernovae explosions is dense and thus it would quickly radiate away the received energy due to its efficient cooling. See \cite{Stinson2013} for further information and an extended feedback parameter search. 

\section{Host galaxies and their sub-haloes}
\label{sec:G_props}

\subsection{Host galaxy selection and properties}

For this work the virial mass, $M_{200}$, of each isolated halo is
defined as the mass of all particles within a sphere containing
$\Delta$ = 200 times the cosmic critical matter density, $\rhocrit$.
The virial radius, $R_{200}$, is defined accordingly as the radius of
this sphere. The haloes in the zoom-in simulations were identified
using the MPI+OpenMP hybrid halo finder \texttt{AHF2}
\citep{Knollmann2009,Gill2004}. The masses for sub-haloes and
satellite galaxies are defined as the sum off all gravitationally
bound particles belonging to these haloes and are denoted with $M_{\rm
  halo}$. They are calculated by \texttt{AHF2} via an iterative
unbinding procedure \citep[for more details see e.g.][]{Knebe2011b}.

The four galaxies used for this work have stellar masses between $2.0\times10^{10}\Msun$ and
$1.59\times10^{11}\Msun$. Galaxy g2.79e12, g8.26e11 and g7.08e11 are disc galaxies as can be seen from the
most right panels of figure \ref{fig:impression}. The stellar disc of galaxy g2.79e12 has a scale length of R$_{\rm d}\sim 5$ kpc and a scale height of H$_{z}\sim750$ pc. The corresponding
parameters for galaxy g8.26e11 are R$_{\rm d}\sim 5.5$ kpc and a scale height of H$_{z}\sim500$ pc and galaxy g7.08e11 has a disc scale length of R$_{\rm d}\sim3.7$ kpc and a scale height of H$_{z}\sim600$. Detailed parameters like the total virial mass, virial radius, gas and stellar mass as well as the number of satellites can be taken from table 2.

\begin{figure}
\includegraphics[width=\columnwidth]{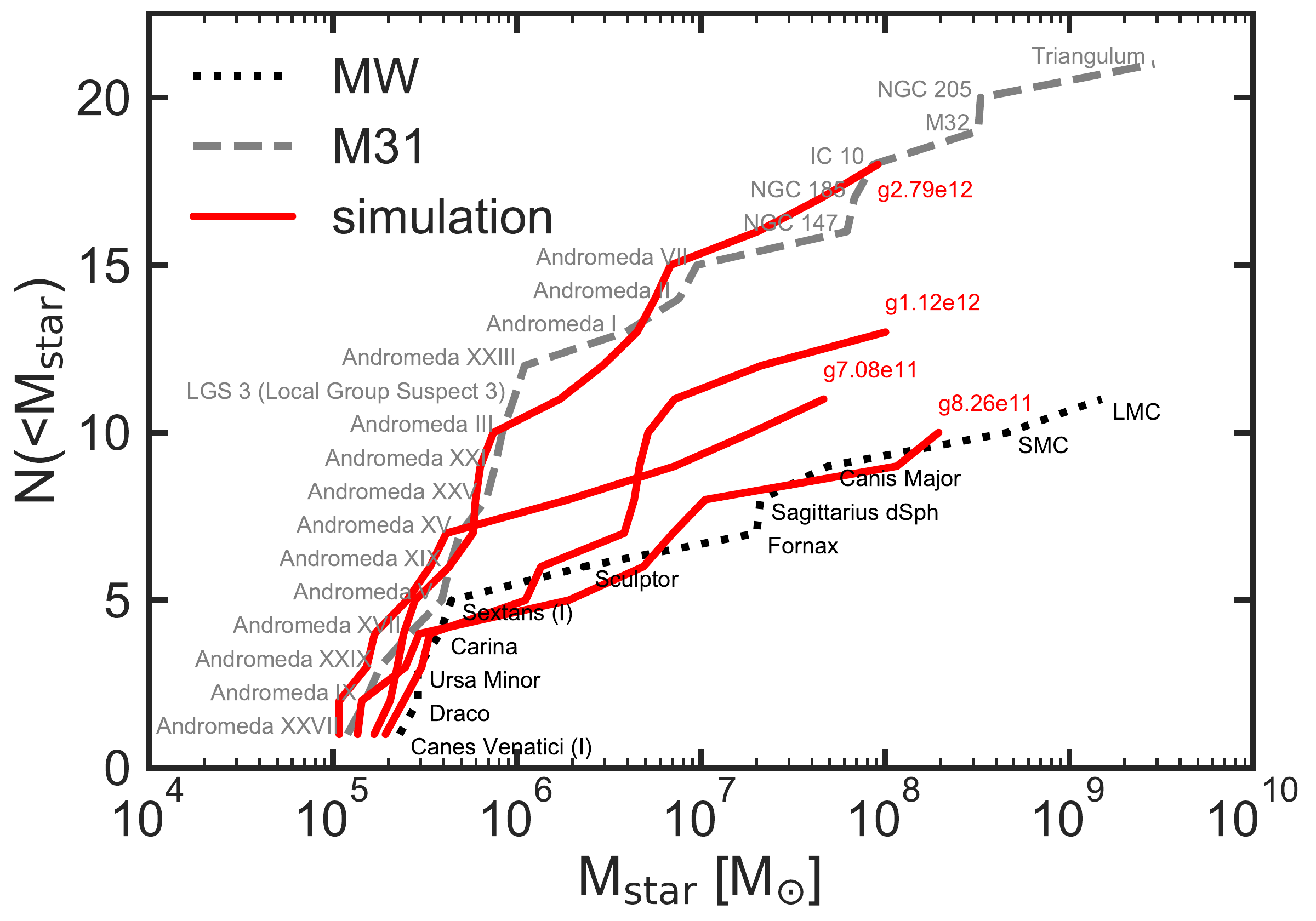}
\vspace{-.35cm}
\caption{Satellite stellar mass function for the three simulations
  (thick red lines) and the observations of Milky Way satellites
  (black dotted line) as well as M31 satellites (gray dashed
  line) within 300 kpc from the host. Two of our simulations perfectly 
  match the observed mass function of Milky Way and M31, except for the most massive satellites, like e.g. LMC, M32 or Triangulum which are known to be rare in cosmological simulations \protect\citep{Boylan2011}. }
\label{fig:mass_func}
\end{figure}

\begin{figure*}
\includegraphics[width=\columnwidth]{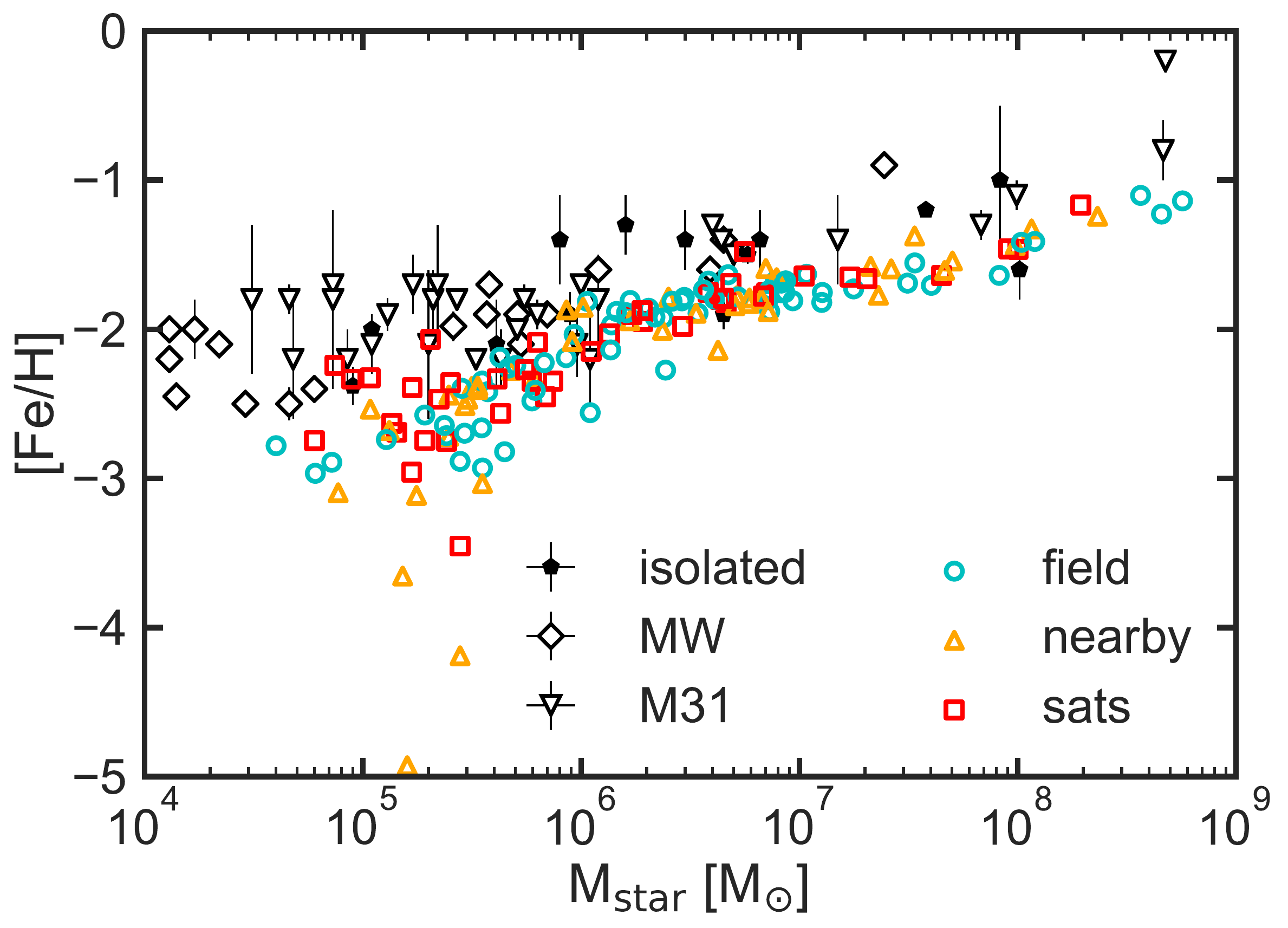}
\includegraphics[width=\columnwidth]{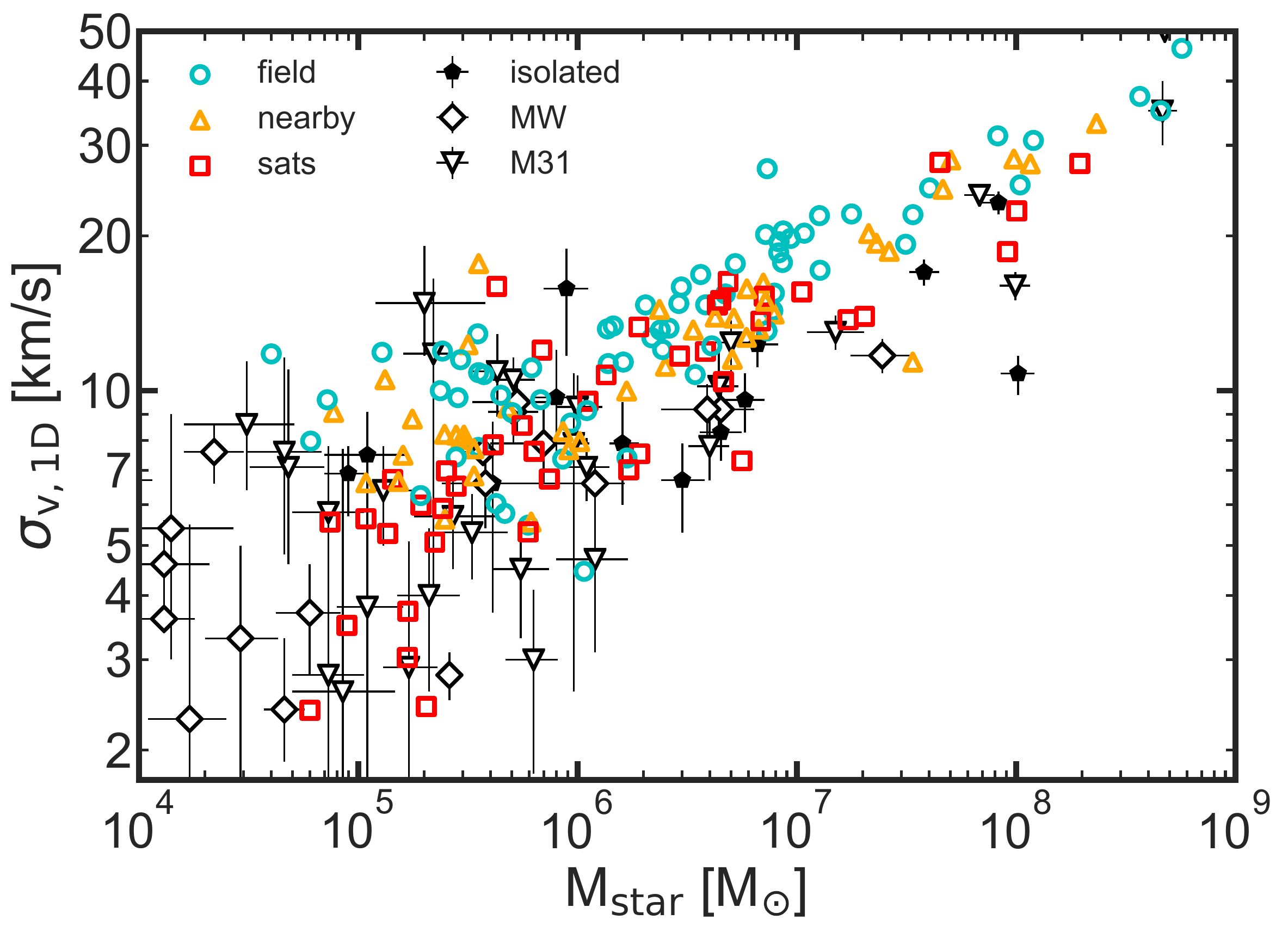}
\vspace{-.25cm}
\caption{Left: Stellar mass vs. 1D velocity dispersion of stars within the half
  light radius. Right: Stellar mass vs. metallicity where the metallicity is
  calculated for all stars within $0.2\times R_{\rm 200}$. Red squares show {\it{satellites}} within the virial
  radius ($\rm{R}_{\rm 200}$), orange triangles show {\it{nearby}} galaxies within
  $\rm{R}_{\rm 200}<R<2.5\rm{R}_{\rm 200}$ and cyan dots show all
  {\it{field}} within $2.5\rm{R}_{\rm 200}<R<3$ Mpc. Black and gray
  points show observations for satellites of the Milky Way and M31 and
  isolated Local Group dwarfs taken from \protect\cite{Fattahi2017} and references therein.}
\label{fig:vdisp}
\end{figure*}

\subsection{Sub halo selection}

We are interested in the satellite system and the dwarf galaxies
around the host galaxies. In order to keep a clean sample we only
select dark matter halos with at least 10 stellar particles for our
luminous galaxy. Given our stellar particle masses this leaves us with
a lower galaxy stellar mass of $\sim10^{5}\Msun$. We additionally require every dwarf galaxy outside the virial radius of the host to reside in a dark matter halo consisting of only high resolution particles. Figure \ref{fig:impression} gives an impression of the large scale morphology (left two columns) as well as the closer environment inside $R_{\rm 200}$ (column 3 and 4) and the host galaxies themselves (column 5 and 6). The first column shows the dark matter surface density of a cube of edge-size 2 Mpc where we highlight every luminous dark matter halo outside the virial radius with a purple circle and the ones inside it with a cyan circle\footnotemark[3]. All the galaxies highlighted that way reside in dark matter halos with masses larger than
$\sim10^8\Msun$. We thus indicate all other dark matter halos with
masses above $\sim10^8\Msun$ with white circles finding several completely dark halos. The second column shows the corresponding stellar
surface density. The third and fourth column show the same for a
smaller aperture of only 700 kpc aside. The right most column shows a
multi-color RGB image of the main galaxies in the wavelength bands
$i,v,u$ in face-on and edge-on view.

\footnotetext[3]{The two luminous satellites of galaxy g7.08e11 and g1.12e12 not circled in cyan can not be identified by AHF in the $z=0$ snapshot. Due to the previous pericenter passage all particles (DM, gas and stars) of these satellites are removed by the unbinding procedure\citep[see also][]{Bosch2018a,Bosch2018b}. We include values for these satellites from the preceeding snapshot $70$ Myr earlier.}

Having selected all (sub) halos meeting our selection criteria we
distinguish between dwarf galaxies inside the host's virial radius at
$z=0$ (red squares in all plots) from now on named {\it{satellites}},
dwarf galaxies between one and 2.5 times the virial radii
($\sim300<r/\rm{kpc}<\sim750$, orange triangles) named {\it{nearby
    galaxies}}, and {\it{field galaxies}} (outside 2.5 virial radii,
cyan dots). This distinction seems to be ad-hoc but we find that
$2.5R_{\rm 200}$ marks a boundary at which dwarf galaxy properties
change significantly \citep[see section \ref{sec:field_dwarfs} and e.g.][]{Behroozi2014}.

Figure \ref{fig:rad_dist} shows the radial distribution of luminous
halos found in all four simulations. We apply our color-coding scheme
outlined above where {\it{satellites}} are shown with a red line which
changes to orange once we reach the radial range where we denote
galaxies as {\it{nearby}} galaxies and then changing to cyan at radii
larger than 2.5$R_{200}$, the range of {\it{field}} galaxies. As we can see from this figure, at $z=0$ there are no luminous satellites closer than 60 kpc to the host. However, at slightly earlier time steps we do find
luminous satellites closer to the host than 60 kpc who's orbits take
them further away from the host at z=0 (compare also Figure
\ref{fig:rmin} for the minimum radii of {\it{satellites}}).

In Figure \ref{fig:mass_func} we show the stellar mass function of the
{\it{satellites}} from our four simulations in comparison to the ones
of Milky Way and Andromeda satellites calculated from the
\cite{McConnachie2012} data. For the observations we apply a stellar
mass cut of M$_{\rm star}>10^{5}\Msun$ in order to match our
resolution limit. As we can see from figure \ref{fig:mass_func} our
simulations match perfectly the observed mass functions of MW and M31. They only miss the most massive satellites LMC and SMC or
Triangulum, NGC 205 and M32, which are known to be rare in cosmological simulations \citep{Boylan2011}. From visual
inspection of the simulation movies we further see that a lot of low mass
satellites get destroyed coming close to the stellar disc of the
host. The debries of these satellites can still be seen in the $z=0$
snapshots as stellar streams, e.g. in figure \ref{fig:impression}.

Finally, for all {\it{satellites}}, {\it{nearby}} and {\it{field}} galaxies we
create merger trees in post processing. In order to link sub-haloes
between two consecutive snapshots we trace their dark matter particles
via their unique particle ID. For each luminous sub-halo we find the
progenitor which shares the most dark matter particles with the
sub-halo in question. From these merger trees we extract several
evolutionary parameters like e.g. the infall time, the peak mass,
defined as the maximum virial mass ever reached, and the peak
$v_{\max}$, defined as the temporal maximum of the maximum of the
circular velocity curve \citep[e.g.][]{Kravtsov2004}. 

\section{Dwarf galaxy properties}
\label{sec:dwarf_props}

In this section we compare the properties of the simulated dwarf
galaxies (inside and outside the virial radius) with observed
properties of Local Group dwarfs. Observational data is taken from the compilation presented in \cite{Fattahi2017} (for further references
see Table \ref{tab:peak_mass}). The velocity dispersion for Phoenix is taken from \citep{Kacharov2017}. Observational data for the star
formation histories (SFH) are taken from \cite{Weisz2015}. For the
simulations we measure the 1D-velocity dispersion for all stars within the 3D half-light radius using a projection along the x-axis of the simulation cube. The metallicity is measured for all stars within $0.2R_{\rm 200}$.

\subsection{Velocity dispersion and metal abundance}

Figure \ref{fig:vdisp} compares the stellar mass-metallicity relation
(left panel) and the one-dimensional velocity dispersion vs. stellar
mass of our galaxies (right panel) to observational data. From the
left panel of Fig. \ref{fig:vdisp} we see that our galaxy sample
recovers the observed mass metallicity relation well, although with a
slight offset to lower metallicities. This offset is no major issue
and can be resolved by using a different/updated yield set for the
chemical enrichment (Buck \etal in prep). For very low mass galaxies
we see an increasing deviation to lower metallicities from the
observed relation. This is a result of our enrichment scheme which is
not handling metal enrichment of single starburst galaxies
properly. For future simulations we will employ a different enrichment
scheme to reduce these shortcomings.

Looking at the right panel of Fig. \ref{fig:vdisp} we see that the
mass-velocity dispersion relations obtained from our simulated
galaxies are in excellent agreement with the observed
relations. Especially at low stellar mass ($M_{\rm star}<10^6\Msun$)
we recover the very low dispersion galaxies observed in the Local
Group as well as the large scatter in the observed velocity dispersion
values. At higher stellar masses ($\sim10^8\Msun$) we notice some
velocity dispersion excess in comparison to observed galaxies. 

The most interesting result of the above two figures is however that
galaxies from all three populations, {\it satellite}, {\it nearby}
and {\it field galaxies}, mix well. It is hard to distinguish
{\it satellite galaxies} from {\it nearby} or {\it field galaxies}
just by measuring their velocity dispersion or metallicity as a
function of stellar mass. Only at very low stellar masses we see a
slight separation between {\it satellite} and
{\it nearby}/{\it field} galaxies. {\it Satellites} tend to have
lower stellar velocity dispersions compared to equal mass
{\it nearby} or {\it field} galaxies. As we will elaborate in
section \ref{sec:field_dwarfs} this is the signature of tidal mass
loss which reduces the stellar velocity dispersion of satellites
\citep[see e.g.][]{Frings2017,Maccio2017}.

\begin{figure}
\vspace{-.35cm}
\includegraphics[width=\columnwidth]{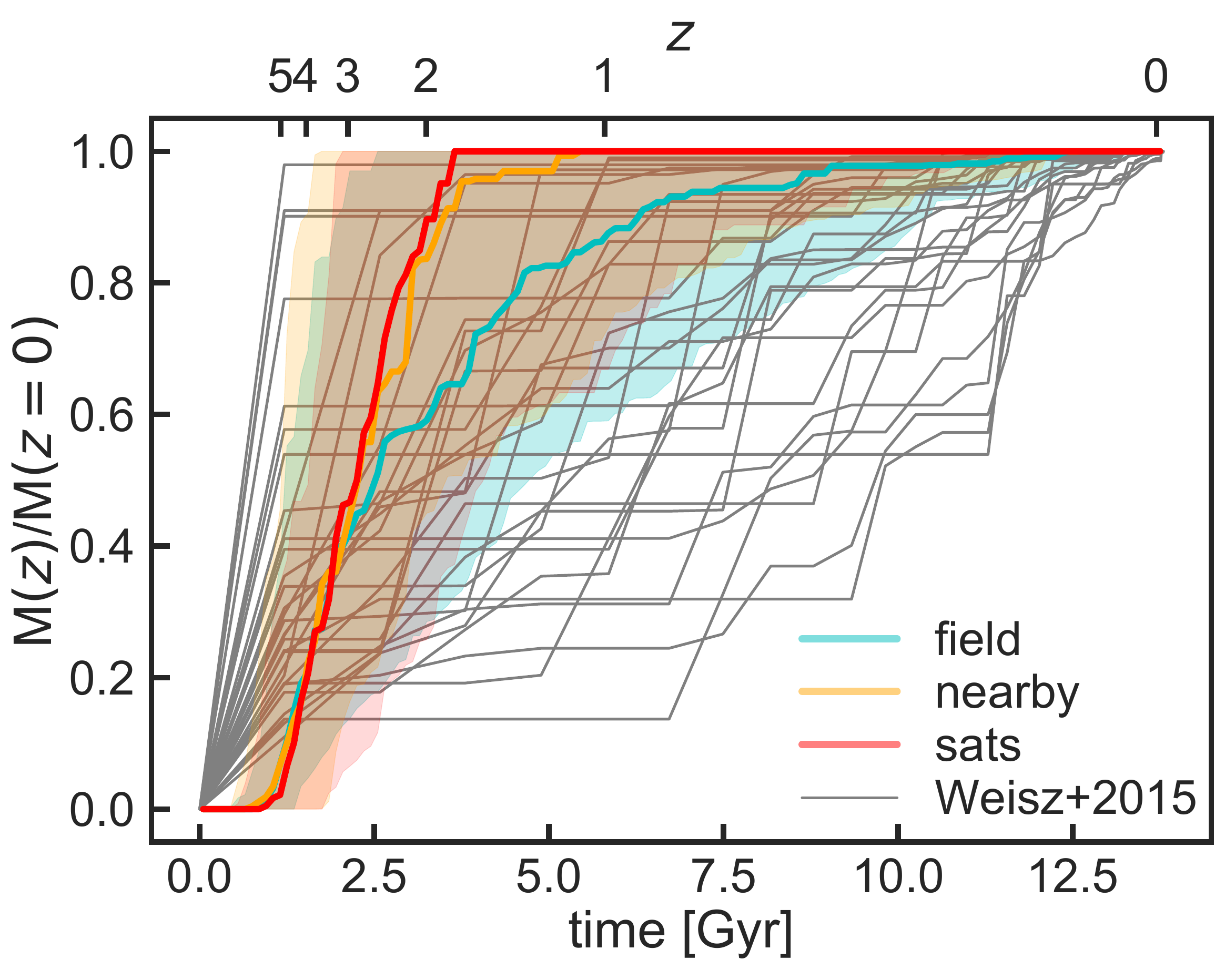}
\vspace{-.35cm}
\caption{Median SFH of our simulated galaxies (colored lines) in
  comparison to observations of Local Group galaxies (gray lines)
  taken from \protect\cite{Weisz2015}. The red line shows the median SFH for
  \textit{satellite galaxies}, the orange line for \textit{nearby
    galaxies} and the cyan line for \textit{field galaxies}. Shaded
  areas indicate the 16th and 84th percentile. Field galaxies show a
  much more extended SFH compared to satellites and dwarf galaxies.  }
\label{fig:sfh}
\end{figure}

\subsection{Star formation history and gas fraction}

Figure \ref{fig:sfh} shows the cumulative SFH of simulated galaxies in
comparison to observational data taken from \citet[][gray
  lines]{Weisz2015}. Our simulated galaxies cover the full range of
observed SFH. Similar to the results from the last subsection we see
not much difference between \textit{satellites} and \textit{nearby
  galaxies} in their SFH. These two types of galaxies show an early
stellar mass growth which then quickly stops at redshift $z\sim2$
which is when \textit{satellites} fell into their host halo and got
stripped of their cold gas and thus rapidly quenched \citep[see
  e.g.][]{Wetzel2013,Fillingham2016}. As we will see later
\textit{nearby dwarfs} are biased towards early stellar mass growth by
the \textit{backsplash} galaxies belonging to this population. \textit{Field
  galaxies} however show a more extended SFH and most are basically
star forming up to today. These findings are in broad agreement with
the results obtained for Local Group galaxies by \cite{Weisz2015} and
\cite{Brown2014} who find similar results for \textit{satellites} and
\textit{field galaxies}.

\begin{figure}
\includegraphics[width=.5\textwidth]{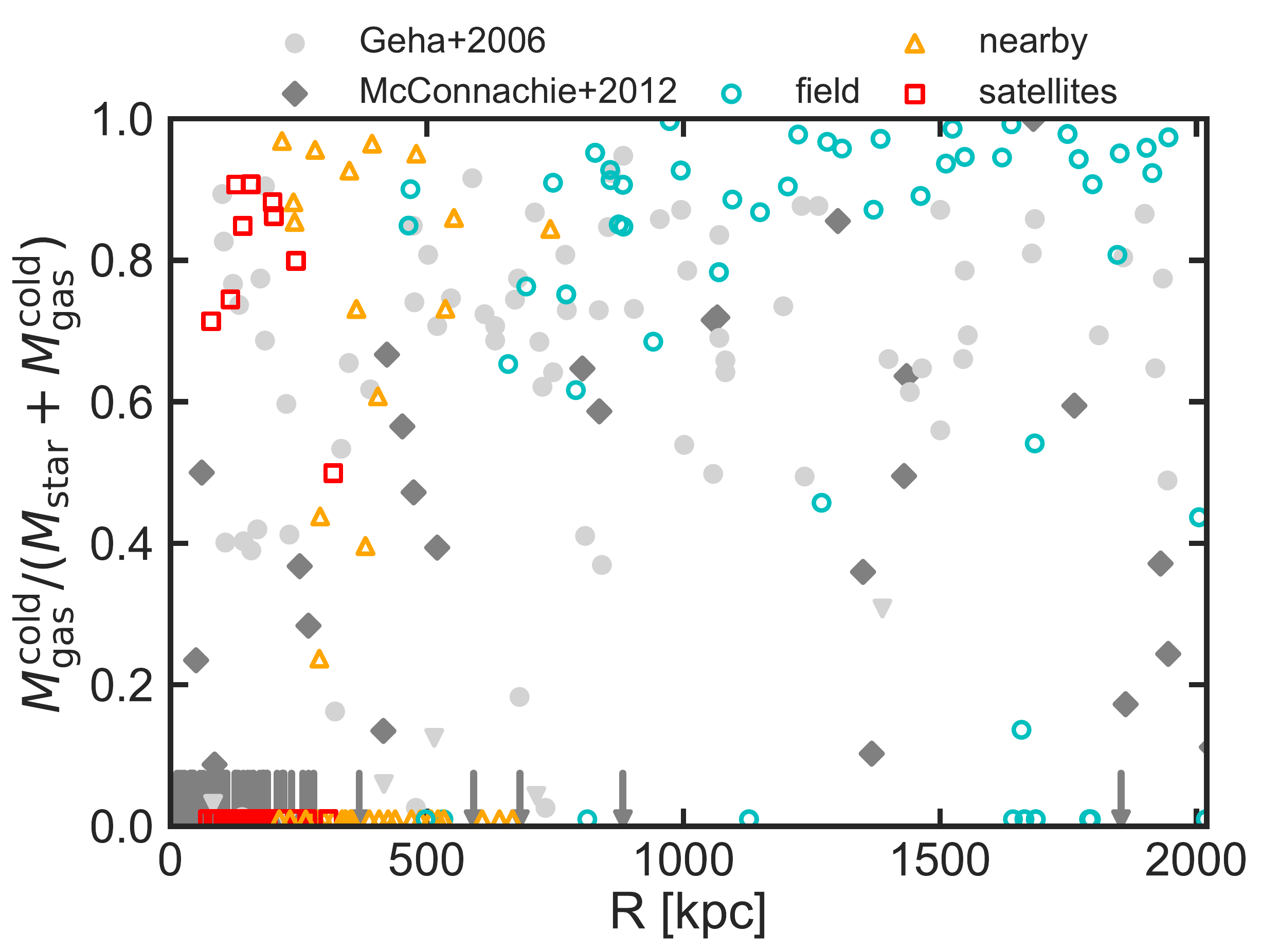}
\vspace{-.35cm}
\caption{Cold gas fraction of \textit{satellites} (red squares),
  \textit{nearby galaxies} (orange triangles) and \textit{field
    galaxies} (cyan dots) as a function of distance from the host. The gray diamonds show Local Group HI data taken from \citet{McConnachie2012} where downwards pointing arrows indicate no gas detection. The lightgray points show measurements taken from \citet{Geha2006} where the lower triangels indicate upper limits. For the observations the distance indicates the distance to the nearest massive galaxy. The overlap between the three galaxy categories of the simulations reflects that the host galaxies have different values for the virial radius which defines the category.} The closer one gets to the host the lower is the fraction of galaxies that still have a cold gas content.
\label{fig:gas_frac}
\end{figure}

That indeed the dwarf galaxies closer to the host are more affected by
stripping off the cold gas can be seen from
Fig. \ref{fig:gas_frac}. In this Figure we show the gas fraction of
our sample of galaxies as a function of distance from the host
(colored dots). The gray points show observational data from \citet{Geha2006} and \citet{McConnachie2012}. Our simulated gas fractions agree very well with the observed gas fractions which even show very gas poor galaxies. We clearly see the
separation into \textit{satellites}, with almost all of them being gas
poor, \textit{nearby dwarfs}, which partly still have some cold gas
left and \textit{field galaxies} which almost all show very high gas
fractions around unity. We ascribe these differences to ram pressure
stripping which quickly removes the gas from the \textit{satellites} and
\textit{nearby galaxies} \citep{Wetzel2013,Fillingham2016,Simpson2017}. Stripping
off the cold gas from the \textit{satellites} removes the fuel for
star formation which shuts down late time star formation. Only \textit{satellites} on their first infall (and at the same time among the most massive dwarf galaxies still have some cold gas left. Overall the fraction of \textit{satellites}
and \textit{nearby dwarfs} with a cold gas fraction larger than 0.1 quickly
drops from $\sim 0.8$ at $R=4R_{\rm 200}$ to $\sim0.3$ at $R<R_{\rm 200}$
(purple dotted line in Fig. \ref{fig:gas_frac}). We like to note that the few \textit{field galaxies} which do not have any cold gas still posses hot gas. We have checked that these \textit{field galaxies} have indeed masses $\lesssim10^9\Msun$ and are thus loosing their gas via evaporation by the UV background which is in perfect agreement with the results obtained by \citet{Sawala2012}. For the \textit{nearby} sample the situation is slightly more complicated. Most of them belong to the category of \textit{backsplash galaxies which loose their gas via ram pressure stripping. However, for four of the \textit{nearby galaxies} (all with stellar masses below $\lesssim10^9\Msun$), the main cause for their lack of cold gas is the UV background. Indeed, when looking at the gas fractions of all the dwarf galaxies at the time when they reach their peak mass (including the satellites) we find that there is a sharp transition at $\sim10^9\Msun$ with dwarfs above that mass having gas fractions of $\sim80-100\%$ and the ones below this mass having $0\%$ cold gas.}

\section{Environmental effects}
\label{sec:field_dwarfs}

We have seen that \textit{satellites} differ from \textit{field galaxies} in the properties strongly affected by the host environment. In this section we quantify the mass loss via stripping for individual galaxies over their lifetime.  

\subsection{Ram Pressure Stripping}
\label{sec:ram_press}

Figure \ref{fig:gas_frac} shows that \textit{satellites} and \textit{nearby galaxies} show a large fraction of gas poor systems while all the \textit{field galaxies} have very high cold gas fractions. As previously suggested \citep[e.g.][]{Simpson2017}, this might be the result of ram pressure stripping off the gas reservoir of galaxies orbiting inside the diffuse gas halo of the host galaxy \citep{Gunn1972}. Following the recent study by \citet{Simpson2017} we estimate the ram pressure felt by the orbiting \textit{satellite} and the restoring force per area of the \textit{satellite's} gravitational potential to test if the latter is able to strip off the \textit{satellite's} gas.

The ram pressure, $P_{\rm ram}$, felt by the \textit{satellite's} gas is given by the product of the density of the circum-galactic medium (CGM), $\rho_{\rm CGM}$, and the squared velocity of the \textit{satellite} relative to the CGM, $v_{\rm sat}^2$. We derive $\rho_{\rm CGM}$ by calculating a spherically averaged density profile of the gas out to $4R_{\rm 200}$ of the host for every snapshot in the sample of the g2.79e12 galaxy. We then interpolate the profile to estimate the gas density at the \textit{satellite's} position.

The restoring force per area felt by the \textit{satellite's} gas is given by $P_{\rm rest}=\vert\frac{\partial\Phi}{\partial z_{h}}\vert_{\rm max} \Sigma_{\rm gas}$ where $z_{h}$ is the direction of motion and $\Sigma_{\rm gas}$ is the \textit{satellite's} gas surface density. $\Phi$ is the \textit{satellite's} gravitational potential and $\vert\frac{\partial\Phi}{\partial z_{h}}\vert_{\rm max}$ represents the maximum of the derivative of $\Phi$ along $z_{h}$ \citep{Roediger2005}. Following \citet{Simpson2017} we adopt a simple estimate for $\Sigma_{\rm gas}$ and $\vert\frac{\partial\Phi}{\partial z_{h}}\vert_{\rm max}$ where the gas surface density is estimated from the total gas mass $M_{\rm gas}$ and the \textit{satellite's} half gas mass radius, $r^{\rm gas}_{\rm half}$ as $\Sigma_{\rm gas}=M_{\rm gas}/2\pi(r^{\rm gas}_{\rm half})^2$. The derivative of the gravitational potential can be approximated by $\vert\frac{\partial\Phi}{\partial z_{h}}\vert_{\rm max}\sim v_{\rm max}^2/r_{\rm max}$ with $v_{\rm max}$ the maximum of the \textit{satellite's} rotation curve and $r_{\rm max}$ the radius where this value occurs.

Using the merger trees, we calculate the ram pressure, the restoring force per area, the galacto-centric distances and the gas fractions as a function of cosmic time for all \textit{satellites} (shown in figure \ref{fig:ram}), \textit{nearby dwarfs} and \textit{field dwarfs} (both shown in the appendix in section \ref{sec:ram}). For visual clarity we only show galaxies with stellar masses above $10^6\Msun$. Figure \ref{fig:ram} shows that indeed the sharp drop in gas fraction corresponds to the \textit{satellites} approaching pericenter and thus to an increased value of ram pressure acting on their gas reservoir. For the three \textit{satellites} (pink/orange lines) of lowest stellar mass the ram pressure spikes up shortly after infall or shortly before pericenter passage and removes quickly all the gas. The highest stellar mass \textit{satellites} (black/blue/purple lines) on the other hand are massive enough to keep some gas after first pericenter passage and interestingly the ratio of ram pressure to restoring force per area does not rise above a value of unity. Still, they show a sharp drop in gas fraction due to enhanced ram pressure at pericenter passage and then a plateau in gas fraction during apocenter passage. This points towards the fact that ram pressure alone might not be able to strip off the gas \citep[see also][]{Emerick2016}. However, gas can be effectively stripped off by the combined effect of ram pressure and tidal forces.
 At second pericenter passage they show again a sharp drop in gas fraction where ram pressure and tidal forces then remove the rest of the gas reservoir leaving them gas free. Only the very late infalling satellites (except for the most massive one, black line) still have some gas at redshift zero. Very similar results are found for the \textit{nearby sample} while for the \textit{field} sample at distances greater than $2.5R_{\rm vir}$ we see that ram pressure is not important in removing their gas.

\begin{figure}
\includegraphics[width=.5\textwidth]{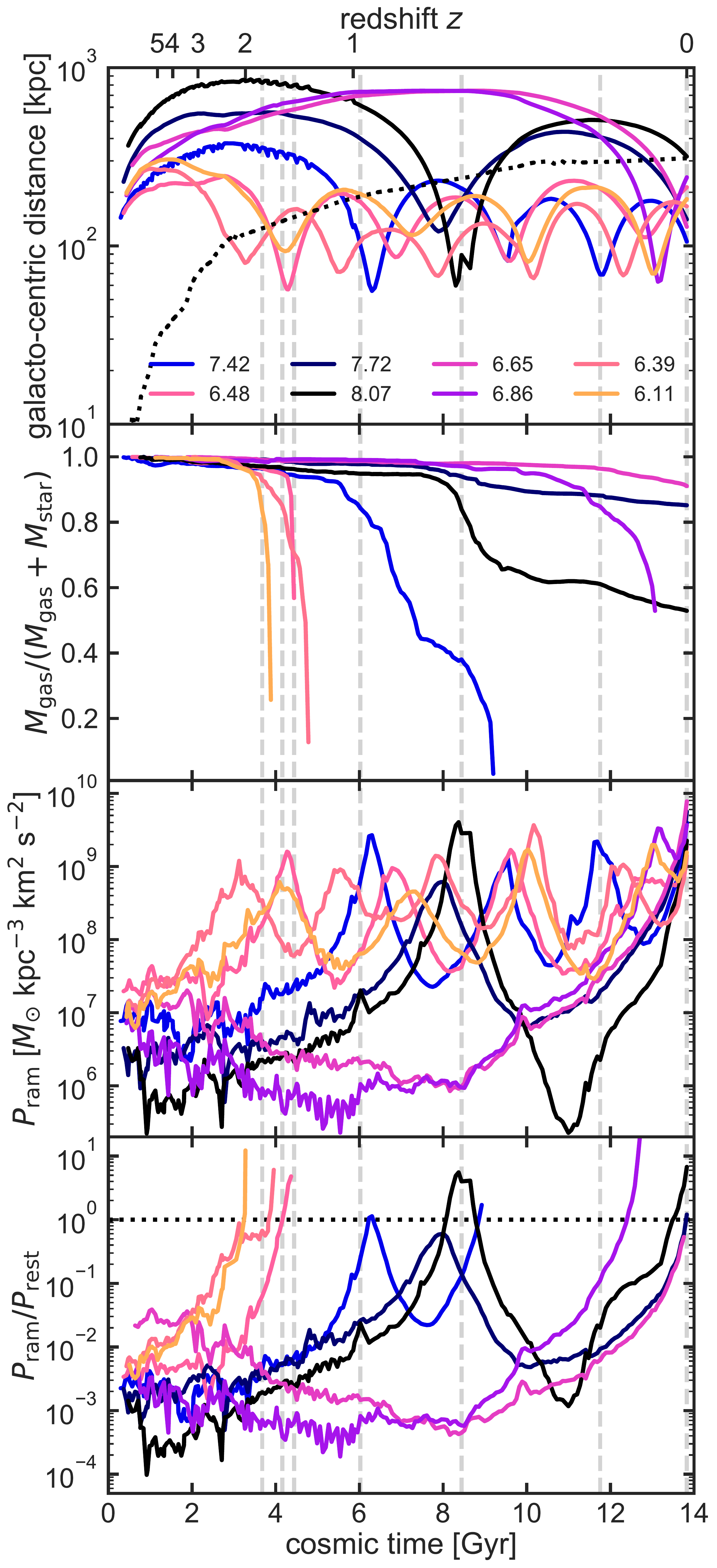}
\vspace{-.35cm}
\caption{The effect of ram pressure stripping on the gas fraction of luminous \textit{satellites} in the galaxy g2.79e12. 
From top to bottom we show the evolution of the galacto-centric distance of the \textit{satellites}, the gas fraction, the ram pressure of the ambient gas of the host exerted on the \textit{satellite's} gas and the ratio of ram pressure to \textit{the satellite's} restoring force as a function of cosmic time. In the top panel the dotted black line shows the evolution of the virial radius of the host. In all panels the gray dashed lines indicated the point in time when the gas fraction drops below a value of 0.85. The lines are colored by the \textit{satellite's} $\log(M_{\rm star})$ as indicated in the top panel. For visual clarity we only show satellites with stellar masses greater than $M_{\rm star}>10^6\Msun$.}
\label{fig:ram}
\end{figure}

\subsection{Abundance matching - $M_{\rm star}$ vs. $M_{\rm halo}$}
\label{sec:abundance}

\begin{figure*}
\includegraphics[width=\textwidth]{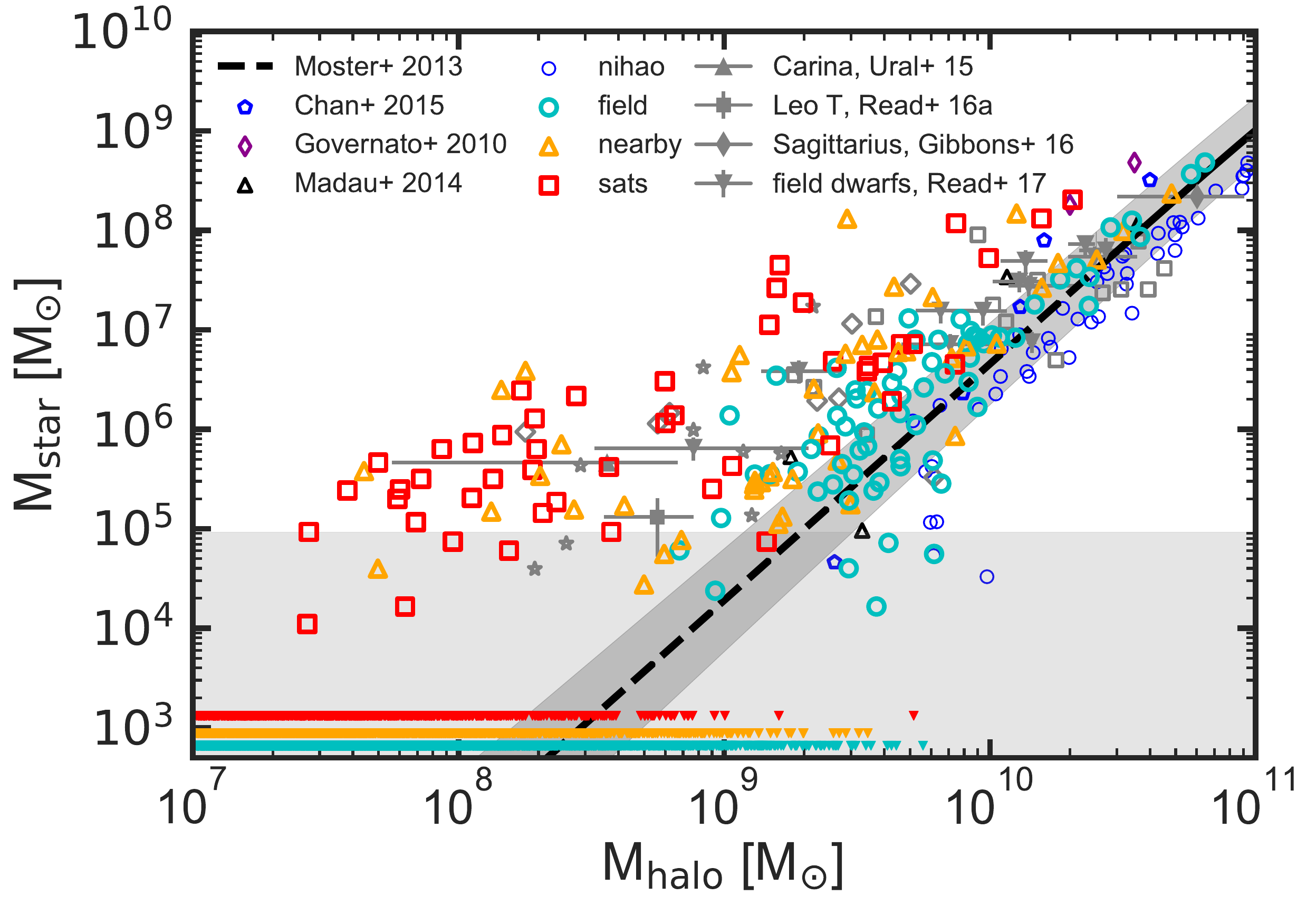}
\vspace{-.25cm}
\caption{ Stellar mass vs. the total halo mass $\rm{M}_{halo}$ for
  satellites, dwarfs and field galaxies. Colored points denote the
  satellites and dwarfs in our simulations with the same color coding
  convention as in previous plots and gray points show observational
  results for Local Group galaxies
  \protect\citep{Ural2015,Read2016,Gibbons2017,Read2017}. Gray open symbol
  measurements are taken from \protect\cite{Sawala2014} (and references
  therein), gray open stars show results for individual Milky Way
  satellites, gray open diamonds for M31 satellites and gray open
  squares for dwarf irregulars. The black line shows the abundance
  matching result from \protect\cite{Moster2013}, we turn the line style to
  dashed where the result is extrapolated. Open blue circles show the
  result from the whole NIHAO sample \protect\citep{Wang2015} and all other
  open colored symbols show results from other simulation suites
  \protect\citep{Chan2015,Governato2010,Madau2014}. Lower triangles at the
  bottom of the plot indicate dark sub-haloes with no stellar
  particles at all. The gray shaded area marks the region where we
  have fewer than 10 stellar particles in our haloes.}
\label{fig:abund1}
\end{figure*}

In \lcdm there exists a well established relation for isolated
galaxies between the mass of their dark matter halo $M_{\rm halo}$ and
the amount of stellar mass $M_{\rm star}$ present in that halo \citep{ Conroy2009, Moster2010, More2011, Behroozi2013}. This relation provides, as
we will see below, a very good tool to tell environmentally affected
and unaffected dwarf galaxies apart. In Fig. \ref{fig:abund1} we present the stellar mass of our simulated galaxies
against their total gravitationally bound mass: for isolated galaxies this is $M_{200}$ and for satellite galaxies this is the bound mass returned by AHF. We indicate our
fiducial resolution limit of 10 stellar particles with the gray shaded
region and show dark sub-haloes (sub-haloes without any stellar
particle) with lower triangles at $M_{\rm star} \sim
10^3\Msun$. Luminous galaxies follow our fiducial color scheme where
\textit{satellites} are shown with red squares, \textit{nearby
  galaxies} with orange triangles and \textit{field galaxies} with
cyan dots. For comparison we show results from the NIHAO sample
\citep[][open blue dots]{Wang2015} as well as results from other
simulation suites (open colored symbols) and measurements for three
Milky Way satellites and several observed field dwarfs taken from the
literature in gray symbols. Finally, the black dashed line with the
shaded dark gray area marks the abundance matching relation from
\citet{Moster2013} and its corresponding scatter.

From Fig. \ref{fig:abund1} we see that, as we expect, the isolated
\textit{field galaxies} follow nicely the observed relation (black
dashed line) and are well in agreement with the higher resolution
versions taken from the NIHAO sample. The \textit{satellite} galaxies strongly
depart from the relation of isolated galaxies. For a given stellar
mass these galaxies have a too low virial mass compared to isolated
galaxies. This deviation is a signature of mass removal from
the \textit{satellites} \citep[see also][for a recent study of the scatter in the abundance matching relation at low halo masses]{Munshi2017}. We have checked that almost all
\textit{satellites} retain their stellar mass after infall
\citep[see also][]{Frings2017}. 

The \textit{nearby galaxy} sample which is located in the radius range
of $R_{200}$ to $2.5 R_{200}$ from the host galaxy also shows signs
of stripping. However, the galaxies of this sample deviate less from
the main relation compared to the \textit{satellite sample}. This can
be explained by two effects: first, the environmental effect of
striping is already present at these radii far away from the host
although less pronounced due to a weaker tidal field. Second, most of
the dwarf galaxies actually spent some time orbiting within the virial
radius of the host (see next section and the appendix).

\begin{figure*}
\includegraphics[width=.51\textwidth]{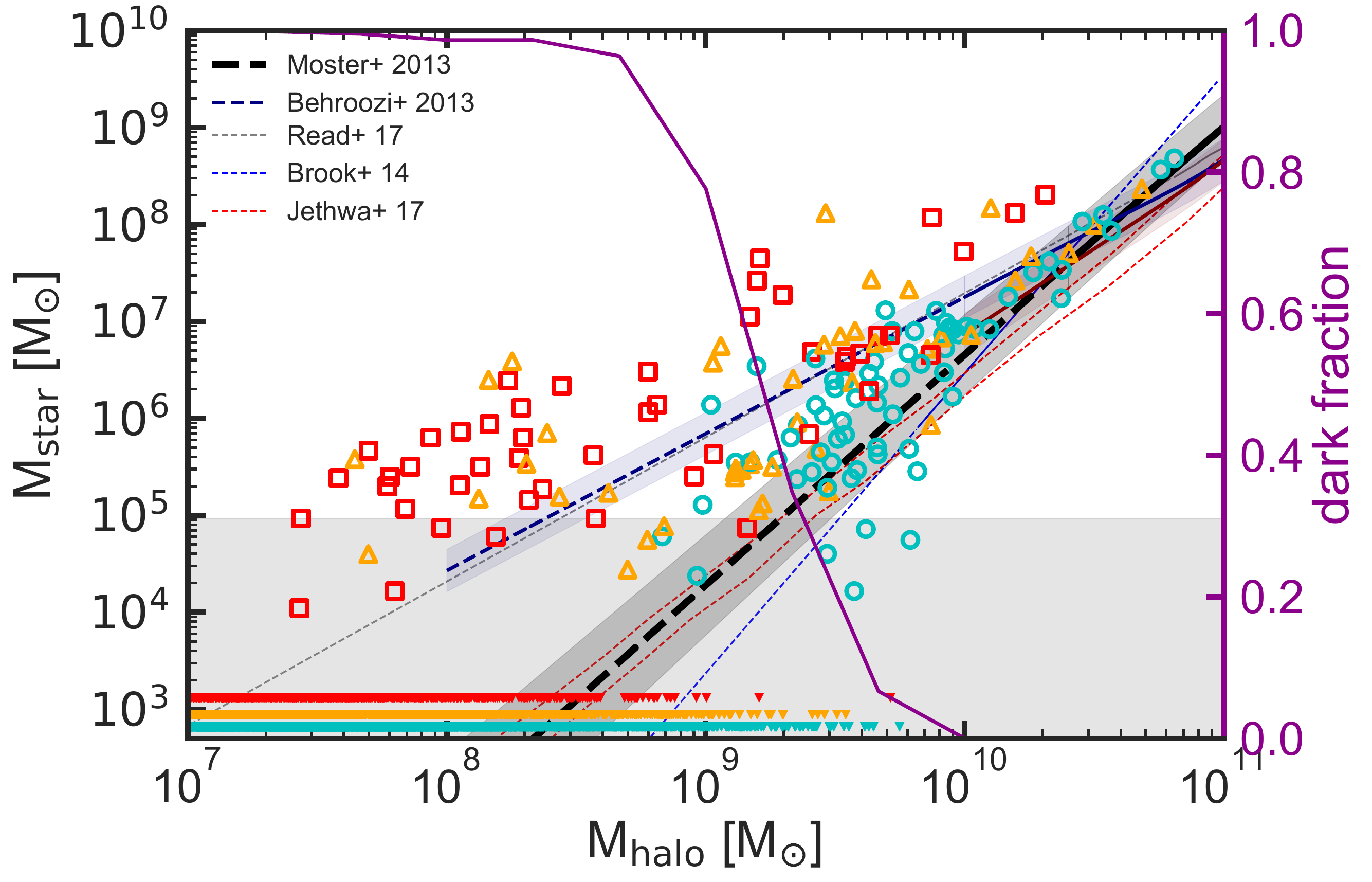}
\includegraphics[width=.48\textwidth]{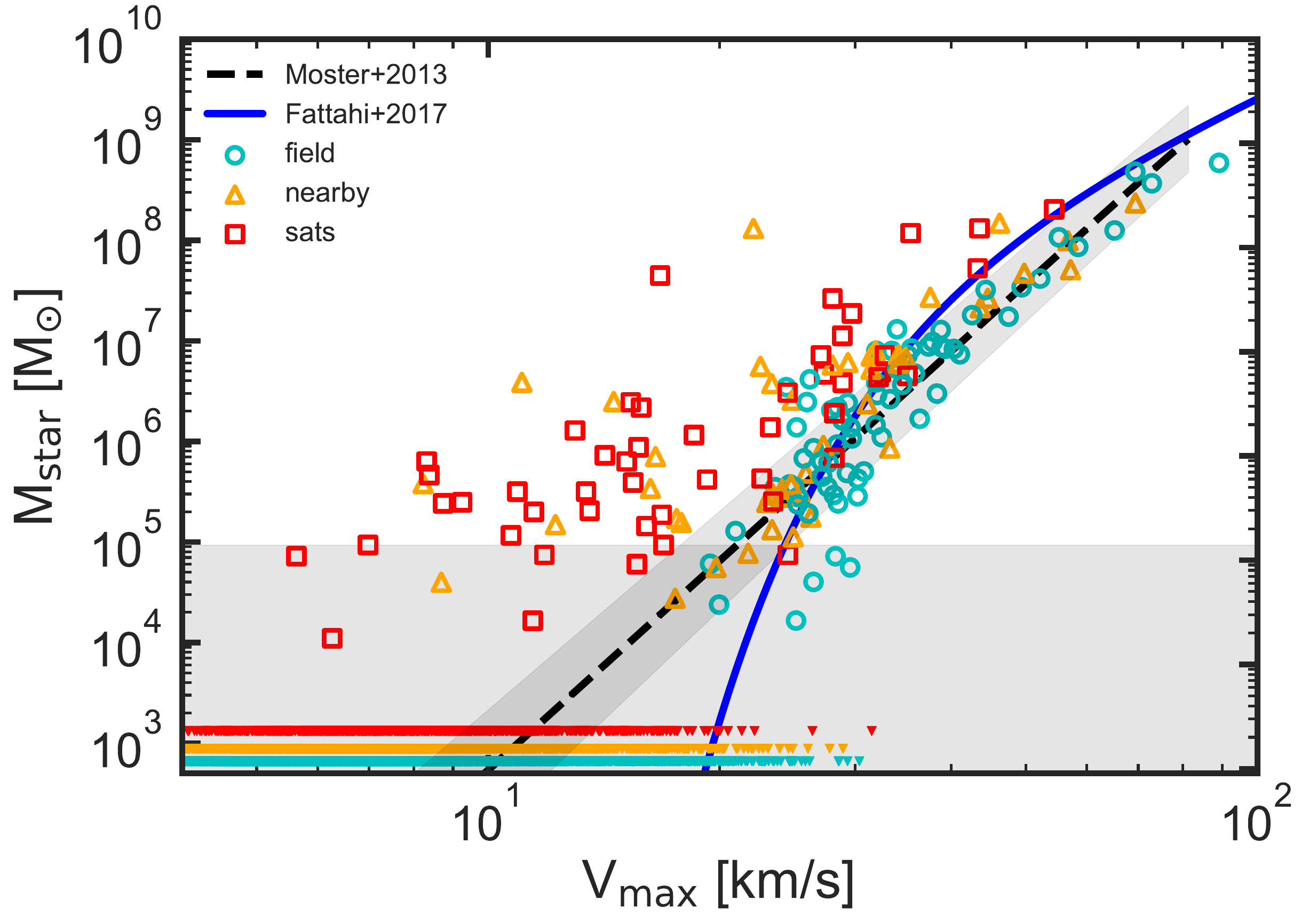}
\vspace{-.25cm}
\caption{ \emph{Left panel}: Stellar mass vs. the total halo mass
  $\rm{M}_{halo}$, the same data as in Fig. \ref{fig:abund1} is shown
  but this time in comparison to several abundance matching relations
  from the literature \protect\citep{Moster2013, Behroozi2013, Brook2014, Read2017, Jethwa2016}. The purple dotted line (right
  y-axis) shows the fraction of dark haloes with masses greater than
  $10^7\Msun$.  \emph{Right panel}: Stellar mass vs. maximum circular
  velocity. Colored points show again the simulation results with the
  same color coding convention as before. The black dashed line shows
  the result for the abundance matching relation from
  \protect\cite{Moster2013} translated to a function of $V_{\rm max}$. The gray shaded area around this
  line marks the uncertainty of this relation. The blue solid line
  shows the $M_{\rm star}$ vs. $V_{\rm max}$ relation obtained for
  central galaxies from the APOSTLE simulations
  \protect\citep{Fattahi2017}. Again the gray shaded area at the bottom of the plot marks the region in which the galaxies contain fewer than $10$
  stellar particles. Lower triangles at the bottom of the plot
  indicate dark sub-haloes with no stellar particles at all.}
\label{fig:abund2}
\end{figure*}

In the left panel of Fig. \ref{fig:abund2} we compare our simulated
galaxies to several abundance matching relations taken from the
literature (colored lines). We turn the line style from solid to
dashed where we extrapolate the relations. The purple dotted line
(right y-axis) shows the fraction of dark haloes (e.g. haloes not able
to form stars). At $\sim10^9\Msun$ we find a fraction of $\sim$50\%
dark haloes in agreement with other simulations \citep[e.g.][]{Fattahi2016}. In the range where we have well resolved
galaxies ($M_{\rm halo}\gsim 10^9\Msun$) our galaxy follow the abundance
matching relations of \citet{Moster2013} and \citet{Jethwa2016}. Compared to our simulations, the relations from \cite{Behroozi2013} and \cite{Read2017} show a slightly too shallow slope while the relation from \cite{Brook2014} shows a
slope slightly too steep. In general we find much larger scatter for
our simulated galaxies than is predicted by the abundance matching
relations (even excluding \textit{satellite galaxies}).

One could argue that using the bound mass for satellites and the
virial mass for isolated galaxies introduces a bias. 
Hence, in the right panel of Fig. \ref{fig:abund2} we use the maximum of the
circular velocity curve $V_{\rm max}$ as a proxy for the total
mass. This quantity is less effected by the host environment and the definition used to set the total mass. We convert the abundance
matching relation from \cite{Moster2013} from being a function of
virial mass to a function of $V_{\rm max}$ by assuming a NFW-profile
\citep{NFW} and using the concentration-mass relation from \cite{Dutton2014}.

 Again, we indicate our fiducial resolution limit of 10 stellar particles with the gray shaded region and we show dark sub-haloes with lower triangles at $M_{\rm star}\sim 10^3\Msun$. The solid blue line shows the scaling relation
derived from isolated dwarf galaxies of the APOSTLE simulations
\citep{Fattahi2017}. Although still in agreement with our findings, we
do not see the bending behavior in our $M_{\rm star}$ vs. $V_{\rm
  max}$ relation. Our simulations are rather well in agreement with
the transformed (linear) Moster relation. As we can see from this
plot, we find no qualitative difference between the results obtained
by using the total bound mass or the maximum circular velocity to
derive the abundance matching relations. \textit{Satellites} deviate
in both versions from the expected abundance matching relation of
isolated galaxies.

One could further argue that the deviation might be partly due to a
resolution issue where we sample the abundance matching relation
unequally (galaxies with stellar masses below $10^5\Msun$ are barely
or not at all resolved). This issue would lead to preferentially
sampling galaxies with stellar masses above the Moster relation for
virial masses lower than $\sim10^9\Msun$ and might explain the
deviation from the abundance matching relation. However, this is not
the case here as we will show in the next subsection (see also section \ref{sec_res} in the appendix for a resolution study). 

\subsection{The peak masses of Local Group galaxies}

It is well established that satellite galaxies loose mass while orbiting inside the potential of the host galaxies \citep[e.g.][]{Kravtsov2004,Frings2017,Fattahi2017}. Especially using the peak mass
$M_{\rm peak}$, the maximum mass of a galaxy ever reached, instead of
the present day virial mass $M_{\rm 200}$ brings \textit{satellites} and \textit{field galaxies} back onto the same relation. Here we use the
merger trees to identify the peak mass for every dwarf galaxy in our
sample and plot in Fig. \ref{fig:abund_peak} the stellar mass at
redshift zero vs. the peak mass. \footnote{We like to add that for this analysis
as well as for the following two subsections we had to exclude 8 dwarf
galaxies. The reason for their exclusion is, either no progenitor
could be found or the unique determination of the peak mass was not
possible due to major merger.} In order to guide the eye, in Fig. \ref{fig:abund_peak} we also show the abundance matching relation at redshift $z=0$ taken from \citet{Moster2013} (black dashed line) and the results from
\cite{Kimmel2017} (purple solid line, green dashed line). We use the
same color-coding as in previous plots to distinguish
\textit{satellites, nearby dwarfs} and \textit{field dwarfs}. The gray
shaded area indicates again our fiducial resolution limit. We find
that all dwarfs follow nicely the predicted abundance matching
relation from \cite{Kimmel2017} with an assumed scatter of zero dex
and a slope of 1.8 (purple line). However, comparison to the Moster
relation (with an approximate slope of 2.5) shows that the simulations
deviate from this at lower peak masses. Similar behavior was discussed by \cite{Sawala2014} in the APOSTLE simulations. 

We fit a simple power law to our simulations in order to estimate the slope and
the scatter of this relation. For the simulations used here we find a
slope of $\alpha=1.87\pm0.02$ fits the data best and the scatter
around the main relation is $0.46$ dex. We have seen from the
simulations that the stellar mass loss is negligible thus we can use
the abundance matching relation of our simulations to estimate the
peak masses of Local Group galaxies using their total stellar mass today
(from \citet{Fattahi2017}, derived from their total luminosity). We present the
result in Table \ref{tab:peak_mass}. 

\subsection{Deviations from the abundance matching relations are caused by mass loss}
\label{sec:mass_loss}

In Fig. \ref{fig:abund_peak} we have seen that we can bring
\textit{satellites} and \textit{nearby dwarfs} and \textit{field
  galaxies} in agreement with abundance matching relations using the
peak mass instead of the present day virial mass.  This shows that it is indeed mass-loss via stripping and thus a reduction in $V_{\rm max}$ which
causes the \textit{satellites} and \textit{nearby dwarfs} to depart
from the abundance matching relation. We investigate this further in
Fig. \ref{fig:mass_loss}. This plot shows the ratio of present day $V_{\rm max}(z=0)$ to its maximum over the whole cosmic time, $V_{\rm peak}$, as a function of distance from the host galaxy. There is a sharp transition around $\sim2 R_{200}$ where most of the \textit{satellites} and some of the \textit{nearby dwarf galaxies} show a $V_{\rm max}/V_{\rm peak}$ ratio of 0.8 while \textit{field objects} on the other hand nearly retain their peak value. The small reduction in $V_{\rm max}$ for the \textit{field} galaxies is mostly due to a loss of gas due to stellar feedback. The four \textit{field} galaxies with $V_{\rm max}/V_{\rm peak}$ ratio around 70\% are objects which had a major dwarf dwarf merger and their $V_{\rm peak}$ value is biased high and thus the ratio of $V_{\rm max}$ to $V_{\rm peak}$ is biased low. We further checked that resolution does not affect our results (see section \ref{sec:peri} and \ref{sec:res} in the appendix).

The sharp transition between \textit{field} galaxies and \textit{nearby} galaxies is a hint that those \textit{nearby} galaxies did experience some tidal stripping by being ``backsplash" galaxies (e.g. they have spent some time inside the host halo). To test this hypothesis, in Fig. \ref{fig:rmin} we show the mimimum distance to the host against the present day distance from the host for \textit{satellites}, \textit{nearby dwarfs} and \textit{field galaxies}. We use the same symbols as in previous plots to distinguish galaxies from different populations and
color-code the points by the mass-loss quantified via the ratio of
present day total mass $M_{\rm halo}$ to peak mass $M_{\rm
  peak}$. This plot shows several distinct regions marked by gray
dashed lines. The plot is cut into half by the diagonal dashed gray
line. Every thing on the upper left side of that line is excluded
because the minimum distance can not be greater than the present day
distance. This line at the same time marks the position of galaxies
that are today at their minimum distance and on their first infall. The lower right part of the plot is separated into three distinct regions: 

\begin{figure}
\includegraphics[width=\columnwidth]{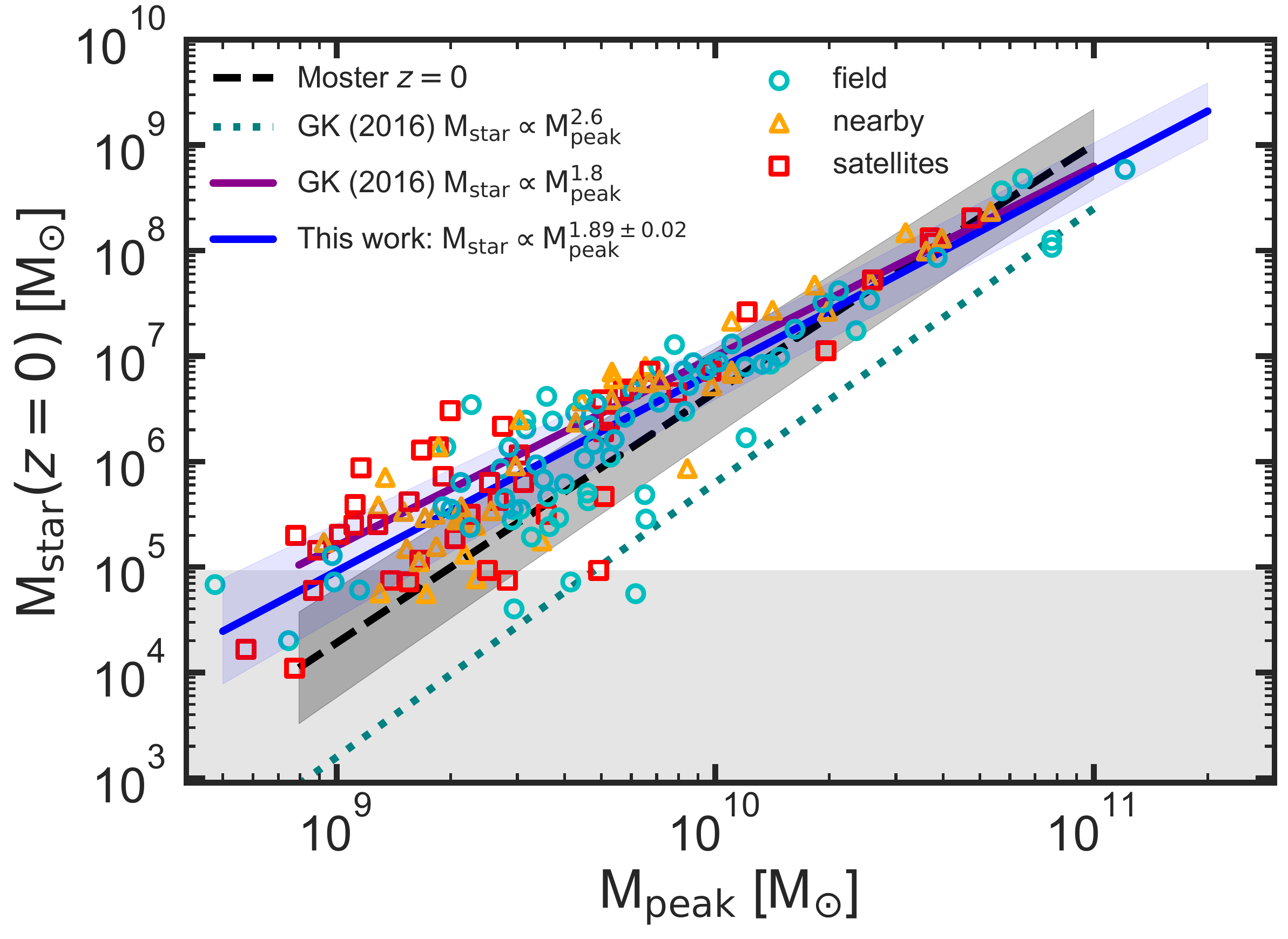}
\vspace{-.35cm}
\caption{Present day stellar mass $M_{\rm star}$ vs. the maximum total
  mass ever reached (peak mass) $M_{\rm peak}$. Colored points show
  the results for simulated galaxies with the same color coding as
  applied to previous plots. The black dashed line shows the result
  from \protect\cite{Moster2013} at redshift $z=0$. The purple and turquoise
  lines are taken from \protect\cite{Kimmel2017}. The blue solid line
  indicates a power law fit to the simulation data. The blue shaded
  area indicates the scatter of the simulation data around this fit
  and the gray shaded area at the bottom of the plot marks again our
  fiducial resolution limit of less than 10 stellar particles.}
\label{fig:abund_peak}
\end{figure}
 
%
\begin{description}
\setlength\itemsep{0em}
\item (1.) Satellites: present day distance is smaller than the host's virial radius (lower left).
\item (2.) Infalling field: present day distance and the minimum distance are larger than the host's virial radius (upper right).
\item (3.) Backsplash field: present day distance is larger than the host's virial radius but the minimum distance was smaller than the host's virial radius at earlier times (lower right).
\end{description}

Part 1) is located on the lower left and is filled, by definition,
with only \textit{satellites}. Part 2) is located on the upper right
part of the allowed region and inhabited only by \textit{field
  objects}. Almost all of them line up on the first infall line. Part
3) is the region where we find the backsplash galaxies which
have a minimum distance smaller than the present day distance. Indeed,
we learn from this plot that about 75\% of the dwarf galaxies are
backsplash galaxies having spent some time within the virial radius.

\begin{figure}
\includegraphics[width=\columnwidth]{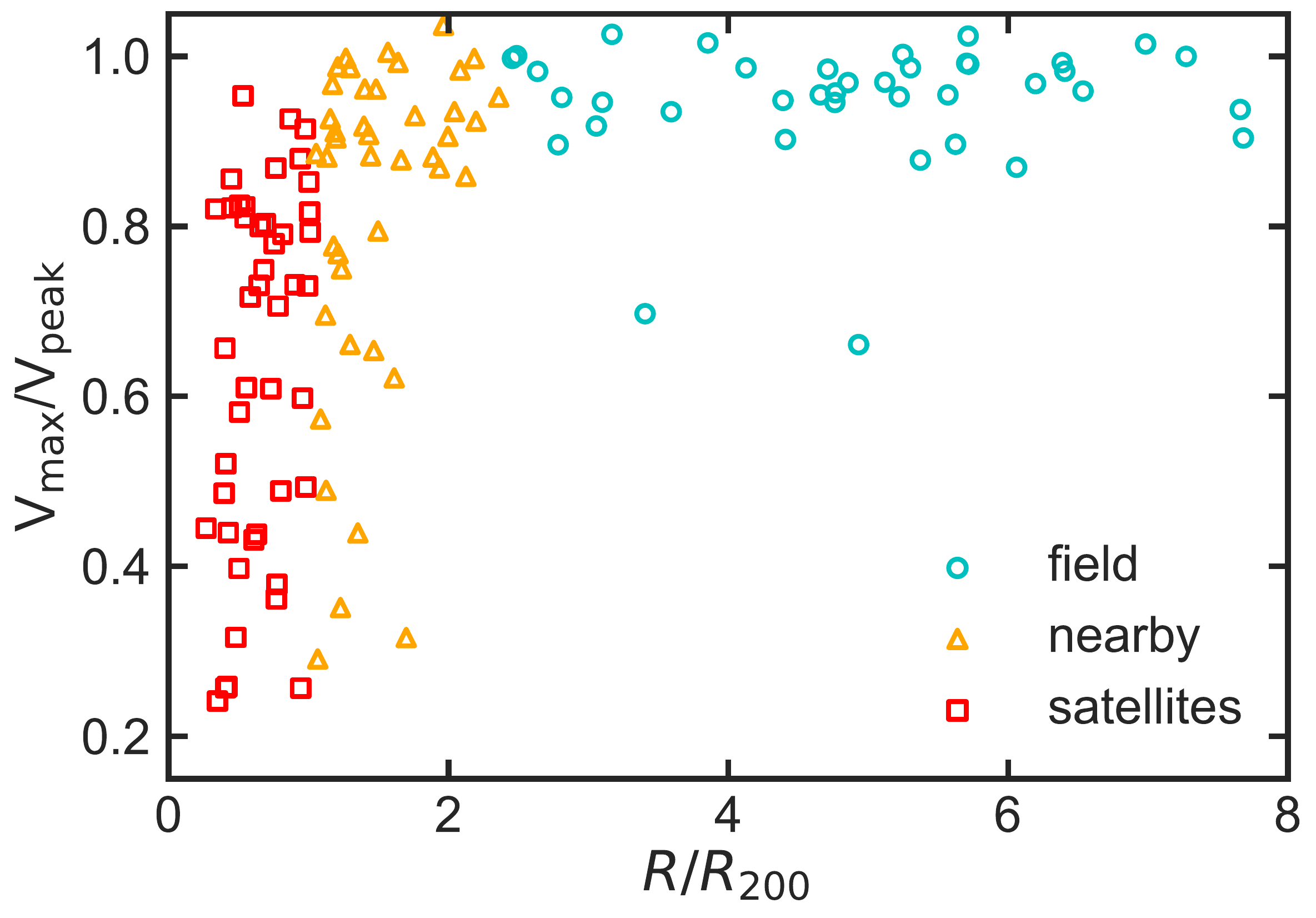}
\vspace{-.35cm}
\caption{The ratio of the present day maximum of the rotation curve
  $V_{\rm max}$ and the maximum of the same ever reached during the
  lifetime of the galaxy $V_{\rm peak}$ as a function of the distance from the host at $z=0$. This ratio is a proxy for the mass lost from the galaxies
  compared to their maximum mass reached. We adopt the same symbols
  and color coding convention as in previous plots
  (\textit{satellites} red squares, \textit{nearby dwarfs} orange
  triangles and \textit{field galaxies} cyan dots). The closer
  galaxies are to the host at present day the more severe is their
  mass loss.  }
\label{fig:mass_loss}
\end{figure}

\begin{figure*}
\centering
\includegraphics[width=.75\textwidth]{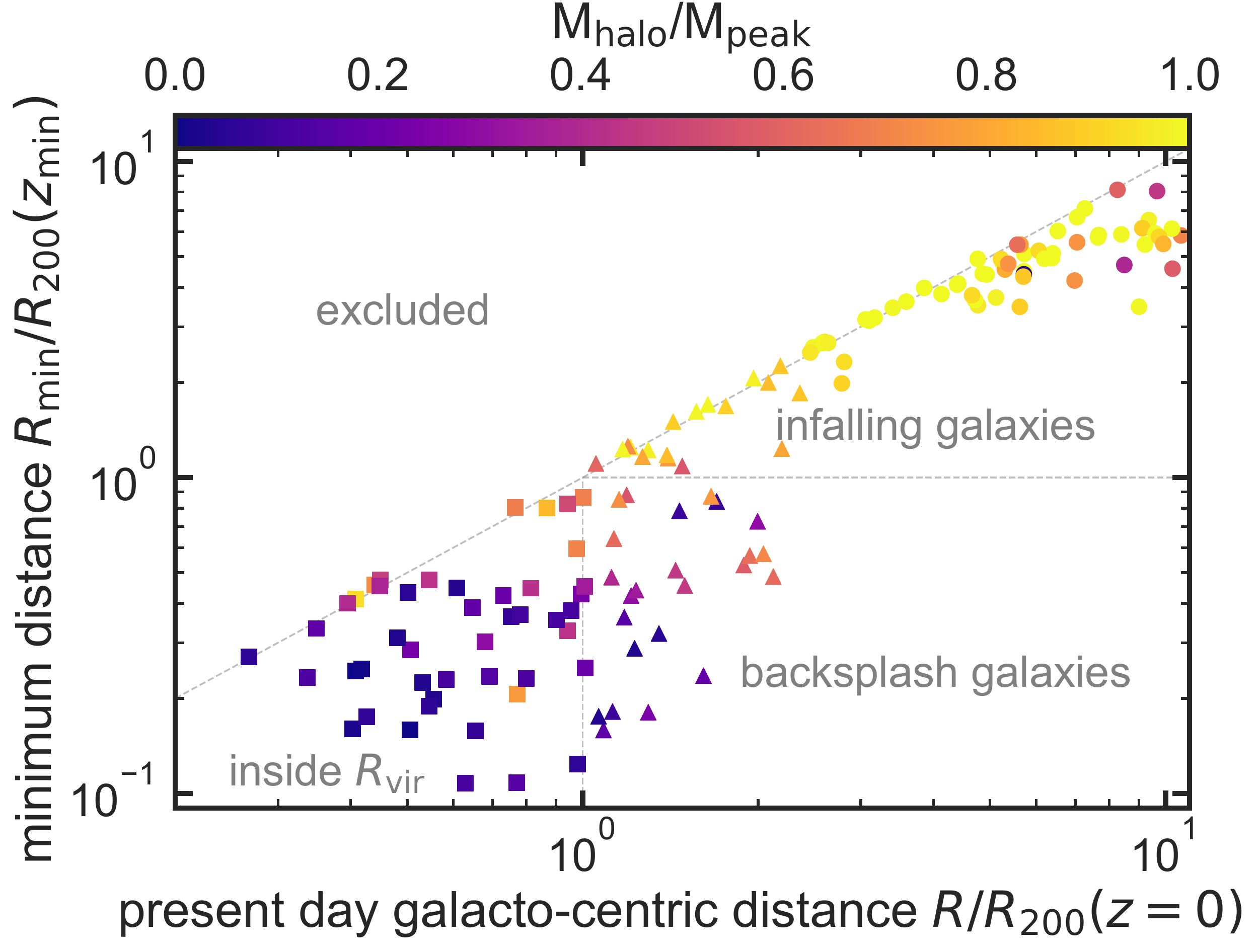}
\vspace{-.35cm}
\caption{The minimum distance to the host galaxy in terms of the
  virial radius at the time this distance was reached vs. the present
  day distance in terms of the present day virial radius. Different
  symbols mark \textit{satellites} (squares), \textit{nearby dwarfs}
  (triangles) and \textit{field galaxies} (dots). The color coding
  shows the fraction of present day virial mass in terms of the peak
  mass $M_{\rm halo}/M_{\rm peak}$. There is a clear color gradient in
  y-direction indicating a correlation of mass loss with minimum
  distance to the host. Gray dashed lines divide the plot into four
  different regimes, separating three different populations of
  galaxies, \textit{satellites} (lower left), \textit{backsplash
    galaxies} (lower right) and \textit{infalling dwarfs} (upper
  right).  }
\label{fig:rmin}
\end{figure*}

Looking at the color-coding of this plot we can draw four conclusions
for the mass loss of the galaxies: 
\begin{enumerate}
\setlength\itemsep{0em}
\item There is only a color gradient in y-direction but not in
  x-direction. Thus, mass loss directly correlates with the minimum
  distance and not with the present day distance. This shows that
  close encounters with the host is the primary cause for mass-loss.
\item Looking only at the \textit{nearby dwarf galaxy} population, we
  see that only galaxies which entered the host's virial radius show
  strong mass-loss. This explains the fact that some \textit{nearby
    dwarfs} show deviations from the Moster relation and some
  not. Their dependence of mass loss on the minimum distance shows
  that it is the pericenter distance which sets the amount of mass
  lost.
\item Almost all \textit{field objects} are on first infall. Only at
  very large distances $R/R_{\rm 200}>5$ we see some deviation from
  this behavior. This suggests that a clear signal for being
  unaffected by the host is a positive radial velocity.
\item At very large distances from the host we see some field objects
  showing mass-loss of about 20\% to 30\%. These objects were part of
  merging systems at earlier times which leads to an artificially high
  peak mass and thus to higher values of mass-loss \citep[see
    also][]{Lee2017}.
\end{enumerate} 

Fig. \ref{fig:mass_loss} together with Fig. \ref{fig:rmin} clearly
shows that the deviation from the abundance matching relation observed
in Fig. \ref{fig:abund1} is caused by (tidally) stripping dark matter
mass from the galaxies which entered the virial radius of the
host. Most remarkably the effects of mass loss are also strongly
visible for the backsplash galaxies which only spent a short
time (one pericenter passage) inside the host. We therefore advocate
to carefully select the dwarf galaxies in the Local Group in order to
be unaffected by the host and to be compared to \LCDM predictions to
test dark matter theories.   

\begin{figure}
\includegraphics[width=\columnwidth]{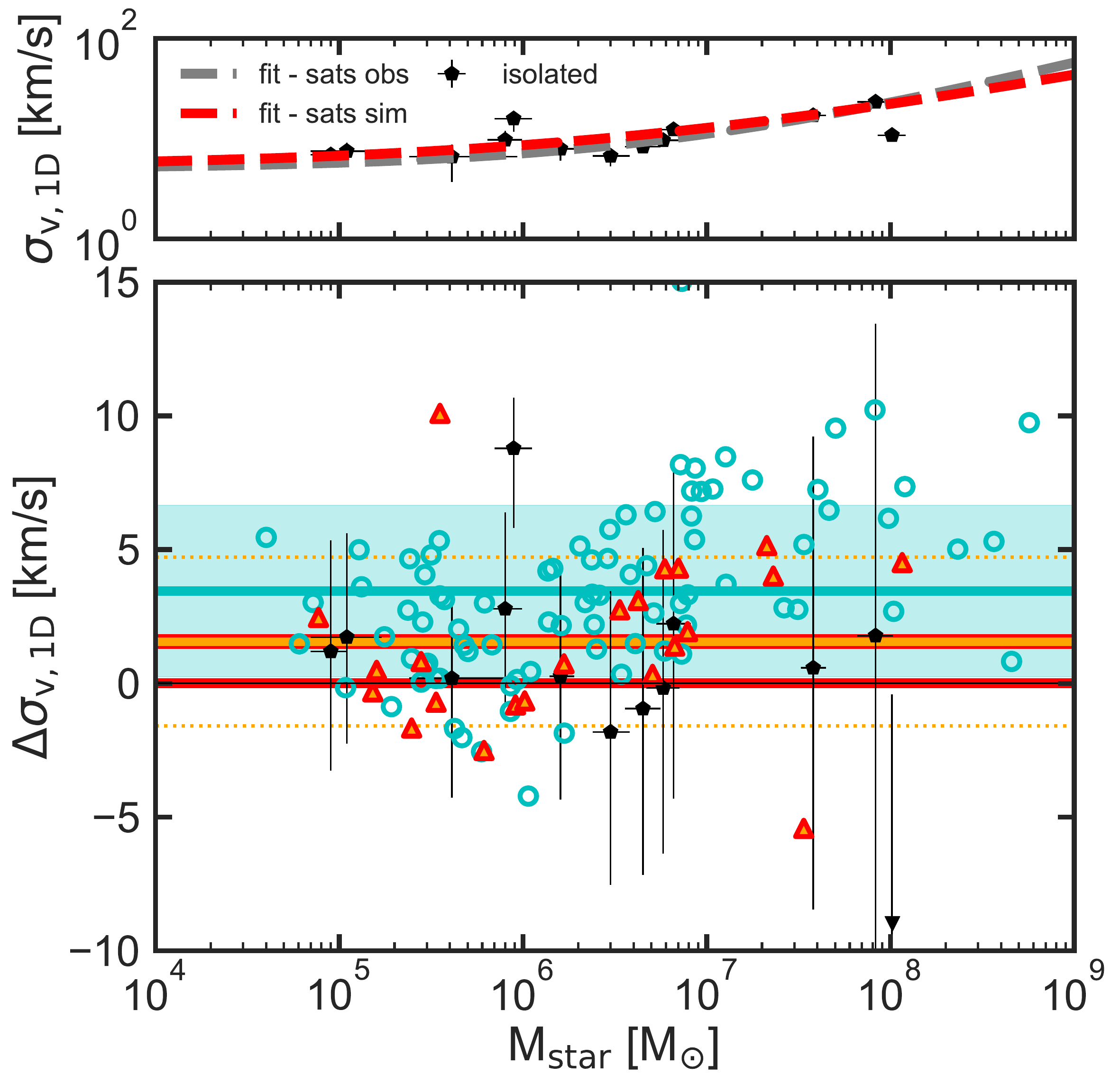}
\vspace{-.35cm}
\caption{\emph{Upper panel:} Stellar mass vs. 1D velocity dispersion for isolated Local Group galaxies similar to Fig. \ref{fig:vdisp}. With dashed lines we show fits to the velocity dispersion of the (\textit{satellites} in the simulations (red) and in the observations (gray).
\emph{Lower panel}: Deviations of the measured velocity dispersion for simulated galaxies from the relation of the simulated \textit{satellite galaxies}. The red horizontal line at $\Delta\sigma_{\rm v, 1D}=0$ shows the average relation for simulated \textit{satellites}, the orange/red line indicates the average deviation for \textit{backsplash} galaxies and the blue line the deviation for \textit{field} galaxies. The blue shaded area indicates the scatter for the \textit{field} population and the orange dotted lines indicate the scatter for the \textit{backsplash galaxies}. The deviation of observed isolated Local Group galaxies from the relation for observed satellites is shown by black pentagons. \textit{Satellite galaxies} show at a given stellar mass lower velocity dispersions compared to the \textit{field} sample which is indicative of having lost dark matter mass due to tidal forces. Individual galaxies from the \textit{field sample} are shown with blue open circles. \textit{Backsplash galaxies}, dwarf galaxies which have spent some time inside the virial radius of their host, are highlighted with orange triangles with red borders.}
\label{fig:vdisp_fit}
\end{figure}

\section{Implications for Local Group galaxies}
\label{sec:prediction}

\begin{figure}
\includegraphics[width=\columnwidth]{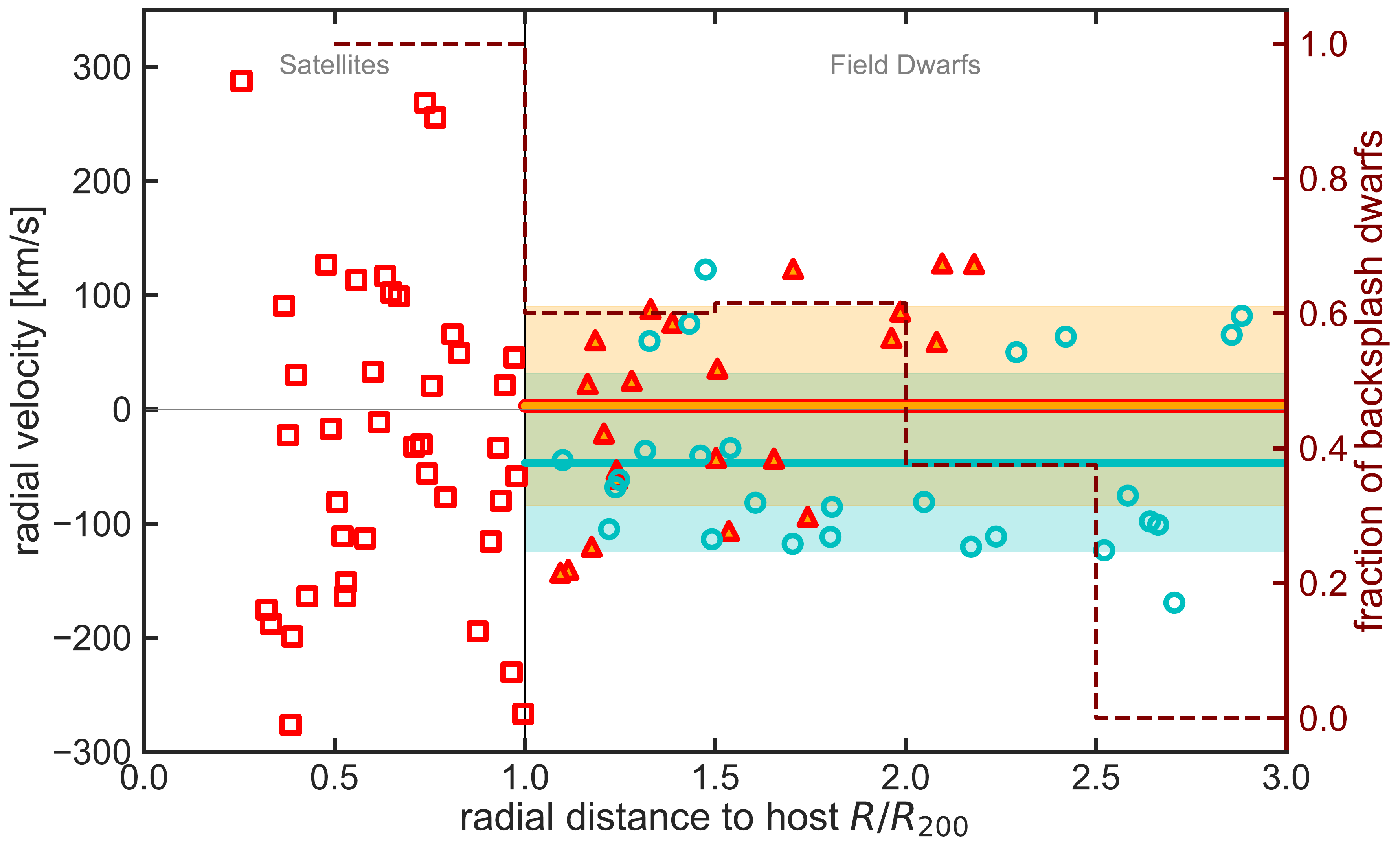}
\vspace{-.35cm}
\caption{Radial velocity vs. distance from the host galaxy for
  simulated galaxies (colored points). 
  For the simulations we adopt the same symbol
  convention as in previous plots but this time we only distinguish
  between \textit{satellites} (red squares), \textit{backsplash
    galaxies} (orange/red triangles) and \textit{field galaxies} (cyan
  dots). The red/orange line shows the mean radial velocity for
  \textit{backsplash galaxies} while the cyan line shows it for the
  \textit{field population}. Qualitatively
  \textit{backsplash galaxies} show more positive radial velocities
  compared to \textit{field galaxies}. The maroon dotted line (right
  y-axis) shows the fraction of \textit{backsplash galaxies} as a
  function of distance from the host for our simulations.}
\label{fig:LG_strip}
\end{figure}

We have established and quantified the environmental influence of the host galaxy on its satellite and dwarf galaxies. We now turn to discuss the implications of our findings for the Local Group. Because of their proximity the small galaxies of the Local Group are among the most well studied systems we know. However, environmental effects of the Milky Way or M31 on their small companions are mostly neglected, especially for dwarf galaxies outside the assumed virial radius of the hosts. The galaxies outside $\sim300$ kpc of MW and M31 were long thought of as clean representatives of field dwarf galaxies. In this work however, we have shown that this zeroth order approximation breaks down for a significant fraction ($\sim50-80\%$, see e.g. right axis of Fig. \ref{fig:LG_strip}) of galaxies as close as 2.5 $R_{200}\sim$ 450 kpc to their host. The dwarf galaxies in this region show similar properties to satellite galaxies with signs of significant mass loss (e.g. shown in Fig. \ref{fig:abund1} and Fig. \ref{fig:rmin}), early quenching of their SFH (Fig. \ref{fig:sfh}) and reduced gas fractions (Fig. \ref{fig:gas_frac}) compared to our clean sample of field galaxies. In the following subsections we will discuss possible signatures to distinguish between \textit{backsplash galaxies} and clean field galaxies.

\subsection{Distinguishing \textit{backsplash galaxies} from unaffected \textit{field galaxies}}

\subsubsection{Reduction in velocity dispersion - One way to identify \textit{backsplash galaxies}?}

We like to get a handle on how to observationally separate
\textit{backsplash galaxies} from unaffected \textit{field
  galaxies}. One possibility was pointed out by
\cite{Knebe2011} who found that \textit{backsplash galaxies} have
lower mass-to-light ratios compared to \textit{field galaxies}. However, getting firm estimates of the dynamical mass of dwarf galaxies is still worrisome and involves careful mass modeling
and accounting for inclination effects \citep[e.g.][]{Oman2015}. One rather easy way around this would be to use the line-of-sight velocity dispersion easily accessible by spectroscopic measurements.

In section \ref{sec:mass_loss} we showed that \textit{backsplash galaxies} show significant mass loss due to their close encounter with their host and as was recently shown by \cite{Frings2017} mass loss results in a drop in stellar velocity dispersion. Thus one obvious choice of discriminating variable would be the stellar
line-of-sight velocity dispersion which should on average be lower for \textit{backsplash galaxies} compared to \textit{field galaxies} at the same stellar mass, or equivalently higher compared to \textit{satellites}. We test this assumption by fitting an exponential of the form
\begin{equation}
\sigma_{\rm v}\left(M_{\rm star}\right) = A + B \exp{C\log\left(M_{\rm star}\right)}
\end{equation}
to the velocity dispersion-$M_{\rm star}$ relation obtained from our simulation in order to take out the mass dependence. In Fig. \ref{fig:vdisp_fit} we show the resulting fits (upper panel) for \textit{satellite} galaxy populations in the simulation (red dashed line) and observations (gray dashed line) and the
average deviation of populations from the fitted main relation (lower panel). From the upper panel of this figure we see that the fits for the \textit{satellite galaxies} in simulations and observations agree well. However, given the large scatter in velocity dispersion, the more robust test is the deviation from the fitted relation. In the lower panel we show the residuals if we subtract the fitted main relation for satellites from the simulated dwarfs (colored symbols) and the observed dwarfs (black symbols). We find that the \textit{backsplash galaxies} (orange triangles with red borders) scatter around zero while the \textit{field galaxies} show on average positive residuals. This means at a given stellar mass \textit{backsplash galaxies} have on average lower velocity dispersions compared to \textit{field galaxies}, well in agreement with the findings of \cite{Frings2017}. However, due to the large scatter \textit{backsplash galaxies} overlap with low velocity dispersion \textit{field galaxies} and there is no clear separation between \textit{field galaxies} and \textit{backsplash galaxies}. Thus, the velocity dispersion alone can not serve as a
discriminating variable on an object by object basis. Having low
velocity dispersion for a given stellar mass is only indicative for mass loss. Given the low number statistics of measurements for Local Group field galaxies (11) it is not possible to make a firm prediction for which galaxies might be \textit{backsplash galaxies} based on the velocity dispersion alone. 

\subsubsection{Distance vs. radial velocity - Another way to separate \textit{backsplash galaxies} from \textit{field objects}?}

\begin{figure}
\includegraphics[width=\columnwidth]{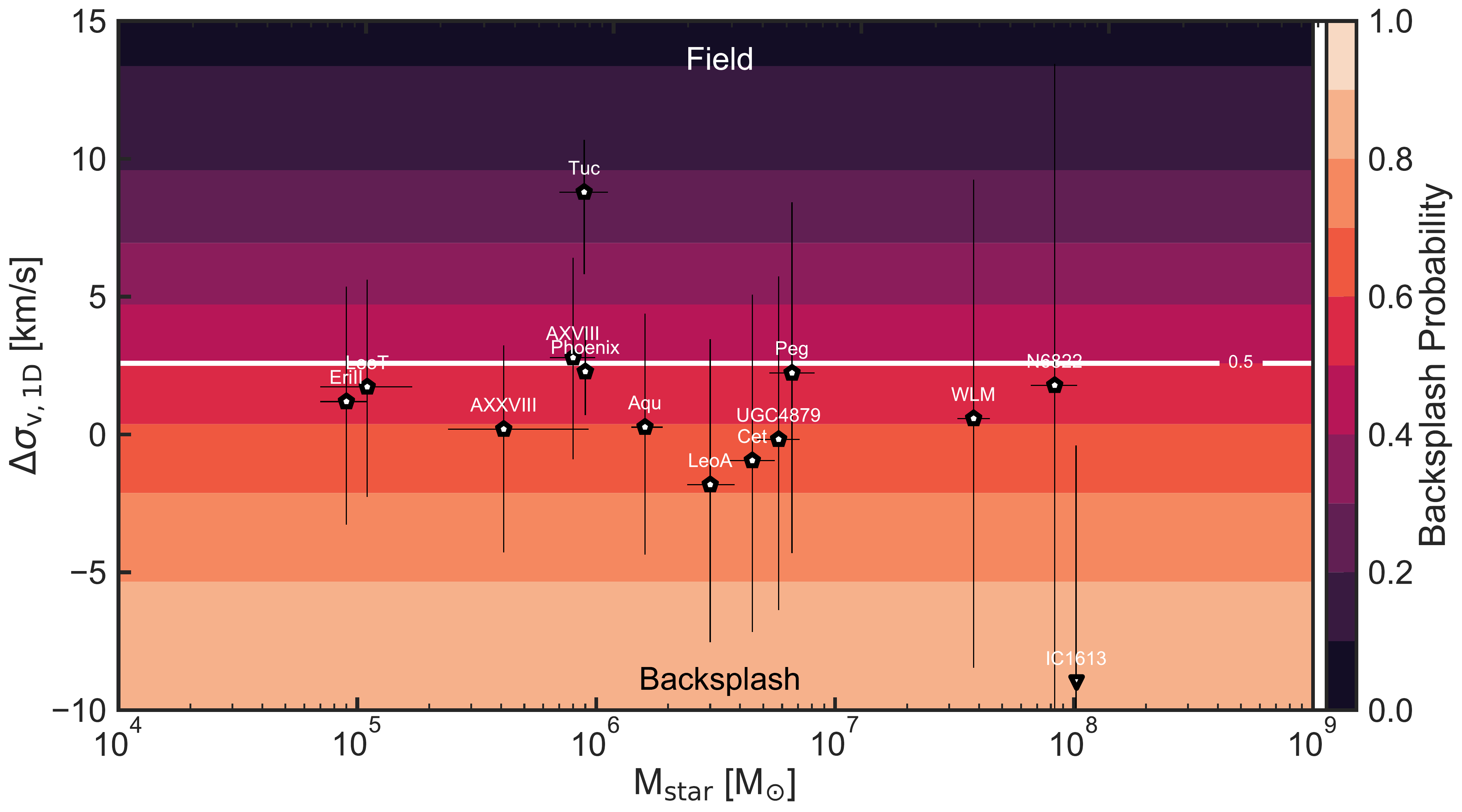}
\vspace{-.35cm}
\caption{Same as the lower panel of Fig. \ref{fig:vdisp_fit} but this time we only show the observed isolated Local Group galaxies. The background color shows the estimated probability to be a \textit{backsplash galaxy} based on the velocity dispersion offset $\Delta\sigma_{\rm v,1D}$ as derived from our simulations.}
\label{fig:vdisp_fit_prob}
\end{figure}

\cite{Teyssier2012} suggested, based on
an analysis of the dark matter only simulation Via Lactea II
\citep{Kuhlen2009}, that \textit{backsplash haloes} and \textit{field haloes} show, at a given distance from the host, different values of radial velocities. While \textit{field haloes} in the vicinity of the host (300kpc up to 900 kpc) show on average negative radial velocities, \textit{backsplash haloes} have positive velocities. These findings are then used to estimate which Local Group galaxy might be a
\textit{backsplash galaxy} \citep[see also][]{Pawlowski2014c}. Here we test this behavior for our set of luminous galaxies. In Fig. \ref{fig:LG_strip} we show the radial velocity as a function of distance from the host for our dwarf galaxy populations (\textit{satellites} in red squares, \textit{field dwarfs} cyan dots, \textit{backsplash galaxies} orange/red triangles). On average \textit{backsplash galaxies} show indeed a higher value of radial velocity compared to \textit{field galaxies} but the scatter is large and our sample size
small. Thus there is a large overlap of the two populations in this plot and again, these results are only suggestive of being a
\textit{backsplash galaxy}. This leaves the only firm measurement/test of being affected by the host through the total mass of the galaxies and their deviation from abundance matching relations. However, we are far from being able to infer the total masses of observed field galaxies robustly.

Another result we can take away from Fig. \ref{fig:LG_strip} is the fact that we predict a large fraction of \textit{backsplash galaxies} in the vicinity of the Milky Way or M31. In our simulations the fraction of \textit{backsplash galaxies} between 1 and 2.5 virial radii is about 80 to 50 per cent. Therefore, in the next subsection we estimate the probability of Local Group dwarf galaxies to be a \textit{backsplash} galaxy based on their radial velocity, distance and velocity dispersion.

\subsection{Probability of being a \textit{backsplash galaxy} for observed Local Group dwarf galaxies}

\begin{figure}
\includegraphics[width=\columnwidth]{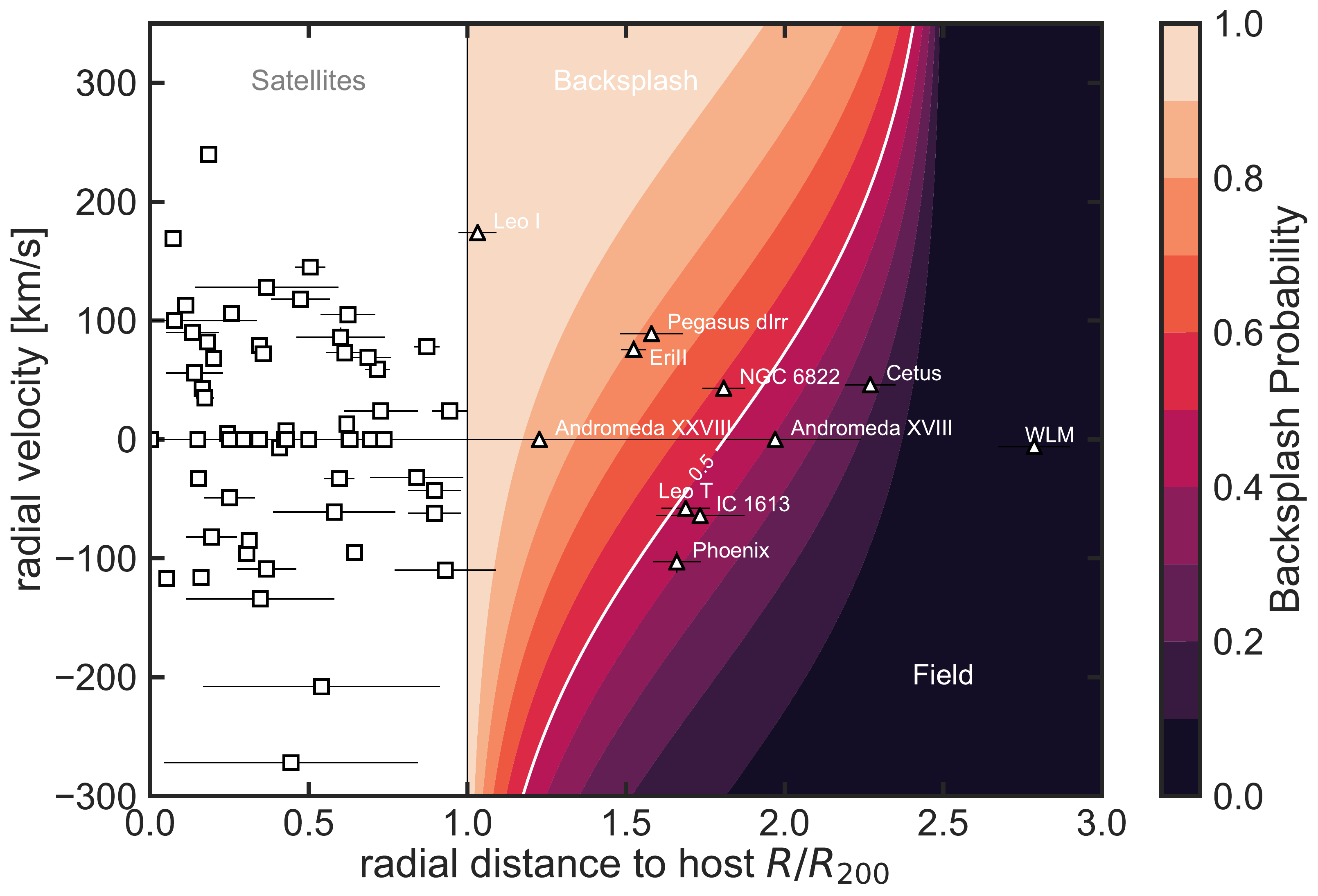}
\vspace{-.35cm}
\caption{Radial velocity vs. distance from the host galaxy for Local Group observations (black open points). The background color shows the probability for given distance and radial velocity to be a \textit{backsplash} galaxy as derived from our simulations. Especially the newly discovered Milky Way dwarf galaxy EriII has a high probability to be a \textit{backsplash} galaxy.}
\label{fig:LG_strip_prob}
\end{figure}

Having quantified the differences between \textit{backsplash galaxies} and unaffected \textit{field galaxies} in the parameters space distance from the host, radial velocity and velocity dispersion of the stars, we calculate the probability for the observed Local Group galaxies to be a \textit{backsplash galaxy} and thus being tidally affected by the host based on these three properties. For the velocity dispersion and radial velocity we assume a Gaussian distribution $\mathcal{G}(\mu,\sigma)$ around the mean relations derived for each population in the previous subsection. We compute the probability by a simple ratio of the distributions:
\begin{equation}
P_{\rm BS}(x_{\rm i})=\frac{\mathcal{G}(x_{\rm i},\mu_{\rm BS},\sigma_{\rm BS})}{\mathcal{G}(x_{\rm i},\mu_{\rm BS},\sigma_{\rm BS}) + \mathcal{G}(x_{\rm i},\mu_{\rm F},\sigma_{\rm F})}
\end{equation}
Here $\mathcal{G}(x_{\rm i}\mu_{\rm BS},\sigma_{\rm BS})$ denotes the \textit{backsplash galaxy} distribution and $\mathcal{G}(x_{\rm i},\mu_{\rm F},\sigma_{\rm F})$ the \textit{field galaxy} distribution. $x_{\rm i}$ denotes either the velocity dispersion offset $\Delta\sigma_{\rm v, 1D}$ or the radial velocity value of galaxies. All relevant values extracted from our simulations and needed to calculate these probabilities are given in table 3. We use the fraction of \textit{backsplash} galaxies as a function of distance as shown in Fig. \ref{fig:LG_strip} in order to derive the probability of being a \textit{backsplash} galaxy as a function of distance from the host. For simplicity we assume a linear decrease with distance, from 100\% at the virial radius (here: 250 kpc for MW and 300 kpc for M31) to 0\% at 2.5 virial radii. 

\begin{figure*}
\includegraphics[width=\textwidth]{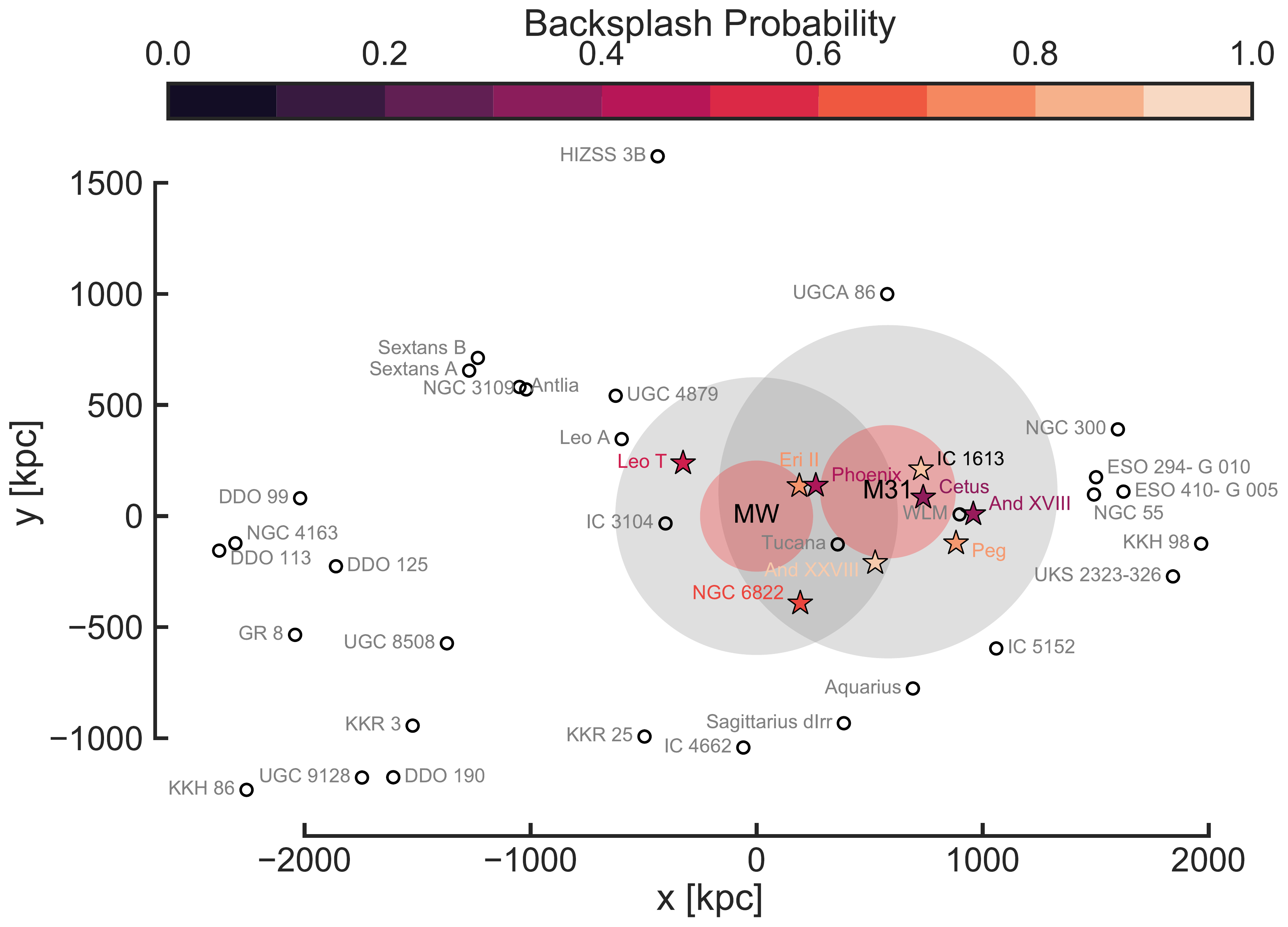}
\vspace{-.35cm}
\caption{Top down view onto the Local Group such that Milky Way's stellar disc lies in the $x$-$y$ plane. Observations are taken from
  \protect\cite{McConnachie2012}. Red circles show the virial radius
  of the Milky Way (250 kpc) and M31 (300 kpc) and gray circles
  show their corresponding backsplash radii of $2.5\times R_{\rm
    200}$. Colored stars (with gray names) mark nearby Local Group galaxies (within $2.5\times R_{\rm 200}$) where the color-coding shows the estimated probability to be a \textit{backsplash galaxy}. Open black circles show Local Group galaxies which we quantify as clean field galaxies. We do not show satellite galaxies on this plot which would fall into the red circles.}
\label{fig:LG}
\end{figure*}

The combined probabilities are calculated by multiplying single probabilities and re-normalizing them in the following way:
\begin{equation}
P_{\rm BS}(x_{\rm i},x_{\rm j}) = \frac{P_{\rm BS}(x_{\rm i})P_{\rm BS}(x_{\rm j})}{P_{\rm BS}(x_{\rm i})P_{\rm BS}(x_{\rm j}) + \left(1-P_{\rm BS}(x_{\rm i})\right)\left(1-P_{\rm BS}(x_{\rm j})\right)}
\end{equation}
Numerical values for the derived probabilities can be found in table 4 and graphical representations are given in in Fig. \ref{fig:vdisp_fit_prob}, Fig. \ref{fig:LG_strip_prob} and Fig. \ref{fig:LG}. In Fig. \ref{fig:vdisp_fit_prob} we show the probability for the velocity dispersion offset. We color-code the $\Delta_{\sigma, \rm 1D}$ vs. $M_{\rm star}$ plane by the probability of being a \textit{backsplash galaxy} which in our assumption only depends on the deviation from the \textit{satellite} relation. Except for IC 1613 all observed \textit{field galaxies} have about 50\% probability to be a \textit{backsplash galaxy} given only their velocity dispersion measurements. This reflects the fact that velocity dispersion alone is not able to discriminate between \textit{backsplash} and \textit{field galaxies}. In \ref{fig:LG_strip_prob} we show the combined probability of being a \textit{backsplash} galaxy in the radial velocity vs. distance plane. Every galaxy which is closer to the host than 1 virial radius is by definition a \textit{satellite} and is excluded. We do not find any \textit{backsplash galaxy} further away than 2.5 virial radii and thus the probability for galaxies with larger distances from the host than this is 0. In between we find that the probability of being a \textit{backsplash galaxy} decreases with distance but increases with more positive radial velocity. We see that Leo I is basically a satellite given its distance and radial velocity. We find that Pegasus, Eri II, NGC 6822 and And XXVIII have a high probability of being a \textit{backsplash galaxy}. Of particular interest is the result for the newly discovered dwarf Eri II. In the abstract of their paper about properties of Eri II \citet{Li2017} state: \textit{The lack of gas and recent star formation in Eri II is surprising given its mass and distance from the Milky Way}. Here however, we find that Eri II is most likely a \textit{backsplash galaxy} and given that fact it is not surprising at all that this dwarf galaxy shows a lack of recent star formation and cold gas. Actually, given Eri II's radial velocity and distance we expect it to show a low cold gas content and a lack of recent star formation. We like to note here that recent studies by \citet{Fritz2018} estimate the peri- and apocenter of Eri II using Gaia DR2 data \citep{GaiaDR2} indicating that Eri II might be currently at pericenter. If this is to be true, this might indicate that Eri II might actually be a dwarf galaxy that had a close encounter with M31 previously to be accreted onto the MW \citep[see e.g.][]{Knebe2011c} which would then explain its properties. Another explanation for Eri II's properties might be group pre-processing by satellite-satellite interactions as suggested by \citet{Wetzel2015}. However, here we do not find that group pre-processing is important as previously found by \citet{DeJong2010} using cosmological simulations.

Finally, we compile our total probability of being a \textit{backsplash galaxy} for Local Group galaxies in Fig. \ref{fig:LG}. Here we show a top-down view of the Local Group. We exclude all satellite galaxies of M31 and MW and indicate the virial radius with red filled circles ($\sim250$ kpc for MW and $\sim300$ kpc for M31). The region of 2.5 virial radii around MW and M31 is shown with gray filled circles and indicates the backsplash radius \citep{More2015,Diemer2017,Mansfield2017}. Colored stars show nearby dwarf galaxies with the color-coding indicating the total probability of being a \textit{backsplash galaxy}. Open black circle show all the dwarfs outside 2.5 virial radii which have by definition a probability of 0\%.

\begin{table}
\label{tab:props}
\begin{center}
\caption{Mean and standard deviation for the velocity dipsersion offset and the radial velocity of \textit{backsplash} and \textit{field} galaxies.}
\begin{tabular}{l c c c c}
		\hline\hline
		 & $\mu_{\rm BS}$ & $\sigma_{\rm BS}$ & $\mu_{\rm F}$ & $\sigma_{\rm F}$ \\
          & [km/s] & [km/s] & [km/s] & [km/s] \\
		\hline
        $\Delta_{\sigma, \rm 1D}$ & 1.56 & 3.16 & 3.45 & 3.21 \\
        radial velocity & 3.03 & 87.16 & -46.74 & 78.06 \\
        \hline
\end{tabular}
\end{center}
\end{table}

\begin{table}
\label{tab:props}
\begin{center}
\caption{Probabilities of being a \textit{backsplash} galaxy for nearby Local Group galaxies. We derive probabilities based on distance ($P_{\rm d}$), radial velocity ($P_{\rm rv}$) and velocity dispersion ($P_{\sigma}$) as well as a combination of distance and radial velocity ($P_{\rm rv, d}$) and all three properties ($P_{\rm tot}$).}
\begin{tabular}{l c c c c c}
		\hline\hline
		galaxy & $P_{\rm d}$ & $P_{\rm rv}$ & $P_{\sigma}$ & $P_{\rm rv, d}$ & $P_{\rm tot}$ \\
		\hline
        Phoenix & 0.56 & 0.36 & 0.51 & 0.41 & 0.43\\
        NGC 6822 &  0.46 & 0.62 & 0.54 & 0.58 & 0.61\\
        Leo T & 0.54 & 0.43 & 0.54 & 0.48 & 0.51\\
        And XXVIII & 0.85 & 0.54 & 0.61 & 0.87 & 0.91\\
        IC 1613 & 0.51 & 0.42 & 0.92 & 0.43 & 0.90\\
        Cetus & 0.15 & 0.62 & 0.66 & 0.23 & 0.36\\
        Pegasus & 0.61 & 0.69 & 0.52 & 0.78 & 0.79\\
        WLM & 0.0 & 0.52 & 0.59 & 0.0 & 0.0\\
        And XVIII & 0.35 & 0.54 & 0.49 & 0.39 & 0.38\\
        EriII & 0.65 & 0.67 & 0.56 & 0.79 & 0.83\\
        \hline
\end{tabular}
\end{center}
\end{table}

\section{Conclusion}
\label{sec:conclusion}

In this study we explored the environmental effects of the host galaxy onto its satellites and dwarf galaxies. According to their distance to the host we distinguish \textit{satellites} which are inside the virial radius at present day, \textit{nearby dwarf galaxies} which are in between 1 and 2.5 virial radii (which is the expected backsplash radius of the host) and \textit{field galaxies}, outside of 2.5 virial radii from the host. The reason for having three different populations of dwarf galaxies lies in the environmental effect of the host on its companions. While \textit{satellite galaxies} and \textit{field dwarf galaxies} share the same scaling relations, we verify that the former suffered stronger mass loss compared to the latter. \textit{Field galaxies} are mostly on their first infall and almost unaffected by the host. \textit{Nearby dwarf galaxies} make up an intermediate population well outside the virial radius at present day but with a large fraction of \textit{backsplash galaxies}. The orbits of these galaxies took them inside the virial radius of the host at earlier cosmic times (see Fig. \ref{fig:rmin}). We show that this population is strongly tidally affected and shows properties more similar to \textit{satellite galaxies} than to \textit{field objects}. Especially careful comparison shows that their velocity dispersion is on average lower compared to similar objects at the same distance and mass. Such effects should be considered when selecting observed field objects to confront with \lcdm predictions.

Our results are summarized as follows:
\begin{itemize}
\item The NIHAO simulation project, at very high resolution, is able to
  produce realistic host satellite galaxy systems. Our simulations
  compare well with observed satellite stellar mass functions
  (Fig. \ref{fig:mass_func}) of Milky Way or M31.
\item The properties of \textit{satellites}, \textit{nearby dwarfs} and \textit{field galaxies}, like e.g. the stellar mass-metallicity relation and the stellar mass-velocity dispersion relation (Fig. \ref{fig:vdisp}) or their star formation histories (see Fig. \ref{fig:sfh}) agree well with observed properties of Local Group galaxies. 
\item Interestingly, \textit{satellite galaxies} and \textit{field galaxies} exhibit almost the same mass-metallicity and stellar mass-velocity dispersion relations. However, they differ in their SFH and cold gas fractions. \textit{Satellite galaxies} show an early stellar mass build up compared to \textit{field objects}, which show a more extended SFH. Furthermore, \textit{field galaxies} are on average more gas rich compared to \textit{nearby dwarf galaxies} or \textit{satellites} (compare Fig. \ref{fig:gas_frac}).
\item \textit{Field galaxies} almost perfectly line up with
  results from abundance matching while \textit{satellite
    galaxies} and \textit{nearby dwarfs} deviate from these
  results (compare Fig. \ref{fig:abund1}, \ref{fig:abund2}). In our simulations the reason for this is significant mass
  loss due to tidal stripping which results in a (strong) reduction of the maximum circular velocity of the dark matter haloes ($V_{\rm max}$, Fig. \ref{fig:mass_loss}).
\item Accounting for the mass loss by using the peak mass (the maximum mass ever reached) of the galaxies, all populations shift back onto the same abundance matching relation (see Fig. \ref{fig:abund_peak}). For our dwarf galaxies we find a power law index of $1.89\pm0.02$ for the relation between halo mass $M_{\rm halo}$ and stellar mass $M_{\rm star}$ which is well in agreement with recent results from \cite{Kimmel2017}. 
\item We use this relation together with the fact that stellar mass loss is negligible in the simulations to estimate the peak masses of Local Group galaxies; our predicted peak masses are tabulated in table \ref{tab:peak_mass}. 
\item Due to the mass loss, \textit{backsplash galaxies}
  show on average lower stellar velocity dispersions compared to \textit{field objects} (see Fig. \ref{fig:vdisp_fit}). However, the differences are small and can not alone serve to judge if observed dwarf galaxies are stripped or not.
\item \textit{backsplash galaxies} differ from \textit{field galaxies} in their position in the radial velocity-distance plot (\ref{fig:LG_strip}). \textit{Backsplash galaxies} show on average positive radial velocities while \textit{field objects} show negative radial velocities.

\item Clean \textit{field dwarfs} can only be found further away than 2.5 virial radii ($\sim750$ kpc) from any host in the Local Group.
\item Given the distance from the host, the radial velocity and the velocity dispersion we derive a probability for Local Group dwarfs to belong to the \textit{backsplash} population. Our findings are compiled in table 4 and graphically represented in Fig. \ref{fig:LG}. The newly discovered dwarf Eri II has a probability of 83\% to be a \textit{backsplash} galaxy. This fits well to its observed properties which are unusual for its distance from the MW.
\end{itemize}

\section*{acknowledgments}
We like to thank Hans-Walter Rix for fruitful discussions and very helpful comments on this work. TB likes to thank Marcel Pawlowski, Anna Bonaca and Nicolas Martin for helpful comments and suggestions on this work.
This research made use of the {\sc{pynbody}} package \citet{pynbody} to analyze the simulations and used the {\sc{python}} package {\sc{matplotlib}} \citep{matplotlib} to display all figures in this work. Data analysis for this work made intensive use of the {\sc{python}}  library  {\sc{SciPy}} \citep{scipy}, in particular {\sc{NumPy and IPython}} \citep{numpy,ipython}.
TB acknowledges support from the  Sonderforschungsbereich SFB 881 “The Milky Way System” (subproject A2) of the German Research Foundation (DFG). AO is funded by the Deutsche Forschungsgemeinschaft (DFG, German Research Foundation) -- MO 2979/1-1. The authors gratefully acknowledge the Gauss Centre for Supercomputing e.V. (www.gauss-centre.eu) for funding this project by providing computing time on the GCS Supercomputer SuperMUC at Leibniz Supercomputing Centre (www.lrz.de). This research was carried out on the High Performance Computing resources at New York University Abu Dhabi; Simulations have been performed on the ISAAC cluster of the Max-Planck-Institut für Astronomie at the Rechenzentrum in Garching and the HYDRA and DRACO clusters at the Rechenzentrum in Garching. We greatly appreciate the contributions of all these computing allocations.
\bibliography{astro-ph.bib}

\appendix

\section*{Appendix}

\newpage
\onecolumn
\section{Properties of simulated dwarf galaxies}

\begin{ThreePartTable}
\begin{TableNotes}
\footnotesize
\item  The minimum distance $R_{\rm min}$ is given in terms of the virial radius at the time this distance is reached $R/R_{200}(z)$
\end{TableNotes}

\begin{longtable}{p{1.3cm}cccccccccccc}
  \caption{Parameters of simulated dwarf galaxies}\\
  \hline
  \hline
  $M_{\rm 200}$  &  $M_{\rm dark}$ & $M_{\rm star}$ & $M_{\rm gas}$ &  $V_{\rm max}$ & $\sigma_{V}^{\rm 1D}$ & $R$ & [Fe/H] & $z_{\rm infall}$ & $z_{\rm peak}$ & $M_{\rm peak}$ & $V_{\rm peak}$ & $R_{\rm min}$ \\
  $(10^9\Msun$) & $(10^9\Msun$) & $(10^6\Msun$) & $(10^6\Msun$) & (km/s) & (km/s) & (kpc) & (dex) &  &  & $(10^9\Msun$) & (km/s) & ($R/R_{200}$) \\
  \endfirsthead
  \multicolumn{5}{l}{{\bfseries\tablename\ \thetable{} -- (continued)}}\\
  \hline
  \hline
  $M_{\rm 200}$  &  $M_{\rm dark}$ & $M_{\rm star}$ & $M_{\rm gas}$ &  $V_{\rm max}$ & $\sigma_{V}^{\rm 1D}$ & $R$ & [Fe/H] & $z_{\rm infall}$ & $z_{\rm peak}$ & $M_{\rm peak}$ & $V_{\rm peak}$ & $R_{\rm min}$ \\
  $(10^9\Msun$) & $(10^9\Msun$) & $(10^6\Msun$) & $(10^6\Msun$) & (km/s) & (km/s) & (kpc) & (dex) &  &  & $(10^9\Msun$) & (km/s) & ($R/R_{200}$) \\
  \hline
  \endhead

 \insertTableNotes  

  \endlastfoot
  
  \hline

  &\\
\hline
      \multicolumn{13}{c}{g7.08e11}\\
\hline
\underline{{\bf satellites}} \\
$ 4.28$ & $4.25$ & $1.90 $ & $18.64$ & $28.2$ & $10.8$ & $156$ & $-2.23$ & $0.1$ & $0.1$ & $5.06$ & $33.3$ & $0.80$  \\
$ 1.98$ & $1.91$ & $17.09 $ & $46.75$ & $29.7$ & $13.7$ & $79$ & $-1.74$ & $0.0$ & $0.1$ & $2.74$ & $30.2$ & $0.46$  \\
$ 0.90$ & $0.90$ & $0.25 $ & $0.0$ & $23.5$ & $7.9$ & $176$ & $-3.75$ & $0.4$ & $1.6$ & $1.28$ & $29.8$ & $0.60$  \\
$ 0.37$ & $0.37$ & $0.41 $ & $0.0$ & $19.2$ & $8.0$ & $179$ & $-3.21$ & $0.1$ & $0.6$ & $1.55$ & $28.3$ & $0.43$  \\
$ 0.15$ & $0.15$ & $0.06 $ & $0.0$ & $15.6$ & $2.5$ & $117$ & $-9.67$ & $2.6$ & $3.1$ & $0.87$ & $24.6$ & $0.39$  \\
$ 0.07$ & $0.07$ & $0.11 $ & $0.0$ & $10.7$ & $6.0$ & $110$ & $-5.45$ & $0.2$ & $2.1$ & $1.65$ & $27.7$ & $0.45$  \\
$ 0.06$ & $0.06$ & $0.17 $ & $0.0$ & $9.2$ & $3.7$ & $90$ & $-4.07$ & $0.0$ & $2.1$ & $1.11$ & $26.6$ & $0.43$  \\
$ 0.01$ & $0.01$ & $0.07 $ & $0.0$ & $5.6$ & $1.9$ & $73$ & $-2.87$ & $0.0$ & $3.4$ & $1.55$ & $30.9$ & $0.24$  \\

& \\
\underline{{\bf backsplash}} \\
$ 0.27$ & $0.27$ & $0.15 $ & $0.0$ & $17.8$ & $6.8$ & $212$ & $-7.98$ & $0.2$ & $1.5$ & $1.83$ & $26.2$ & $0.36$  \\
$ 0.20$ & $0.20$ & $0.34 $ & $0.0$ & $16.2$ & $6.6$ & $264$ & $-3.46$ & $0.5$ & $1.8$ & $2.56$ & $31.1$ & $0.78$  \\
& \\
\underline{{\bf fields}} \\
$ 57.10$ & $54.75$ & $3660.17 $ & $5917.68$ & $72.9$ & $37.0$ & $790$ & $-1.16$ & $-$ & $0.0$ & $57.10$ & $86.9$ & $4.08$  \\
$ 36.78$ & $34.65$ & $823.73 $ & $7778.83$ & $58.5$ & $32.5$ & $468$ & $-1.70$ & $-$ & $0.3$ & $38.57$ & $66.3$ & $2.69$  \\
$ 34.25$ & $33.24$ & $1198.20 $ & $2718.88$ & $65.2$ & $30.5$ & $941$ & $-1.54$ & $-$ & $1.5$ & $77.46$ & $91.4$ & $4.51$  \\
$ 28.45$ & $27.72$ & $1041.61 $ & $2018.22$ & $55.2$ & $25.2$ & $658$ & $-1.47$ & $-$ & $1.5$ & $77.46$ & $91.4$ & $3.78$  \\
$ 18.04$ & $16.55$ & $463.10 $ & $10247.41$ & $49.8$ & $26.6$ & $282$ & $-1.73$ & $-$ & $0.1$ & $18.29$ & $54.0$ & $1.62$  \\
$ 10.59$ & $10.23$ & $71.99 $ & $2239.76$ & $34.1$ & $17.7$ & $217$ & $-2.02$ & $-$ & $0.1$ & $11.04$ & $40.2$ & $1.25$  \\
$ 8.45$ & $8.06$ & $93.41 $ & $1894.82$ & $37.8$ & $19.6$ & $1850$ & $-2.15$ & $-$ & $0.6$ & $14.79$ & $44.6$ & $4.58$  \\
$ 8.30$ & $8.28$ & $29.70 $ & $168.21$ & $38.3$ & $15.9$ & $464$ & $-2.25$ & $-$ & $0.0$ & $8.30$ & $40.9$ & $2.66$  \\
$ 7.75$ & $7.51$ & $126.72 $ & $1899.00$ & $38.8$ & $21.9$ & $1512$ & $-1.95$ & $-$ & $0.1$ & $7.80$ & $48.6$ & $5.88$  \\
$ 7.40$ & $7.35$ & $8.58 $ & $236.12$ & $33.3$ & $8.5$ & $393$ & $-1.84$ & $-$ & $0.8$ & $8.43$ & $36.1$ & $2.26$  \\
$ 6.77$ & $6.70$ & $36.39 $ & $228.33$ & $34.5$ & $16.9$ & $2528$ & $-2.00$ & $-$ & $0.2$ & $7.10$ & $38.6$ & $5.68$  \\
$ 6.56$ & $6.55$ & $2.85 $ & $9.16$ & $30.2$ & $9.7$ & $694$ & $-5.11$ & $-$ & $0.0$ & $6.56$ & $34.3$ & $3.98$  \\
$ 4.64$ & $4.57$ & $21.80 $ & $280.48$ & $28.4$ & $12.7$ & $857$ & $-2.31$ & $-$ & $0.0$ & $4.64$ & $32.0$ & $4.92$  \\
$ 4.58$ & $4.56$ & $14.54 $ & $153.25$ & $31.9$ & $13.4$ & $858$ & $-2.16$ & $-$ & $0.7$ & $4.77$ & $39.9$ & $3.51$  \\
$ 4.15$ & $4.15$ & $0.72 $ & $0.61$ & $28.3$ & $9.6$ & $1269$ & $-3.01$ & $-$ & $0.0$ & $4.15$ & $30.5$ & $6.66$  \\
$ 3.80$ & $3.77$ & $16.15 $ & $194.62$ & $28.9$ & $11.4$ & $1914$ & $-1.96$ & $-$ & $0.2$ & $5.41$ & $34.5$ & $5.83$  \\
$ 3.84$ & $3.83$ & $2.93 $ & $50.59$ & $28.1$ & $11.5$ & $1621$ & $-6.24$ & $-$ & $0.1$ & $3.86$ & $32.8$ & $3.47$  \\
$ 3.45$ & $3.41$ & $6.78 $ & $7.99$ & $25.7$ & $9.6$ & $1685$ & $-2.94$ & $-$ & $0.0$ & $3.52$ & $29.1$ & $6.53$  \\
$ 2.95$ & $2.95$ & $1.92 $ & $0.0$ & $26.1$ & $6.3$ & $1642$ & $-2.97$ & $-$ & $0.4$ & $3.27$ & $30.2$ & $6.15$  \\
$ 2.94$ & $2.92$ & $0.40 $ & $9.13$ & $26.5$ & $11.8$ & $1309$ & $-9.43$ & $-$ & $0.0$ & $2.94$ & $29.8$ & $7.10$  \\
$ 2.66$ & $2.65$ & $13.67 $ & $50.44$ & $29.6$ & $12.8$ & $2160$ & $-3.12$ & $-$ & $0.3$ & $2.85$ & $37.5$ & $8.31$  \\
$ 2.25$ & $2.25$ & $2.36 $ & $0.0$ & $25.3$ & $9.4$ & $2715$ & $-7.95$ & $-$ & $0.0$ & $2.25$ & $28.2$ & $8.64$  \\
$ 2.12$ & $2.12$ & $6.17 $ & $26.61$ & $27.1$ & $10.4$ & $1846$ & $-6.17$ & $-$ & $0.0$ & $2.12$ & $33.0$ & $6.13$  \\
$ 1.60$ & $1.60$ & $1.08 $ & $0.0$ & $24.9$ & $6.7$ & $233$ & $-8.67$ & $-$ & $0.3$ & $1.65$ & $30.3$ & $1.22$  \\
$ 0.97$ & $0.97$ & $1.28 $ & $0.20$ & $21.0$ & $11.7$ & $1659$ & $-5.97$ & $-$ & $0.0$ & $0.97$ & $24.3$ & $5.46$  \\
&\\
\hline
      \multicolumn{13}{c}{g8.26e11}\\
\hline
\underline{{\bf satellites}} \\
$ 20.48$ & $19.37$ & $194.99 $ & $593.41$ & $54.4$ & $27.6$ & $117$ & $-1.20$ & $0.7$ & $1.7$ & $47.56$ & $78.4$ & $0.47$  \\
$ 2.57$ & $2.53$ & $4.85 $ & $30.47$ & $32.4$ & $15.9$ & $201$ & $-1.88$ & $0.4$ & $1.7$ & $5.97$ & $44.9$ & $0.33$  \\
$ 1.47$ & $1.46$ & $10.53 $ & $0.0$ & $28.9$ & $15.7$ & $161$ & $-1.78$ & $1.6$ & $1.7$ & $19.63$ & $54.8$ & $0.36$  \\
$ 0.28$ & $0.28$ & $1.92 $ & $0.0$ & $15.8$ & $7.6$ & $167$ & $-2.30$ & $1.5$ & $1.5$ & $2.74$ & $30.6$ & $0.37$  \\
$ 0.21$ & $0.21$ & $0.14 $ & $0.0$ & $16.1$ & $5.1$ & $108$ & $-6.80$ & $1.7$ & $1.7$ & $0.89$ & $21.9$ & $0.29$  \\
$ 0.06$ & $0.06$ & $0.14 $ & $0.0$ & $11.5$ & $6.7$ & $140$ & $-6.89$ & $0.4$ & $1.5$ & $0.78$ & $21.8$ & $0.16$  \\
$ 0.04$ & $0.04$ & $0.09 $ & $0.0$ & $8.7$ & $3.6$ & $148$ & $-3.05$ & $0.4$ & $1.7$ & $0.27$ & $15.2$ & $0.23$  \\
& \\
\underline{{\bf backsplash}} \\
$ 12.59$ & $11.03$ & $115.91 $ & $1104.73$ & $46.2$ & $27.5$ & $240$ & $-1.36$ & $1.7$ & $0.2$ & $31.83$ & $74.7$ & $0.48$  \\
$ 3.77$ & $3.74$ & $7.87 $ & $12.37$ & $31.9$ & $14.5$ & $404$ & $-1.73$ & $0.3$ & $0.4$ & $6.54$ & $38.0$ & $0.53$  \\
$ 3.31$ & $3.25$ & $7.02 $ & $41.36$ & $31.5$ & $15.8$ & $242$ & $-1.92$ & $0.7$ & $1.1$ & $5.33$ & $37.5$ & $0.64$  \\
& \\
\underline{{\bf fields}} \\
$ 23.55$ & $22.77$ & $10.80 $ & $262.84$ & $47.5$ & $19.8$ & $2056$ & $-1.90$ & $-$ & $-0.0$ & $23.55$ & $49.0$ & $5.92$  \\
$ 9.04$ & $9.00$ & $8.29 $ & $27.48$ & $40.3$ & $20.2$ & $2086$ & $-1.76$ & $-$ & $0.6$ & $9.83$ & $43.4$ & $5.78$  \\
$ 6.39$ & $6.24$ & $7.89 $ & $75.03$ & $33.5$ & $18.4$ & $1203$ & $-1.80$ & $-$ & $0.9$ & $7.10$ & $42.5$ & $3.47$  \\
$ 6.05$ & $5.98$ & $4.72 $ & $26.24$ & $35.9$ & $13.5$ & $883$ & $-1.67$ & $-$ & $0.0$ & $6.05$ & $39.4$ & $3.81$  \\
$ 4.46$ & $4.40$ & $3.84 $ & $29.96$ & $31.9$ & $13.6$ & $1095$ & $-1.74$ & $-$ & $0.7$ & $4.54$ & $37.9$ & $3.71$  \\
$ 4.61$ & $4.58$ & $0.42 $ & $18.70$ & $30.2$ & $7.3$ & $1224$ & $-1.95$ & $-$ & $-0.0$ & $4.61$ & $31.3$ & $4.49$  \\
$ 4.28$ & $4.25$ & $2.89 $ & $19.63$ & $32.0$ & $13.7$ & $1371$ & $-1.82$ & $-$ & $-0.0$ & $4.28$ & $33.7$ & $5.12$  \\
$ 2.27$ & $2.27$ & $0.85 $ & $0.0$ & $26.5$ & $9.1$ & $2121$ & $-2.33$ & $-$ & $0.4$ & $2.71$ & $28.7$ & $5.49$  \\
$ 1.89$ & $1.89$ & $0.37 $ & $0.0$ & $24.7$ & $10.7$ & $532$ & $-4.44$ & $-$ & $0.1$ & $1.91$ & $41.9$ & $2.58$  \\
$ 1.81$ & $1.80$ & $0.32 $ & $0.0$ & $25.4$ & $11.3$ & $351$ & $-4.85$ & $-$ & $0.1$ & $1.82$ & $27.0$ & $1.70$  \\
$ 1.57$ & $1.56$ & $3.44 $ & $3.06$ & $24.4$ & $12.1$ & $2887$ & $-2.35$ & $-$ & $0.4$ & $2.27$ & $31.4$ & $2.42$  \\
$ 1.37$ & $1.37$ & $0.29 $ & $0.0$ & $23.2$ & $10.7$ & $271$ & $-2.55$ & $-$ & $1.0$ & $1.71$ & $24.0$ & $1.16$  \\
$ 1.30$ & $1.30$ & $0.25 $ & $0.0$ & $23.0$ & $7.6$ & $317$ & $-3.32$ & $-$ & $0.7$ & $2.33$ & $26.3$ & $1.09$  \\
$ 1.05$ & $1.05$ & $1.38 $ & $0.0$ & $25.2$ & $11.3$ & $1990$ & $-2.93$ & $-$ & $0.9$ & $1.94$ & $29.0$ & $6.09$  \\
&\\
\hline
        \multicolumn{13}{c}{g1.12e12}\\
\hline
\underline{{\bf satellites}} \\
$ 15.64$ & $14.25$ & $100.47 $ & $981.23$ & $43.5$ & $22.4$ & $199$ & $-1.48$ & $0.5$ & $0.4$ & $36.90$ & $59.9$ & $0.45$  \\
$ 4.64$ & $4.63$ & $7.09 $ & $0.0$ & $32.8$ & $15.2$ & $110$ & $-2.07$ & $0.5$ & $0.9$ & $9.74$ & $48.3$ & $0.47$  \\
$ 3.95$ & $3.91$ & $4.63 $ & $18.57$ & $27.3$ & $10.4$ & $245$ & $-1.94$ & $0.1$ & $1.7$ & $5.72$ & $37.9$ & $0.87$  \\
$ 3.50$ & $3.49$ & $4.33 $ & $0.0$ & $32.2$ & $14.7$ & $92$ & $-1.95$ & $0.7$ & $0.6$ & $9.35$ & $42.7$ & $0.29$  \\
$ 3.44$ & $3.44$ & $3.82 $ & $0.0$ & $28.9$ & $11.9$ & $187$ & $-1.85$ & $0.2$ & $0.5$ & $5.01$ & $37.0$ & $0.80$  \\
$ 0.65$ & $0.65$ & $1.35 $ & $0.0$ & $23.3$ & $10.7$ & $246$ & $-2.19$ & $0.7$ & $2.7$ & $1.86$ & $32.7$ & $0.45$  \\
$ 0.60$ & $0.60$ & $1.11 $ & $0.0$ & $18.5$ & $9.5$ & $178$ & $-2.80$ & $0.7$ & $1.6$ & $3.03$ & $35.3$ & $0.42$  \\
$ 0.11$ & $0.11$ & $0.19 $ & $0.0$ & $13.6$ & $6.0$ & $220$ & $-7.62$ & $0.5$ & $3.0$ & $1.02$ & $29.0$ & $0.35$  \\
& \\
\underline{{\bf backsplash}} \\
$ 6.09$ & $6.06$ & $21.23 $ & $6.63$ & $43.6$ & $20.2$ & $290$ & $-1.80$ & $0.5$ & $0.6$ & $11.04$ & $55.8$ & $0.88$  \\
$ 0.69$ & $0.69$ & $0.08 $ & $0.0$ & $21.8$ & $9.1$ & $487$ & $-14.44$ & $0.4$ & $2.0$ & $2.33$ & $27.8$ & $0.73$  \\
$ 0.42$ & $0.42$ & $0.16 $ & $0.0$ & $17.6$ & $7.5$ & $352$ & $-18.96$ & $0.3$ & $2.0$ & $0.92$ & $23.1$ & $0.51$  \\
$ 0.04$ & $0.04$ & $0.25 $ & $0.0$ & $8.2$ & $5.6$ & $300$ & $-4.49$ & $0.3$ & $3.1$ & $1.29$ & $29.4$ & $0.29$  \\
& \\
\underline{{\bf fields}} \\
$ 25.23$ & $23.86$ & $50.30 $ & $981.03$ & $57.2$ & $28.1$ & $479$ & $-1.66$ & $-$ & $0.1$ & $25.74$ & $58.5$ & $2.06$  \\
$ 15.68$ & $15.56$ & $26.36 $ & $72.22$ & $44.6$ & $18.7$ & $536$ & $-1.79$ & $-$ & $0.5$ & $19.77$ & $60.1$ & $1.23$  \\
$ 8.74$ & $8.32$ & $8.59 $ & $256.96$ & $38.9$ & $17.8$ & $1280$ & $-2.04$ & $-$ & $-0.0$ & $8.74$ & $42.2$ & $4.82$  \\
$ 7.17$ & $7.16$ & $5.15 $ & $4.06$ & $31.5$ & $13.8$ & $292$ & $-2.03$ & $-$ & $0.3$ & $9.78$ & $41.1$ & $1.26$  \\
$ 4.94$ & $4.88$ & $12.75 $ & $43.61$ & $34.0$ & $17.1$ & $2360$ & $-1.96$ & $-$ & $0.9$ & $11.05$ & $47.8$ & $8.06$  \\
$ 4.84$ & $4.65$ & $5.89 $ & $76.90$ & $29.4$ & $12.7$ & $349$ & $-1.95$ & $-$ & $0.3$ & $5.39$ & $39.2$ & $1.50$  \\
$ 3.15$ & $3.10$ & $2.04 $ & $20.54$ & $27.9$ & $14.7$ & $746$ & $-2.14$ & $-$ & $0.0$ & $3.16$ & $34.6$ & $3.16$  \\
$ 3.12$ & $3.09$ & $2.44 $ & $2.65$ & $26.0$ & $12.0$ & $2961$ & $-2.63$ & $-$ & $0.1$ & $3.16$ & $37.0$ & $7.93$  \\
$ 3.06$ & $3.03$ & $0.35 $ & $2.32$ & $27.6$ & $7.8$ & $1149$ & $-7.48$ & $-$ & $-0.0$ & $3.06$ & $31.9$ & $3.65$  \\
$ 2.99$ & $2.98$ & $0.18 $ & $0.0$ & $26.3$ & $8.8$ & $508$ & $-5.76$ & $-$ & $0.3$ & $3.48$ & $31.2$ & $2.00$  \\
$ 2.67$ & $2.67$ & $0.48 $ & $0.0$ & $25.9$ & $9.3$ & $422$ & $-2.29$ & $-$ & $0.5$ & $8.73$ & $38.0$ & $1.32$  \\
$ 1.66$ & $1.66$ & $0.13 $ & $0.0$ & $23.4$ & $10.5$ & $342$ & $-5.94$ & $-$ & $0.3$ & $2.18$ & $26.5$ & $1.15$  \\
$ 1.48$ & $1.48$ & $0.34 $ & $0.0$ & $24.1$ & $7.7$ & $286$ & $-4.13$ & $-$ & $0.1$ & $1.50$ & $28.1$ & $1.23$  \\
$ 1.30$ & $1.30$ & $0.30 $ & $0.0$ & $23.4$ & $8.2$ & $257$ & $-3.42$ & $-$ & $0.6$ & $2.15$ & $28.3$ & $1.11$  \\
$ 0.68$ & $0.68$ & $0.06 $ & $0.0$ & $19.4$ & $8.0$ & $2018$ & $-10.93$ & $-$ & $0.7$ & $1.15$ & $25.3$ & $8.14$  \\
&\\
\hline
        \multicolumn{13}{c}{g2.79e12}\\
\hline
\underline{{\bf satellites}} \\
$ 9.89$ & $9.54$ & $44.88 $ & $295.15$ & $24.5$ & $28.2$ & $141$ & $-1.79$ & $0.7$ & $0.7$ & $26.01$ & $29.9$ & $0.45$  \\
$ 7.46$ & $7.21$ & $91.17 $ & $117.44$ & $16.9$ & $18.5$ & $317$ & $-1.51$ & $0.6$ & $0.6$ & $37.88$ & $27.8$ & $0.25$  \\
$ 7.38$ & $7.33$ & $4.49 $ & $44.03$ & $28.0$ & $15.3$ & $128$ & $-2.21$ & $0.1$ & $0.2$ & $7.90$ & $60.1$ & $0.41$  \\
$ 5.13$ & $5.13$ & $6.83 $ & $0.0$ & $16.7$ & $13.7$ & $243$ & $-1.87$ & $0.1$ & $1.3$ & $6.71$ & $66.3$ & $0.21$  \\
$ 1.58$ & $1.55$ & $0.69 $ & $0.0$ & $43.3$ & $11.8$ & $106$ & $-4.81$ & $1.0$ & $1.5$ & $12.13$ & $64.0$ & $0.23$  \\
$ 1.07$ & $1.07$ & $20.36 $ & $0.0$ & $15.1$ & $14.0$ & $124$ & $-1.89$ & $0.5$ & $0.5$ & $2.67$ & $34.1$ & $0.40$  \\
$ 0.60$ & $0.59$ & $0.43 $ & $0.0$ & $27.1$ & $16.0$ & $214$ & $-3.23$ & $1.7$ & $2.1$ & $1.99$ & $40.1$ & $0.30$  \\
$ 0.20$ & $0.20$ & $0.65 $ & $0.0$ & $14.2$ & $8.4$ & $308$ & $-2.42$ & $0.7$ & $2.5$ & $3.11$ & $31.6$ & $0.12$  \\
$ 0.17$ & $0.17$ & $2.92 $ & $0.0$ & $35.1$ & $11.8$ & $167$ & $-2.63$ & $2.2$ & $2.7$ & $5.25$ & $40.2$ & $0.22$  \\
$ 0.19$ & $0.19$ & $1.71 $ & $0.0$ & $24.5$ & $8.8$ & $183$ & $-1.93$ & $1.7$ & $2.5$ & $1.68$ & $38.3$ & $0.23$  \\
$ 0.19$ & $0.19$ & $0.58 $ & $0.0$ & $8.4$ & $7.6$ & $110$ & $-4.89$ & $1.0$ & $2.7$ & $1.12$ & $46.7$ & $0.33$  \\
$ 0.15$ & $0.14$ & $0.28 $ & $0.0$ & $15.3$ & $6.7$ & $251$ & $-6.50$ & $0.1$ & $2.6$ & $1.16$ & $39.8$ & $0.23$  \\
$ 0.13$ & $0.13$ & $0.74 $ & $0.0$ & $22.7$ & $6.7$ & $171$ & $-2.58$ & $1.2$ & $1.7$ & $2.25$ & $30.5$ & $0.19$  \\
$ 0.11$ & $0.11$ & $0.24 $ & $0.0$ & $13.0$ & $6.5$ & $84$ & $-9.65$ & $0.3$ & $2.7$ & $1.91$ & $32.6$ & $0.27$  \\
$ 0.09$ & $0.09$ & $0.60 $ & $0.0$ & $15.7$ & $5.3$ & $174$ & $-2.71$ & $0.5$ & $2.2$ & $2.53$ & $31.4$ & $0.20$  \\
$ 0.07$ & $0.07$ & $0.20 $ & $0.0$ & $15.4$ & $2.4$ & $126$ & $-7.35$ & $1.0$ & $2.1$ & $3.57$ & $29.5$ & $0.16$  \\
$ 0.05$ & $0.05$ & $0.22 $ & $0.0$ & $8.3$ & $5.1$ & $131$ & $-2.60$ & $2.1$ & $3.0$ & $5.09$ & $35.8$ & $0.25$  \\
& \\
\underline{{\bf backsplash}} \\
$ 8.13$ & $8.03$ & $6.70 $ & $18.58$ & $34.9$ & $13.1$ & $363$ & $-1.81$ & $0.2$ & $0.4$ & $11.08$ & $41.8$ & $0.85$  \\
$ 4.35$ & $4.30$ & $23.06 $ & $17.79$ & $37.5$ & $19.2$ & $380$ & $-2.04$ & $0.9$ & $1.3$ & $14.16$ & $53.1$ & $0.42$  \\
$ 4.52$ & $4.51$ & $5.91 $ & $0.0$ & $34.2$ & $16.0$ & $608$ & $-1.88$ & $0.4$ & $0.5$ & $7.14$ & $45.2$ & $0.57$  \\
$ 2.90$ & $2.77$ & $33.80 $ & $0.0$ & $22.1$ & $11.4$ & $533$ & $-1.45$ & $0.7$ & $1.0$ & $39.63$ & $76.3$ & $0.84$  \\
$ 2.86$ & $2.85$ & $4.24 $ & $0.0$ & $28.0$ & $13.9$ & $470$ & $-2.18$ & $0.6$ & $0.5$ & $6.21$ & $37.9$ & $0.46$  \\
$ 2.26$ & $2.25$ & $0.91 $ & $0.0$ & $27.3$ & $7.7$ & $521$ & $-3.50$ & $0.3$ & $0.5$ & $2.97$ & $35.3$ & $0.87$  \\
$ 1.52$ & $1.52$ & $0.35 $ & $0.0$ & $24.8$ & $17.2$ & $641$ & $-7.42$ & $0.3$ & $0.4$ & $2.13$ & $30.0$ & $0.57$  \\
$ 1.29$ & $1.29$ & $0.28 $ & $0.0$ & $24.2$ & $8.2$ & $667$ & $-9.63$ & $0.4$ & $0.4$ & $2.07$ & $32.4$ & $0.49$  \\
$ 1.14$ & $1.14$ & $5.08 $ & $0.0$ & $22.6$ & $11.5$ & $341$ & $-2.29$ & $1.2$ & $1.1$ & $6.67$ & $42.5$ & $0.16$  \\
$ 1.07$ & $1.06$ & $3.36 $ & $0.0$ & $23.4$ & $13.1$ & $407$ & $-2.06$ & $1.1$ & $1.4$ & $4.43$ & $39.8$ & $0.18$  \\
$ 0.24$ & $0.24$ & $0.61 $ & $0.0$ & $16.5$ & $6.1$ & $506$ & $-4.60$ & $1.0$ & $1.3$ & $1.34$ & $29.8$ & $0.24$  \\
$ 0.18$ & $0.17$ & $1.02 $ & $0.0$ & $11.1$ & $7.9$ & $334$ & $-1.90$ & $1.9$ & $2.4$ & $5.32$ & $44.5$ & $0.18$  \\
$ 0.14$ & $0.14$ & $1.67 $ & $0.0$ & $14.6$ & $9.5$ & $424$ & $-2.19$ & $1.2$ & $3.4$ & $3.04$ & $42.6$ & $0.32$  \\
& \\
\underline{{\bf fields}} \\
$ 120.97$ & $109.98$ & $569.31 $ & $5768.46$ & $9.2$ & $46.4$ & $1797$ & $-1.23$ & $-$ & $-0.0$ & $120.97$ & $23.5$ & $5.11$  \\
$ 64.46$ & $60.50$ & $456.27 $ & $1736.70$ & $23.7$ & $35.6$ & $1069$ & $-1.27$ & $-$ & $0.2$ & $64.85$ & $38.9$ & $3.45$  \\
$ 48.26$ & $45.61$ & $232.43 $ & $1439.81$ & $69.4$ & $33.1$ & $553$ & $-1.31$ & $-$ & $0.4$ & $53.41$ & $82.3$ & $1.69$  \\
$ 31.76$ & $30.62$ & $97.55 $ & $536.17$ & $56.7$ & $28.2$ & $741$ & $-1.51$ & $-$ & $0.3$ & $36.09$ & $67.2$ & $1.85$  \\
$ 23.71$ & $23.12$ & $33.80 $ & $278.32$ & $44.3$ & $21.9$ & $1462$ & $-1.62$ & $-$ & $0.6$ & $25.57$ & $50.0$ & $3.77$  \\
$ 21.19$ & $18.86$ & $40.27 $ & $1425.28$ & $69.4$ & $24.8$ & $1384$ & $-1.86$ & $-$ & $-0.0$ & $21.19$ & $89.2$ & $4.11$  \\
$ 18.37$ & $18.08$ & $31.32 $ & $97.32$ & $29.6$ & $19.3$ & $771$ & $-1.76$ & $-$ & $0.3$ & $19.24$ & $32.9$ & $2.49$  \\
$ 14.73$ & $14.57$ & $17.74 $ & $102.41$ & $25.0$ & $22.0$ & $874$ & $-1.93$ & $-$ & $0.3$ & $16.23$ & $33.1$ & $1.99$  \\
$ 12.47$ & $12.35$ & $8.26 $ & $81.05$ & $42.6$ & $18.5$ & $882$ & $-2.40$ & $-$ & $0.4$ & $13.33$ & $50.6$ & $2.32$  \\
$ 10.88$ & $10.42$ & $7.30 $ & $138.73$ & $52.2$ & $12.8$ & $1770$ & $-1.72$ & $-$ & $0.3$ & $13.99$ & $53.2$ & $5.46$  \\
$ 10.17$ & $9.74$ & $8.66 $ & $154.29$ & $13.3$ & $20.4$ & $1548$ & $-2.36$ & $-$ & $0.1$ & $10.18$ & $32.2$ & $4.39$  \\
$ 9.48$ & $8.79$ & $7.32 $ & $519.79$ & $49.4$ & $27.0$ & $1524$ & $-1.96$ & $-$ & $-0.0$ & $9.48$ & $56.7$ & $4.42$  \\
$ 8.97$ & $8.83$ & $1.68 $ & $72.54$ & $8.1$ & $7.4$ & $2214$ & $-1.81$ & $-$ & $0.3$ & $12.06$ & $36.2$ & $5.55$  \\
$ 8.40$ & $8.21$ & $5.23 $ & $108.30$ & $35.5$ & $17.6$ & $2412$ & $-1.94$ & $-$ & $0.3$ & $8.51$ & $44.3$ & $5.85$  \\
$ 8.25$ & $8.00$ & $7.19 $ & $91.62$ & $30.8$ & $20.1$ & $2052$ & $-2.06$ & $-$ & $-0.0$ & $8.25$ & $36.6$ & $6.03$  \\
$ 6.09$ & $5.92$ & $0.47 $ & $62.87$ & $29.7$ & $5.8$ & $1639$ & $-2.95$ & $-$ & $0.8$ & $6.51$ & $35.0$ & $4.91$  \\
$ 5.64$ & $5.57$ & $2.61 $ & $52.13$ & $34.0$ & $13.2$ & $828$ & $-1.85$ & $-$ & $0.7$ & $5.77$ & $40.8$ & $2.67$  \\
$ 5.27$ & $5.22$ & $1.10 $ & $41.02$ & $33.3$ & $9.2$ & $1945$ & $-4.62$ & $-$ & $0.0$ & $5.28$ & $39.1$ & $4.93$  \\
$ 4.59$ & $4.53$ & $0.50 $ & $36.34$ & $28.5$ & $9.0$ & $2405$ & $-2.26$ & $-$ & $0.0$ & $4.59$ & $33.0$ & $5.76$  \\
$ 3.68$ & $3.67$ & $2.36 $ & $0.0$ & $31.1$ & $15.0$ & $438$ & $-2.83$ & $-$ & $0.4$ & $4.28$ & $37.0$ & $1.18$  \\
$ 3.65$ & $3.64$ & $0.24 $ & $0.19$ & $27.2$ & $11.9$ & $2005$ & $-9.24$ & $-$ & $0.0$ & $3.65$ & $31.5$ & $4.95$  \\
$ 3.45$ & $3.35$ & $2.38 $ & $56.66$ & $26.2$ & $13.1$ & $1903$ & $-2.57$ & $-$ & $0.5$ & $3.72$ & $35.1$ & $5.22$  \\
$ 3.36$ & $3.33$ & $0.93 $ & $11.79$ & $36.4$ & $8.7$ & $995$ & $-2.24$ & $-$ & $0.0$ & $3.36$ & $41.1$ & $3.21$  \\
$ 3.23$ & $3.23$ & $0.60 $ & $0.0$ & $32.5$ & $5.5$ & $1665$ & $-2.47$ & $-$ & $1.2$ & $3.99$ & $35.3$ & $4.55$  \\ 
$ 2.86$ & $2.79$ & $1.07 $ & $48.90$ & $29.3$ & $4.5$ & $1749$ & $-1.69$ & $-$ & $0.5$ & $4.51$ & $34.7$ & $5.46$  \\
$ 2.65$ & $2.46$ & $4.09 $ & $67.76$ & $41.0$ & $12.2$ & $2192$ & $-2.14$ & $-$ & $0.4$ & $3.59$ & $41.8$ & $4.20$  \\
$ 2.77$ & $2.77$ & $0.45 $ & $0.0$ & $28.4$ & $9.8$ & $1128$ & $-3.32$ & $-$ & $0.0$ & $2.78$ & $34.8$ & $3.60$  \\
$ 2.57$ & $2.56$ & $0.28 $ & $0.0$ & $29.3$ & $7.4$ & $1791$ & $-6.86$ & $-$ & $0.8$ & $2.90$ & $33.4$ & $4.32$  \\
$ 1.49$ & $1.48$ & $0.35 $ & $0.0$ & $27.7$ & $10.8$ & $1687$ & $-3.01$ & $-$ & $0.9$ & $2.00$ & $33.0$ & $4.75$  \\

  \hline
  \label{tab:dwarf}
\end{longtable}
\end{ThreePartTable}

\newpage
\section{Predicted peak masses for Local Group galaxies}
\def\arraystretch{1.3}
\begin{ThreePartTable}
\begin{TableNotes}
\footnotesize
\item {\bf References}: The stellar masses are adopted from \citet{Fattahi2017} who derived these mostly from parameters taken from the updated
  (October 2015) version of the tables from \citet{McConnachie2012} but also used parameters from other references, including the following: 1: \citet{McConnachie2012}, 2: \citet{Torrealba2016},
  3: \citet{Caldwell2017}, 4: \citet{Koposov2015b},
  5:\citet{Walker2016}, 6: \citet{Martin2016}, 7: \citet{Kirby2015},
  8: \citet{Kirby2017}, 9: \citet{Tollerud2012}, 10: \citet{Ho2012},
  11: \citet{Collins2013}, 12: \citet{Tollerud2013}, 13:
  \citet{Collins2010}, 14: \citet{Martin2016b}, 15:
  \citet{Kirby2014}, 16: \citet{Hunter2006}, 17: \citet{Leaman2012},
  18: \citet{Bellazzini2011}, 19: \citet{Kirby2017a}, 20:
  \citet{McConnachie2006b}, 21: \citet{Crnojevic2016}, 22: \citet{Li2017}.
\end{TableNotes}

\begin{longtable}{{l} cc {r}}
  \caption{Stellar mass and peak total mass of dwarf galaxies in the Local Group}\\
  \hline
  \hline
  Gal. Name  &  $M_{\rm star}$ & $M_{\rm peak}$ & references \\
  & $(10^5 \Msun$) & $(10^9 \Msun$) & \\
  \endfirsthead
  \multicolumn{3}{l}{{\bfseries\tablename\ \thetable{} -- (continued)}}\\
  \hline
  \hline
  Gal. Name  &  $M_{\rm star}$ & $M_{\rm peak}$ & references \\
  & $(10^5 \Msun$) & $(10^9 \Msun$) & \\
  \hline
  \endhead

  \insertTableNotes  
  \endlastfoot
  
  \hline

  \underline{{\bf MW satellites}} \\
     For  & $  245  _{-   69} ^{+   96}$ & $ 19.10 _ {- 3.38} ^{+ 12.39} $ &      1  \\
    LeoI  & $   45  _{-   13} ^{+   19}$ & $ 7.81 _ {- 1.36} ^{+ 5.21} $ &      1  \\
     Scl  & $   39  _{-   15} ^{+   25}$ & $ 7.24 _ {- 1.03} ^{+ 5.78} $ &      1  \\
   LeoII  & $   12  _{-    3} ^{+    4.4}$ & $ 3.89 _ {- 0.73} ^{+ 2.46} $ &      1  \\
    SexI  & $    7.0  _{-    3} ^{+    4.3}$ & $ 2.93 _ {- 0.38} ^{+ 2.29} $ &      1  \\
     Car  & $    3.8  _{-    1.4} ^{+    2.3}$ & $ 2.12 _ {- 0.31} ^{+ 1.65} $ &      1  \\
     Dra  & $    5.1  _{-    1.2} ^{+    1.5}$ & $ 2.48 _ {- 0.48} ^{+ 1.45} $ &      1  \\
     Umi  & $    5.3  _{-    2.0} ^{+    3.3}$ & $ 2.53 _ {- 0.37} ^{+ 1.99} $ &     23  \\
 CanVenI  & $    3.7  _{-    0.8} ^{+    0.9}$ & $ 2.09 _ {- 0.42} ^{+ 1.16} $ &      1  \\
   CraII  & $    2.6  _{-    0.3} ^{+    0.4}$ & $ 1.73 _ {- 0.41} ^{+ 0.86} $ &    2,3  \\
     Her  & $  0.60  _{-  0.18 } ^{+  0.23}$ & $ 0.80 _ {- 0.14} ^{+ 0.51} $ &      1  \\
    BooI  & $  0.46  _{-  0.09 } ^{+  0.11}$ & $ 0.70 _ {- 0.14} ^{+ 0.38} $ &      1  \\
   LeoIV  & $  0.29  _{-  0.09 } ^{+  0.14}$ & $ 0.55 _ {- 0.09} ^{+ 0.38} $ &      1  \\
    UMaI  & $  0.22  _{-  0.06 } ^{+  0.08}$ & $ 0.47 _ {- 0.08} ^{+ 0.30} $ &      1  \\
    LeoV  & $  0.17  _{-  0.06 } ^{+  0.08}$ & $ 0.41 _ {- 0.06} ^{+ 0.29} $ &      1  \\
   PisII  & $  0.14  _{-  0.07 } ^{+  0.13}$ & $ 0.37 _ {- 0.04} ^{+ 0.36} $ &      7  \\
CanVeniI  & $  0.13  _{-  0.05 } ^{+  0.08}$ & $ 0.36 _ {- 0.05} ^{+ 0.28} $ &      1  \\
   HydII  & $  0.13  _{-  0.03 } ^{+  0.05}$ & $ 0.36 _ {- 0.07} ^{+ 0.23} $ &      7  \\
   UMaII  & $  0.065  _{-  0.03 } ^{+  0.05}$ & $ 0.25 _ {- 0.03} ^{+ 0.22} $ &      1  \\
  ComBer  & $  0.060  _{-  0.02 } ^{+  0.04}$ & $ 0.24 _ {- 0.04} ^{+ 0.19} $ &      1  \\
    Tuc2  & $  0.045  _{-  0.01 } ^{+  0.012}$ & $ 0.20 _ {- 0.04} ^{+ 0.12} $ &      5  \\
    Hor1  & $  0.031  _{-  0.006 } ^{+  0.008}$ & $ 0.17 _ {- 0.03} ^{+ 0.09} $ &      4  \\
    Gru1  & $  0.031  _{-  0.009 } ^{+  0.013}$ & $ 0.17 _ {- 0.03} ^{+ 0.11} $ &      5  \\
   DraII  & $  0.020  _{-  0.011 } ^{+  0.025}$ & $ 0.13 _ {- 0.01} ^{+ 0.15} $ &      6  \\
   BooII  & $  0.016  _{-  0.009 } ^{+  0.021}$ & $ 0.12 _ {- 0.01} ^{+ 0.14} $ &      1  \\
    Ret2  & $  0.016  _{-  0.003 } ^{+  0.004}$ & $ 0.12 _ {- 0.02} ^{+ 0.07} $ &      4  \\
   Will1  & $  0.016  _{-  0.008 } ^{+  0.018}$ & $ 0.12 _ {- 0.01} ^{+ 0.13} $ &      1  \\
   SegII  & $  0.014  _{-  0.004 } ^{+  0.005}$ & $ 0.11 _ {- 0.02} ^{+ 0.07} $ &      1  \\
   TriII  & $  0.0071 _{-  0.0027 } ^{+  0.0045}$ & $ 0.08 _ {- 0.01} ^{+ 0.06} $ &      8  \\
    SegI  & $  0.0054  _{-  0.0029 } ^{+  0.0063}$ & $ 0.07 _ {- 0.01} ^{+ 0.07} $ & 1\\
  & \\
\underline{{\bf M31 satellites}} \\
   N205  & $ 4650  _{-  650} ^{+  760}$ & $ 90.29 _ {- 20.42} ^{+ 45.13} $ &      1  \\
     M32  & $ 4760  _{- 1010} ^{+ 1260}$ & $ 91.41 _ {- 18.34} ^{+ 51.86} $ &      1  \\
    N185  & $  680  _{-   100} ^{+  118}$ & $ 32.74 _ {- 7.32} ^{+ 16.58} $ &      1  \\
    N147  & $  990  _{-  150} ^{+  164}$ & $ 39.91 _ {- 8.86} ^{+ 20.01} $ &      1  \\
    AVII  & $  150  _{-   39} ^{+   52}$ & $ 14.74 _ {- 2.71} ^{+ 9.14} $ &      9  \\
     AII  & $   40.  _{-    8} ^{+    9}$ & $ 7.34 _ {- 1.50} ^{+ 3.97} $ &  10,14  \\
      AI  & $   44.  _{-    9} ^{+   10}$ & $ 7.72 _ {- 1.57} ^{+ 4.19} $ &   9,14  \\
     AVI  & $   50.  _{-    10} ^{+   12}$ & $ 8.26 _ {- 1.69} ^{+ 4.55} $ &     11  \\
  AXXIII  & $   11.  _{-    2} ^{+    3}$ & $ 3.71 _ {- 0.78} ^{+ 2.13} $ &  11,14  \\
    AIII  & $   10.  _{-    2} ^{+    3}$ & $ 3.53 _ {- 0.72} ^{+ 2.08} $ &   9,14  \\
    LGS3  & $    9.6  _{-    1} ^{+    1.6}$ & $ 3.46 _ {- 0.83} ^{+ 1.73} $ &      1  \\
    AXXI  & $    5.5  _{-    1.4} ^{+    1.9}$ & $ 2.58 _ {- 0.48} ^{+ 1.60} $ &  11,14  \\
    AXXV  & $    6.3  _{-    1.6} ^{+    1.8}$ & $ 2.77 _ {- 0.52} ^{+ 1.61} $ &  11,14  \\
      AV  & $    5.1  _{-    1.0} ^{+    1.3}$ & $ 2.48 _ {- 0.51} ^{+ 1.39} $ &   9,14  \\
     AXV  & $    2.1  _{-    0.6} ^{+    0.8}$ & $ 1.55 _ {- 0.27} ^{+ 0.99} $ &   9,14  \\
    AXIX  & $   12.  _{-    4} ^{+    5}$ &$ 3.89 _ {- 0.62} ^{+ 2.58} $ &  11,14  \\
    AXIV  & $    3.3  _{-    1.2} ^{+    1.5}$ &$ 1.97 _ {- 0.29} ^{+ 1.35} $ &   9,14  \\
   AXXIX  & $    2.7  _{-    0.9} ^{+    1.5}$ & $ 1.77 _ {- 0.28} ^{+ 1.32} $ &     12  \\
     AIX  & $    4.3  _{-    1.1} ^{+    1.5}$ & $ 2.26 _ {- 0.42} ^{+ 1.41} $ &   9,14  \\
    AXXX  & $    2.2  _{-    0.6} ^{+    0.7}$ &$ 1.59 _ {- 0.29} ^{+ 0.96} $ &  11,14  \\
  AXXVII  & $  2.0  _{-  0.8 } ^{+  1.8}$ & $ 1.51 _ {- 0.21} ^{+ 1.42} $ &  11,14  \\
   AXVII  & $  1.7  _{-  0.4 } ^{+  0.6}$ & $ 1.39 _ {- 0.27} ^{+ 0.87} $ &  11,14  \\
      AX  & $  1.3  _{-  0.3 } ^{+  0.5}$ & $ 1.20 _ {- 0.23} ^{+ 0.77} $ &   9,14  \\
    AXVI  & $  1.1  _{-  0.3 } ^{+  0.5}$ &$ 1.10 _ {- 0.20} ^{+ 0.76} $ &   9,14  \\
   AXII  & $  0.85  _{-  0.35 } ^{+  0.62}$ & $ 0.96 _ {- 0.13} ^{+ 0.82} $ &  13,14  \\
   AXIII  & $  0.73  _{-  0.23 } ^{+  0.33}$ &$ 0.89 _ {- 0.15} ^{+ 0.61} $ &   9,14  \\
   AXXII  & $  0.73  _{-  0.23 } ^{+  0.33}$ & $ 0.89 _ {- 0.15} ^{+ 0.61} $ &  11,14  \\
     AXX  & $  0.48  _{-  0.16 } ^{+  0.22}$ & $ 0.71 _ {- 0.11} ^{+ 0.49} $ &  11,14  \\
     AXI  & $  0.46  _{-  0.15 } ^{+  0.22}$ & $ 0.70 _ {- 0.11} ^{+ 0.49} $ &  11,14  \\
   AXXVI  & $  0.31  _{-  0.15 } ^{+  0.20}$ & $ 0.56 _ {- 0.06} ^{+ 0.45} $ &  11,14  \\
 & \\
\underline{{\bf LG field dwarfs}} \\
   N6822  & $  830  _{-  170} ^{+  200}$ & $ 36.37 _ {- 7.39} ^{+ 20.06} $ &  15,16  \\
  IC1613  & $ 1020  _{-  170} ^{+  190}$ & $ 40.55 _ {- 8.78} ^{+ 20.89} $ &  15,16  \\
     WLM  & $  380  _{-   55} ^{+   65}$ & $ 24.08 _ {- 5.40} ^{+ 12.16} $ &     17  \\
 UGC4879  & $   58  _{-   11} ^{+   13}$ &$ 8.93 _ {- 1.86} ^{+ 4.83} $ &  15,18  \\
     Peg  & $   66  _{-   13} ^{+   16}$ & $ 9.56 _ {- 1.97} ^{+ 5.28} $ &  15,16  \\
    LeoA  & $   30  _{-    6} ^{+    8}$ & $ 6.31 _ {- 1.29} ^{+ 3.59} $ &  19,16  \\
     Cet  & $   45  _{-    9} ^{+   11}$ & $ 7.81 _ {- 1.60} ^{+ 4.33} $ &  20,16  \\
     Aqu  & $   16  _{-    2} ^{+    3}$ & $ 4.53 _ {- 1.05} ^{+ 2.34} $ &  20,16  \\
     Tuc  & $    8.9  _{-    1.9} ^{+    2.3}$ & $ 3.32 _ {- 0.66} ^{+ 1.87} $ &      1  \\
  AXVIII  & $    8.0  _{-    1.6} ^{+    1.9}$ & $ 3.14 _ {- 0.64} ^{+ 1.72} $ &      9  \\
 AXXVIII  & $    4.1  _{-    1.7} ^{+    5.2}$ & $ 2.21 _ {- 0.29} ^{+ 2.50} $ &     11  \\
    LeoT  & $    1.1  _{-    0.4} ^{+    0.6}$ &$ 1.10 _ {- 0.16} ^{+ 0.82} $ &      1  \\
   EriII  & $    0.9  _{-    0.2} ^{+    0.2}$ &$ 0.99 _ {- 0.20} ^{+ 0.53} $ &  21,22  \\

  \hline
  \label{tab:peak_mass}
\end{longtable}
\end{ThreePartTable}

\section{Ram Pressure Stripping}
\label{sec:ram}
\begin{figure*}
\includegraphics[width=.49\textwidth]{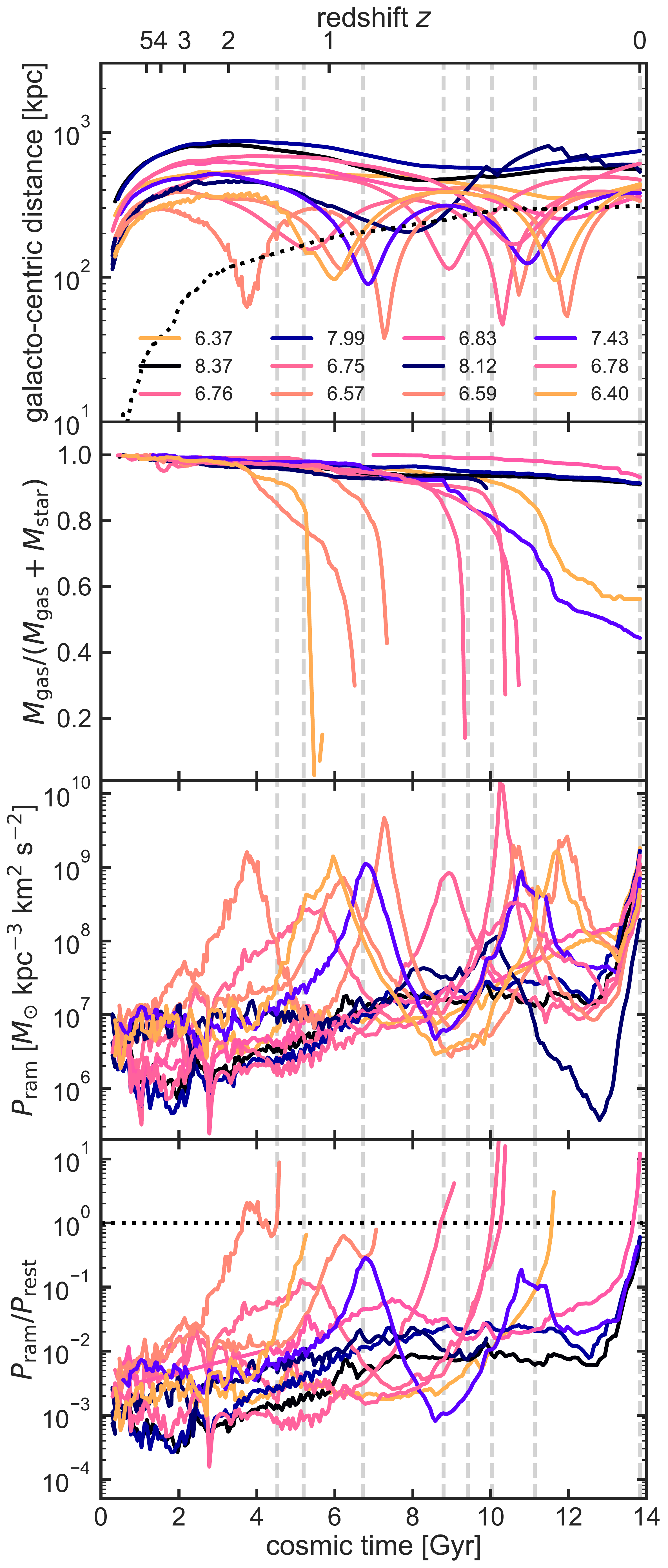}
\includegraphics[width=.49\textwidth]{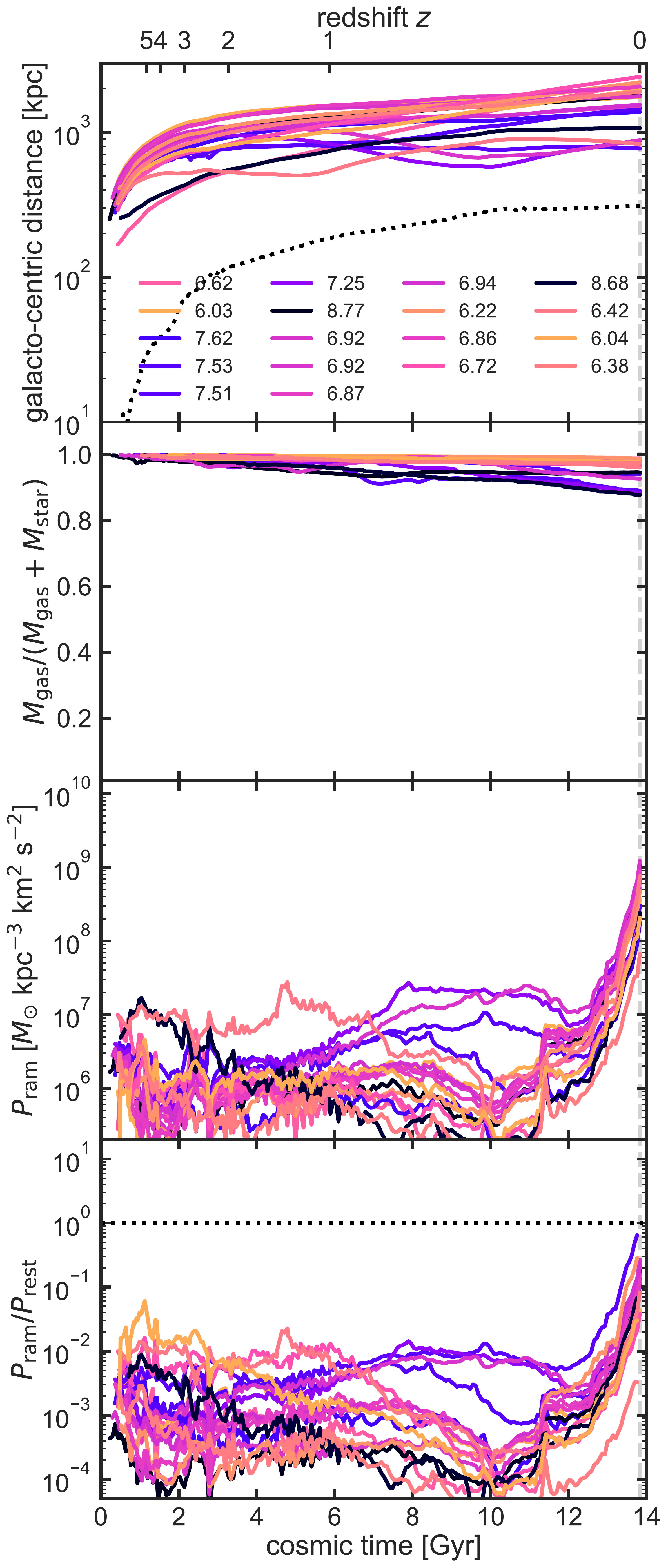}
\vspace{-.35cm}
\caption{Same as figure \ref{fig:ram} but for the \textit{nearby} sample (left panels) and the \textit{field} sample (right panels).}
\label{fig:ram1}
\end{figure*}
The left panels of figure \ref{fig:ram1} shows galacto-centric distance, gas fractions, ram pressure and ratio of ram pressure to restoring force per area for the \textit{nearby galaxies} and the right panel shows the same for the \textit{field sample}. We see that for the \textit{backsplash galaxies} ram pressure is indeed important while for the field sample ram pressure is not effective.

\section{Abundance matching highlighting backsplash galaxies}
\begin{figure}
\centering
\includegraphics[width=0.5\columnwidth]{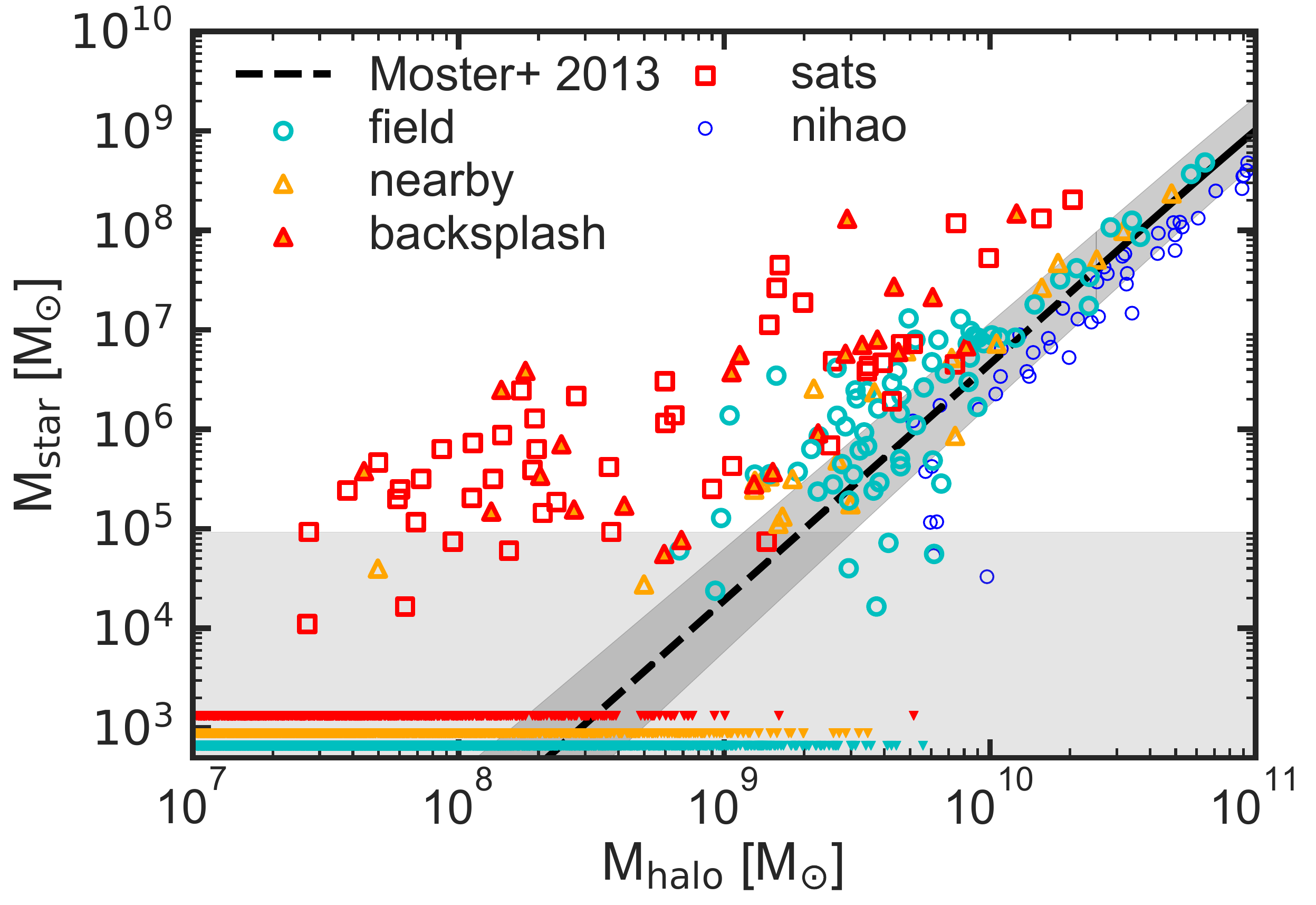}
\vspace{-.25cm}
\caption{ Same as Fig. \ref{fig:abund1}. Stellar mass vs. the total halo mass $\rm{M}_{halo}$ for
  satellites, dwarfs and field galaxies. This time we highlight all \textit{backsplash} galaxies with a red border triangle filled with orange color. Indeed only \textit{backsplash} galaxies deviate from the Moster relation.}
\label{fig:abund_bs}
\end{figure}
Figure \ref{fig:abund_bs} shows the abundance matching plot (already shown in figure \ref{fig:abund1}) for our dwarf galaxy samples where we highlight \textit{backsplash galaxies} as orange triangles with red borders. We see that indeed \textit{backsplash galaxies} deviate from the theoretical expectation in the same way satellites do. This is due to the mass lost by stripping. 

\section{Resolution Tests}
\label{sec:res}
\subsection{Minimum number of particles per halo}

\begin{figure*}
\includegraphics[width=.8\columnwidth]{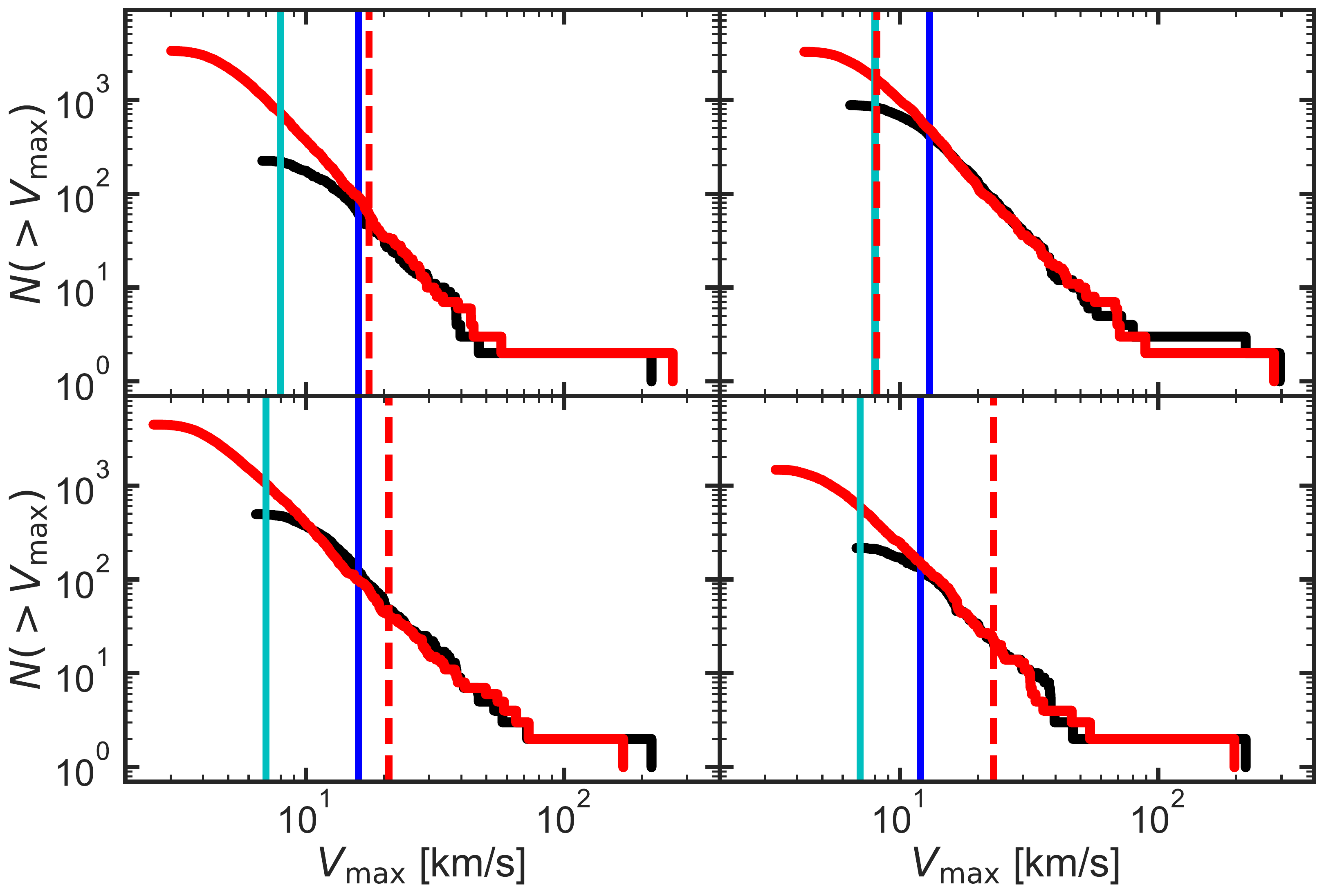}
\vspace{-.25cm}
\caption{Velocity function for low- (black line) and high-resolution (red line) runs. The blue vertical line indicates the velocity where galaxy counts in the low- and high-resolution simulations deviate more than $20\%$ from each other. The cyan vertical line indicates our derived lower $V_{\rm max}$ limit on dwarf galaxies in the high-resolution run (see text for more details) and the dashed red line indicates our stellar resolution limit of $M_{\rm star}>10^5\Msun$.}
\label{fig:res_1}
\end{figure*}

\begin{figure*}
\includegraphics[width=.8\columnwidth]{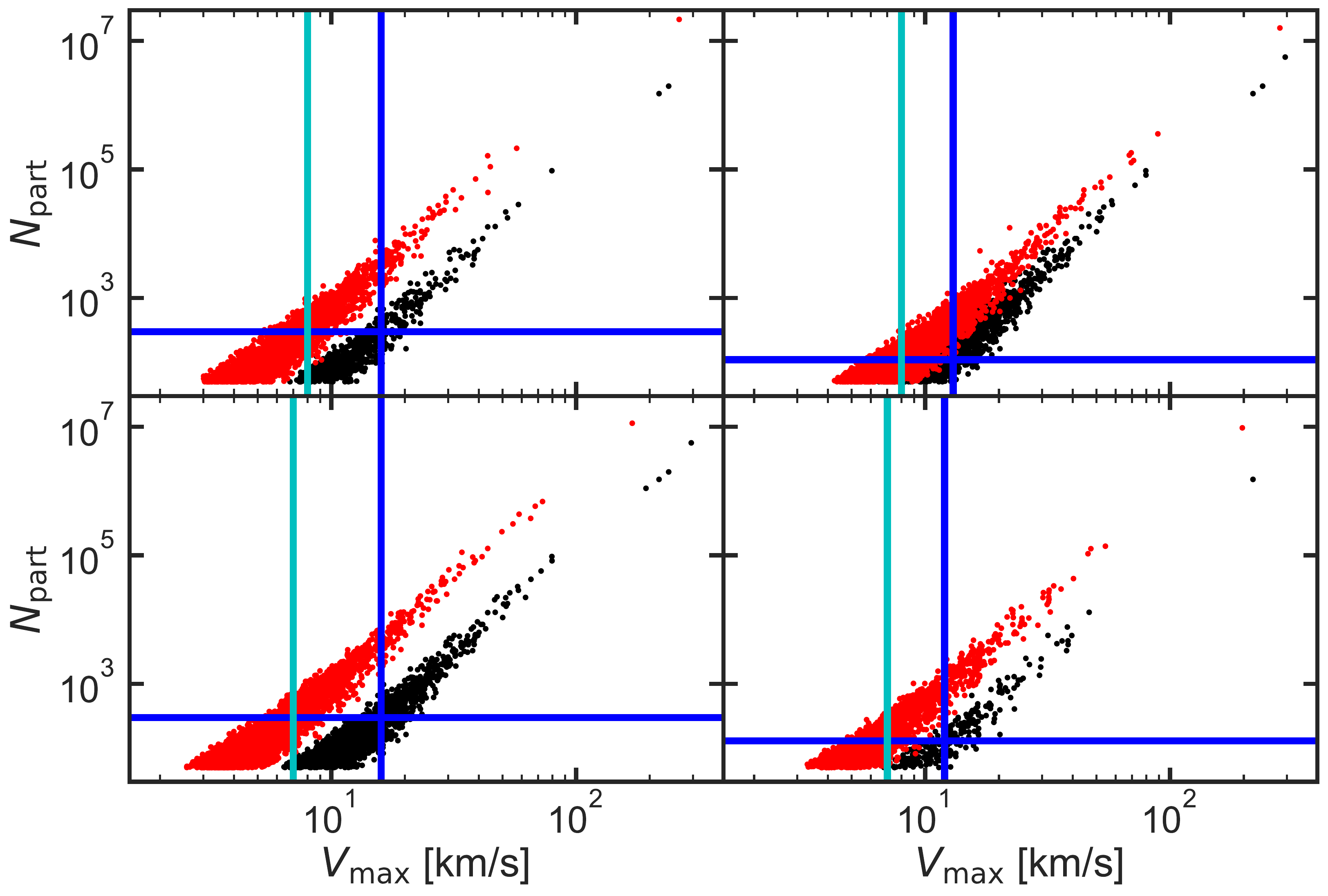}
\vspace{-.25cm}
\caption{Number of particles in the halo vs $V_{\rm max}$ for low- (black points) and high-resolution simulations (red points). The blue vertical line indicates the $V_{\rm max}$ value where low- and high-resolution simulations deviate from each other. The horizontal blue line indicates the derived minimum number of particles in the low resolution run where we still trust the results and the cyan line indicates down to which $V_{\rm max}$ we trust our results in the high-resolution runs derived from the same minimum number of particles in a halo.}
\label{fig:res_2}
\end{figure*}

In figure \ref{fig:res_1} we show the velocity function of low- and high-resolution runs. We derive a lower resolution limit for our simulations by requiring that the number counts of haloes of a given $V_{\rm max}$ should not deviate more than $20\%$ from each other indicated by the blue vertical line. Combining this with figure \ref{fig:res_2} where we show the number of particles in a halo as a function of the halo's $V_{\rm max}$ we can derive a lower limit on the number of particles required to properly resolve the halo. This number of particles is again indicated by the blue horizontal line in figure \ref{fig:res_2}. From this we can then derive a lower limit of $V_{\rm max}$ in the high-resolution runs by adopting the same minimum number of particles indicated by the cyan vertical lines in both figures. Finally, we compare this to our luminous resolution limit of 10 stellar particles per halo which we show as dashed vertical red line in figure \ref{fig:res_1}. We clearly see that all the galaxies we deal with in the main text of this paper are well resolved.

\subsection{Gaseous and stellar properties}

\begin{figure*}
\includegraphics[width=\columnwidth]{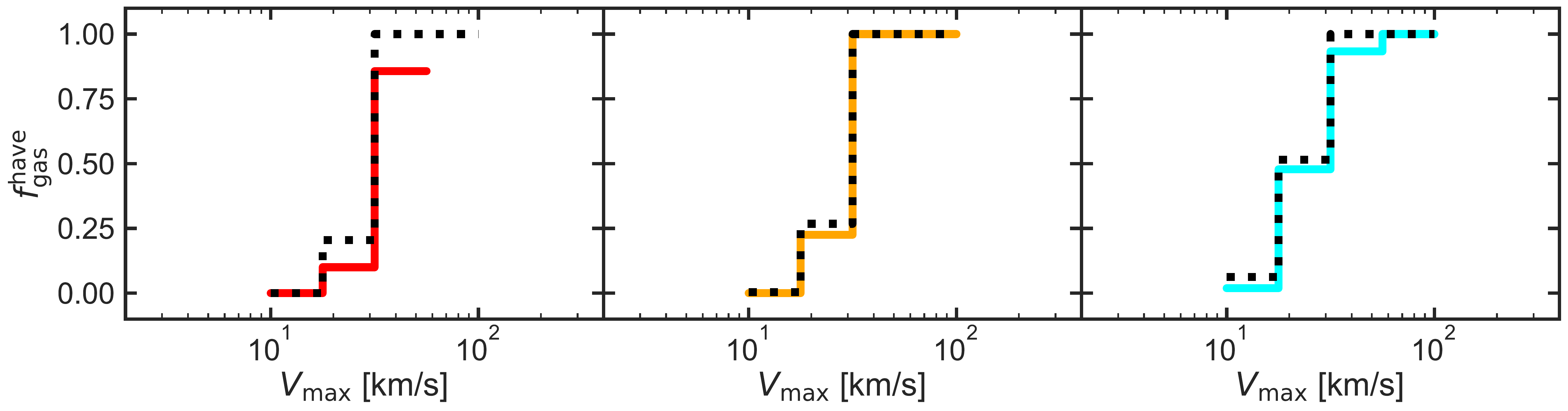}
\vspace{-.25cm}
\caption{The fraction of dwarf galaxies with gas as a function of $V_{\rm max}$ for \textit{satellites} (left panel), \textit{nearby dwarfs} and \textit{field galaxies}. The solid colored lines show the high-resolution runs with the same color-coding as applied in the main text and the dotted black line shows the low resolution simulations.}
\label{fig:gas_frac_res}
\end{figure*}

\begin{figure*}
\includegraphics[width=\columnwidth]{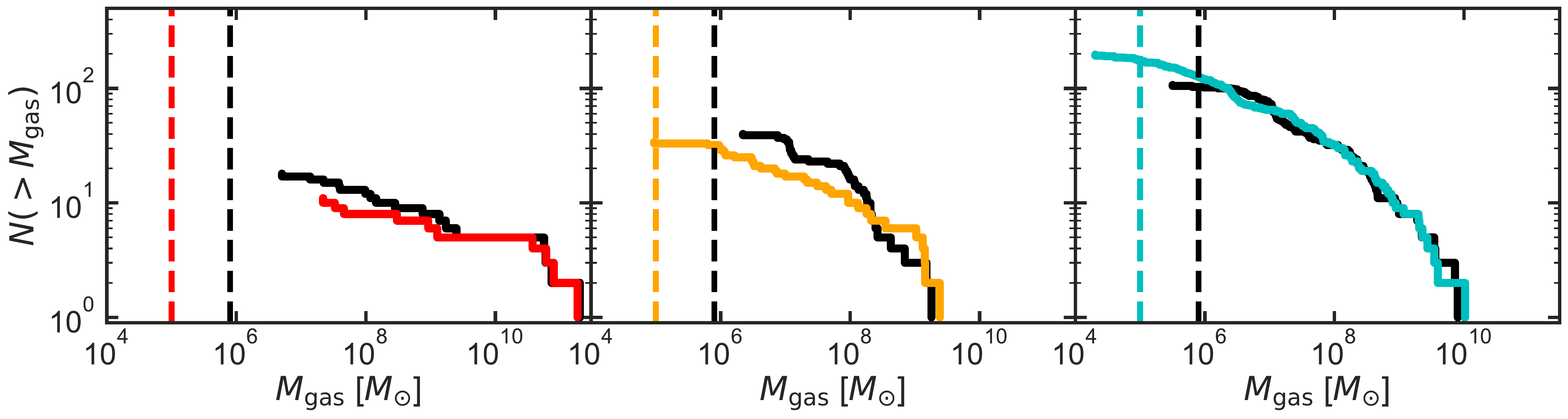}
\vspace{-.25cm}
\caption{Gas mass function for low- (black lines) and high-resolution runs (colored lines) for \textit{satellites} (left panel), \textit{nearby dwarfs} and \textit{field galaxies}. The vertical black and colored dashed lines indicate the resolution limit of 10 gas particles.}
\label{fig:gas_mass_res}
\end{figure*}

\begin{figure*}
\includegraphics[width=\columnwidth]{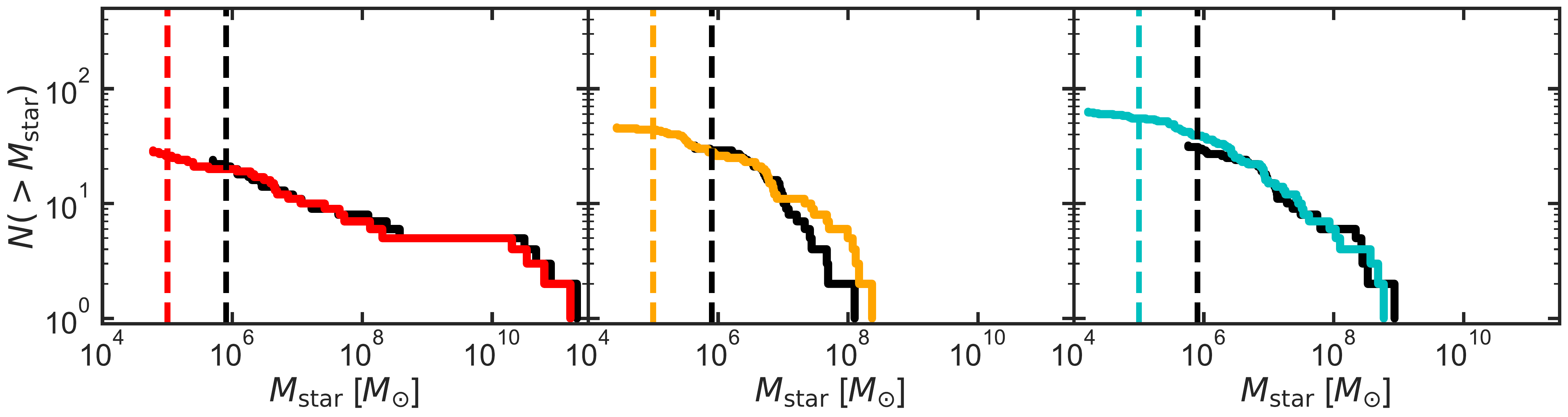}
\vspace{-.25cm}
\caption{Stellar mass function for low- (black lines) and high-resolution runs (colored lines) for \textit{satellites} (left panel), \textit{nearby dwarfs} and \textit{field galaxies}. The vertical black and colored dashed lines indicate the resolution limit of 10 stellar particles.}
\label{fig:star_mass_res}
\end{figure*}

We further test the convergence of the hydrodynamical properties of our dwarf galaxy samples. In figure \ref{fig:gas_frac_res} we compare the fraction of dwarf galaxies that posses gas as a function of the dwarf galaxy's $V_{\rm max}$ for low- (black dotted line) and high-resolution runs (colored lines) for all three dwarf galaxy categories. We average over all four simulations shown in the main text and find almost perfect agreement between different resolution levels. Figure \ref{fig:gas_mass_res} and \ref{fig:star_mass_res} show the gas and stellar mass function for high- (colored lines) and low-resolution simullations (black lines). We find again almost perfect agreement between resolution levels when comparing \textit{satellites} (left panels, red lines) and \textit{field galaxies} respectively. For the \textit{nearby sample} we see some deviations at around $10^8\Msun$ in gas mass with $\sim 10$ more dwarf galaxies of this mass in the low-resolution run. On the other hand, we find a few more ($\sim3$) \textit{nearby galaxies} at gas masses of $\sim10^9\Msun$ in the high resolution runs. Looking at the stellar mass function diagram we find slightly more dwarf galaxies (4) with stellar masses of $\sim10^8\Msun$ in the high resolution run but almost perfect agreement at stellar masses below $\sim10^7\Msun$. Given the very good agreement for \textit{satellites} and \textit{fields} we ascribe these difference between resolution levels to a combination of different effects: slight shifts in formation times of the halos which cause the UV feedback to act differently on single haloes, the stochasticity of star formation which in turn effects the heating of the gas and its susceptibility to stripping and slight changes in the potential of the host which slightly effects dwarf galaxy orbits such that \textit{nearby galaxies} at the edge of the boundary between becoming a satellite or staying a field galaxy might be different in between resolution levels. However, it is hard to identify a unique cause for this discrepancy and we leave a detailed study for future work.

\section{Mass loss vs. Pericenter passages}
\label{sec:peri}
\begin{figure*}
\includegraphics[width=.49\columnwidth]{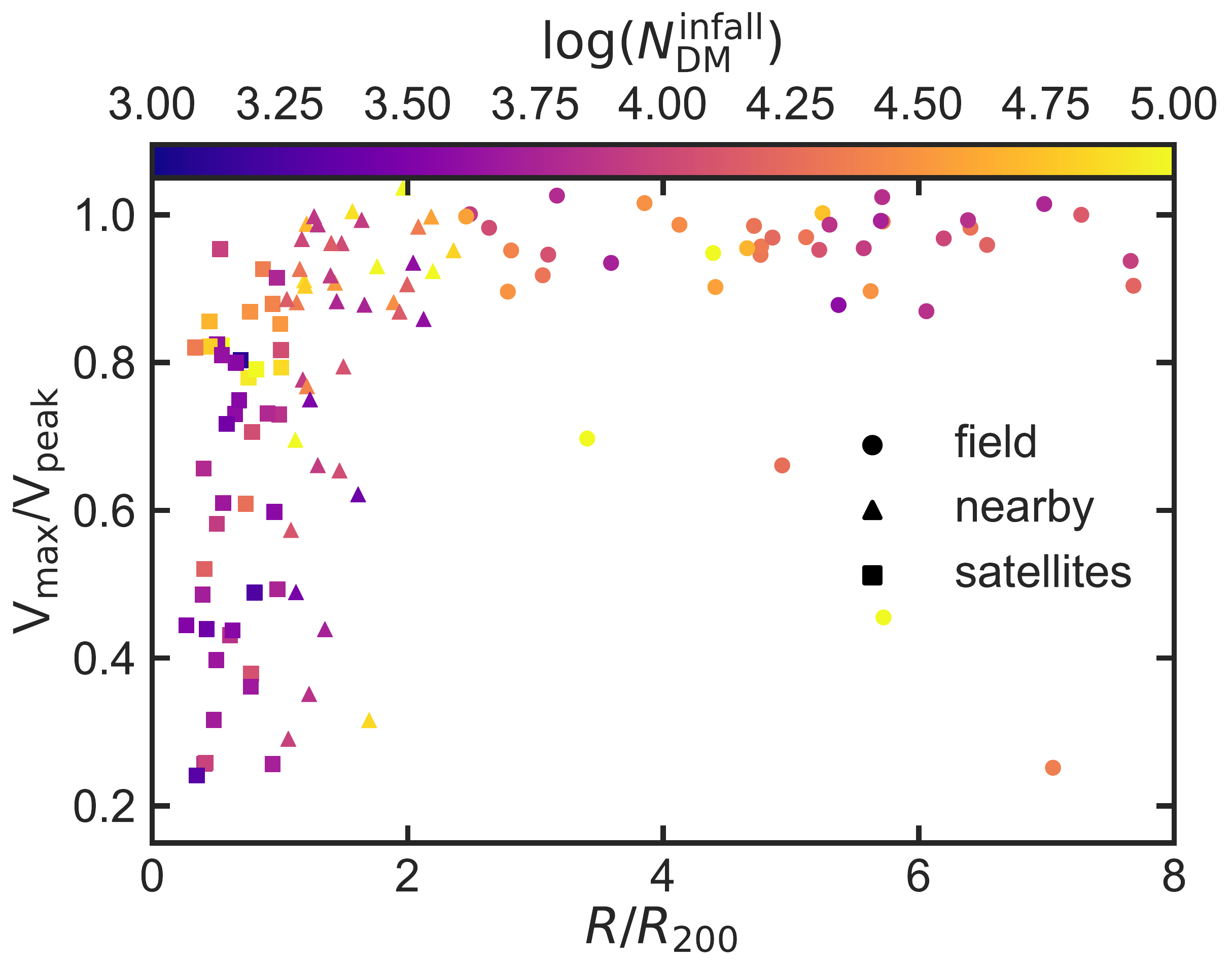}
\includegraphics[width=.49\columnwidth]{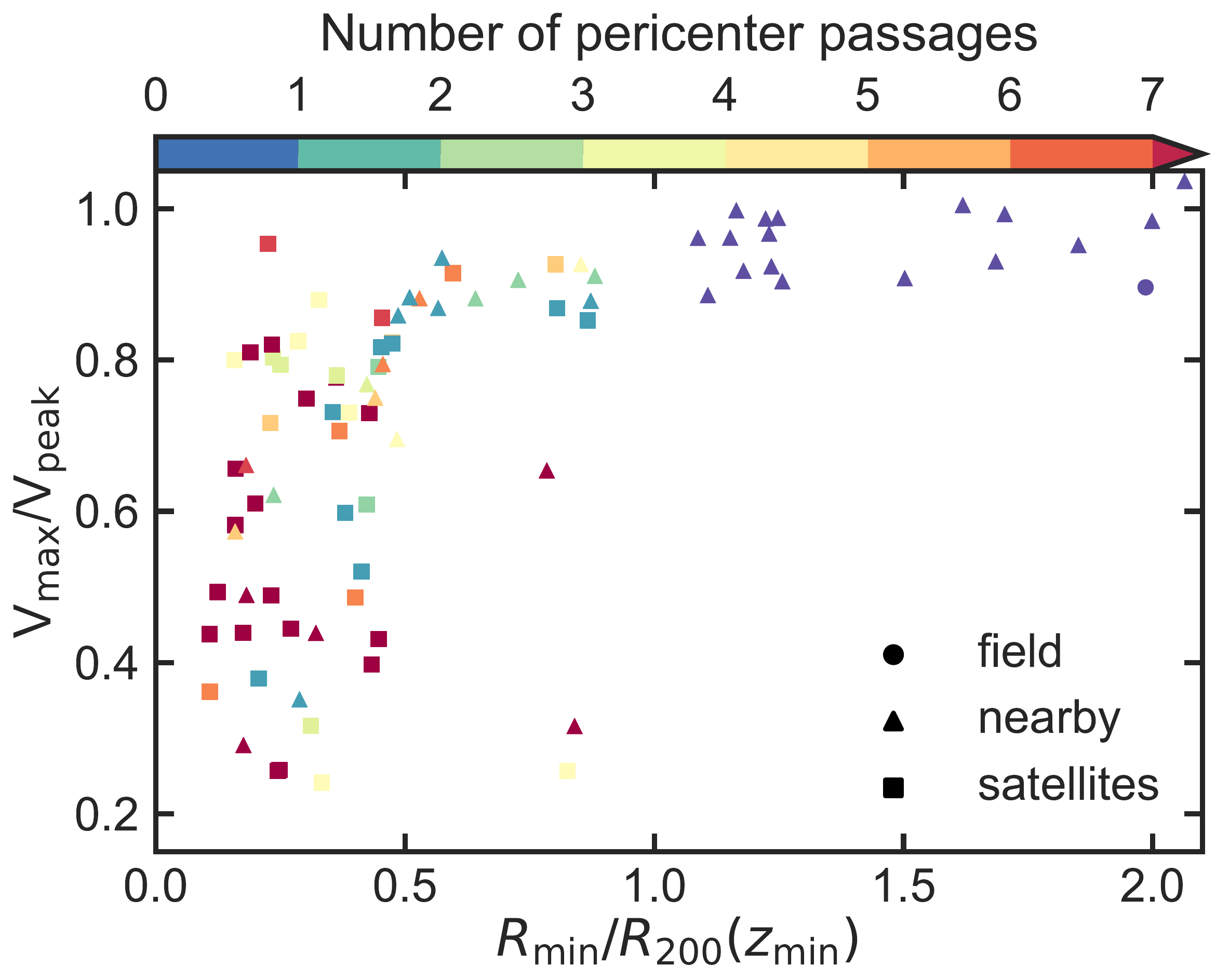}
\vspace{-.25cm}
\caption{\emph{Left panel:} Same as Fig. \ref{fig:mass_loss}. The ratio of the present day maximum of the rotation curve
  $V_{\rm max}$ and the maximum of the same ever reached during the
  lifetime of the galaxy $V_{\rm peak}$ as a function of the distance from the host at $z=0$ color coded by the number of particles at infall. \emph{Right panel:} The ratio of $V_{\rm max}$ and $V_{\rm peak}$ as a function of the minimum distance from the host color coded by the number of pericenter passages.}
\label{fig:res}
\end{figure*}

The left panel of figure \ref{fig:res} shows the ratio of  $V_{\rm max}$ and $V_{\rm peak}$ as a function of the distance from the host color coded by the number of dark matter particles at the time of infall. There is no dependence of the ratio $V_{\rm max}$ and $V_{\rm peak}$ on the number of particles at infall. The right panel of the same figure shows a slightly different version where the ratio of $V_{\rm max}$ and $V_{\rm peak}$ is plotted against the minimum distance of the galaxies ever reached color coded by the number of pericenter passages. This figure shows that satellites which lost the most mass via tidal stripping have done several ($>7$) pericenter passages and penetrate the host halo to smaller distances. Similar results are found for the backsplash galaxies. They have either had very close encounters with the host galaxy or had several pericenter passages. However, we find some deviations from this with either satellites loosing much mass although having only a small number of pericenter passages or the opposite, satellites having lots of pericenter passages but almost retaining all their mass loosing at most 20\%. We ascribe this to the importance of the individual orbits of the galaxies and their internal structure which was studied in depth by \citet{Frings2017} using much higher resolution satellites. These authors find that satellites with a central density core are more susceptible to mass loss orbiting the central galaxy and similar, satellites on more radial orbits loose much more mass in a short time compared to satellites on more gentle orbits.

\label{lastpage}
\end{document}